# The Pluto Energetic Particle Spectrometer Science Investigation (PEPSSI) on the New Horizons Mission

Ralph L. McNutt, Jr.[1,6], Stefano A. Livi[2], Reid S. Gurnee[1], Matthew E. Hill[1], Kim A. Cooper[1], G. Bruce Andrews[1], Edwin P. Keath[3], Stamatios M. Krimigis[1,4], Donald G. Mitchell[1], Barry Tossman[3], Fran Bagenal[5], John D. Boldt[1], Walter Bradley[3], William S. Devereux[1], George C. Ho[1], Stephen E. Jaskulek[1], Thomas W. LeFevere[1], Horace Malcom[1], Geoffrey A. Marcus[1], John R. Hayes[1], G. Ty Moore[1], Bruce D. Williams, Paul Wilson IV[3], L. E. Brown[1], M. Kusterer[1], J. Vandegriff[1]

[1]*The Johns Hopkins University Applied Physics Laboratory, 11100 Johns Hopkins Road, Laurel, MD 20723, USA*

[2]*Southwest Research Institute, 6220 Culebra Road, San Antonio, TX 78228, USA*

[3]*The Johns Hopkins University Applied Physics Laboratory, retired*

[4] *Academy of Athens, 28 Panapistimiou, 10679 Athens, Greece*

[5]*The University of Colorado, Boulder, CO 80309, USA*

[6]443-778-5435, 443-778-0386, Ralph.mcnutt@jhuapl.edu

The Pluto Energetic Particle Spectrometer Science Investigation (PEPSSI) comprises the hardware and accompanying science investigation on the New Horizons spacecraft to measure pick-up ions from Pluto's outgassing atmosphere. To the extent that Pluto retains its characteristics similar to those of a "heavy comet" as detected in stellar occultations since the early 1980s, these measurements will characterize the neutral atmosphere of Pluto while providing a consistency check on the atmospheric escape rate at the encounter epoch with that deduced from the atmospheric structure at lower altitudes by the ALICE, REX, and SWAP experiments on New Horizons. In addition, PEPSSI will characterize any extended ionosphere and solar wind interaction while also characterizing the energetic particle environment of Pluto, Charon, and their associated system. First proposed for development for the Pluto Express mission in September 1993, what became the PEPSSI instrument went through a number of development stages to meet the requirements of such an instrument for a mission to Pluto while minimizing the required spacecraft resources. The PEPSSI instrument provides for measurements of ions (with compositional information) and electrons from 10s of keV to ~1 MeV in a 120° x 12° fan-shaped beam in six sectors for 1.5 kg and ~2.5 W.

## Abbreviations

1PPS - One Pulse Per Second

ADC – Analog-to-digital converter

APL – Applied Physics Laboratory

ASIC – Application specific integrated circuit

C&DH - Command and Data Handling

CCSDS - Consultative Committee for Space Data Systems

CFD - Constant Fraction Discriminator

CSA - Charge Sensitive Amplifier

eV- Electron Volt

FITS – Flexible Image Transport System

FOV - Field of View

FWHM - Full Width Half Maximum

GSE – Ground support equipment

GSFC - Goddard Space Flight Center

HDU – Header Data Unit

HV - High Voltage Section of HVPS

HVPS - High Voltage Power Supply (HV and Bias Supply Sections)

IEM - Integrated Electronics Module

IGSE - Instrument Ground Support Equipment

ICD – Interface Control Document

ITF - Instrument Transfer Frame

LED – Leading Edge Discriminator

MCP – Micro-channel plate

MIDL – Mission Independent Data Layer

MDM – Master Data Manager

MET - Mission Elapsed Time

MOI - Moment of inertia

NA - Not applicable

NASA – National Aeronautics and Space Administration

NH – New Horizons

ns – nanosecond = $10^{-9}$ s

PDS – Planetary Data System

PEPSSI - Pluto Energetic Particle Spectrometer Science Investigation

PFF - Pluto Fast Flyby

PHA - Pulse height analysis

ps – picosecond = $10^{-12}$ s

psi - Pounds per square inch

RTG – Radioisotope Thermoelectric Generator

SQL – Structured Query Language

SSD – Solid-state detector

SSR – Solid-state recorder

STP – Supplemented Telemetry Packet

SwRI - Southwest Research Institute

TDC - Time-to-digital chip

TOF - Time of flight

TRIO – Temperature remote input/output

T-V Thermal-vacuum

UART - Universal asynchronous receive and transmit

# 1 Introduction

The Pluto Energetic Particle Spectrometer Science Investigation (PEPSSI) is one of seven scientific instruments/experiments (Weaver *et al.* 2007) on board the New Horizons spacecraft (Fountain *et al.* 2007), now on its way to Pluto (Stern 2007). While it is doubtful that Pluto has an intrinsic magnetic field and magnetospshere that accelerates charged particles to high energies, Pluto does have (or has had in the very recent past) a substantial atmosphere (Brosch 1995; Elliot *et al.* 2003; Elliot *et al.* 1989; Elliot *et al.* 2007; Sicardy *et al.* 2003) that is escaping into the solar wind in a comet-like interaction (Bagenal *et al.* 1997; Bagenal & McNutt 1989; Delamere & Bagenal 2004; Kecskemety & Cravens 1993; Krasnopolsky 1999; McNutt 1989; Tian & Toon 2005; Trafton *et al.* 1997). Measured interactions at comets show that the outgassing cometary neutral atoms and molecules charge-exchange with the solar wind and are accelerated in the process (Coates *et al.* 1993a; Coates *et al.* 1993b; Galeev 1987; Galeev *et al.* 1985; Huddleston *et al.* 1993; Mendis *et al.* 1986; Motschmann & Glassmeier 1993; Neugebauer 1990). By measuring the in situ energetic particle population, identifying those ions from the emitting body, and noting their variation with distance to the emitting body, the outgassing source strength may be deduced (Gloeckler *et al.* 1986); it is also important to note that energization beyond what one would naively expect from pick-up alone is also observed at comets (McKenna- Lawlor *et al.* 1986; Richardson *et al.* 1986; Sanderson *et al.* 1986; Somogyi *et al.* 1986). For example, shock acceleration of particles at Venus can elevate some of the particles to substantial (~100 keV) energies (Williams *et al.* 1991).

Making these measurements to determine the "outer boundary" of the influence of Pluto's atmosphere is the primary objective of the PEPSSI instrument. The extent of the interaction with the solar wind will be determined by comparing the PEPSSI measurements with those obtained of the solar wind by the SWAP instrument (McComas *et al.* 2007). A deduced atmospheric profile from the surface to the edge of Pluto's atmosphere will be assembled from combining PEPSSI, the New Horizons ultraviolet imaging spectrometer Alice (Stern *et al.* 2007) and the New Horizons radio experiment (REX) (Tyler *et al.* 2007) measurements.

PEPSSI combines energy and time-of-flight measurements in a low-mass (1490g), low-power (2.3 W) unit. At the same time, the instrument has a relatively large total geometric factor of ~0.1 $cm^2$ ster and enables directional information of the particle distribution across a ~12° × 160° swath in six ~25°-wide angular bins. The instrument also discriminates between electrons and ions without the use of (relatively) heavy permanent magnets (or power-hungry electromagnets). While the use of silicon solid state detectors (SSDs) for measuring energetic particles date back to the near-beginning of the space program, the additional species discrimination made possible by including time-of-flight measurements (thus independently, but simultaneously, measuring both an ion energy and speed) is a relatively newer development. Further, packaging of this capability into such a compact instrument providing a variety of engineering and programmatic challenges. This story leading to PEPSSI is of sufficient import that we summarize it here.

**1.1. Beginnings**

During the 1990s, the Pluto Fast Flyby (PFF) had evolved into the Pluto Express (PE) mission but continued to be under active consideration as the next outer planets mission by the Solar System Exploration Division (Stern 2007). As early as 26 Feb. 1993, the Space Physics Subcommittee adopted a recommendation that noted during the Pluto flyby itself "Data on fields and particles are essential to the understanding of atmospheric scavenging, surface darkening, and to important inferences of internal structure." They noted that in addition, en route to Pluto, fields and particles instruments could provide unique new information on the solar wind and a baseline for termination shock studies with minimal impact on mission resources. They further noted that the data would be central to the investigation of heliospheric structure, as the Pluto flyby trajectory would be almost aligned with the symmetry axis of the interstellar wind and could a gap between the trajectories of the Pioneer and Voyager spacecraft.

A Science Definition Team (SDT) for Space Physics Objectives for the Pluto Fast Flyby Mission (SPOPFFM) (Neugebauer *et al.* 1993) was established by Dr. George Withbroe to establish science objectives for both cruise (heliospheric) science as well as particles and fields science during a Pluto-Charon flyby, establish a conceptual science payload, and report out from NASA's Space Physics Division on the prospects for scientific contributions that would accrue to the Pluto mission from such additions.

The SDT noted that any such instrumentation must fit into a very small mass and power profile for the mission concept under discussion at that time; nominal values of 1 kg and 1 W for all fields and particles investigations (as a package) were discussed. Combinations of miniaturized plasma, suprathermal particle and energetic particles sensors were considered along with a plasma wave instrument and a magnetometer. Although a realistic choice of instrumentation had to be based upon maximizing the science return while minimizing the impact on the already tightly constrained spacecraft resources, the overriding constraints of low mass and low power remain were recognized as the principal design drivers.

These initiatives in the space physics community, along with a good showing of space physics topics at the Pluto/Charon meeting in Flagstaff in July 1993 led to the inclusion of Dr. Ralph McNutt on the Science Definition Team for what was then Pluto Express, chaired by Professor Jonathan Lunine. Meetings of the SDT were held in Tucson 5–7 Apr 1995 and 21–23 Jun 1995. The SDT report (Lunine *et al.* 1995), while documenting the potentially unique interaction between Pluto's escaping atmosphere and the solar wind, noted that there were no in situ instruments for making such measurements included as part of the baseline mission, although such a payload was mentioned as part of a potential international collaboration.

Subsequent evolution of the idea into what finally became New Horizons is discussed in the introductory paper in this volume. Suffice it to say, the push for instrumentation on a dedicated Pluto mission to directly measure the in situ environment there began its concerted push toward what became the PEPSSI and SWAP (McComas *et al.* 2007) investigations began in 1992. The rest is history.

## 1.2. Previous Similar Instrumentation

PEPSSI combines time-of-flight (TOF) and total-energy measurements in six angular sectors packaged into a compact arrangement (Fig. 1). Unlike plasma instruments incorporating a TOF system at lower energies, no pre-acceleration, high-voltages ($\geq$ 10s of kilovolts, kV) are used in these devices.

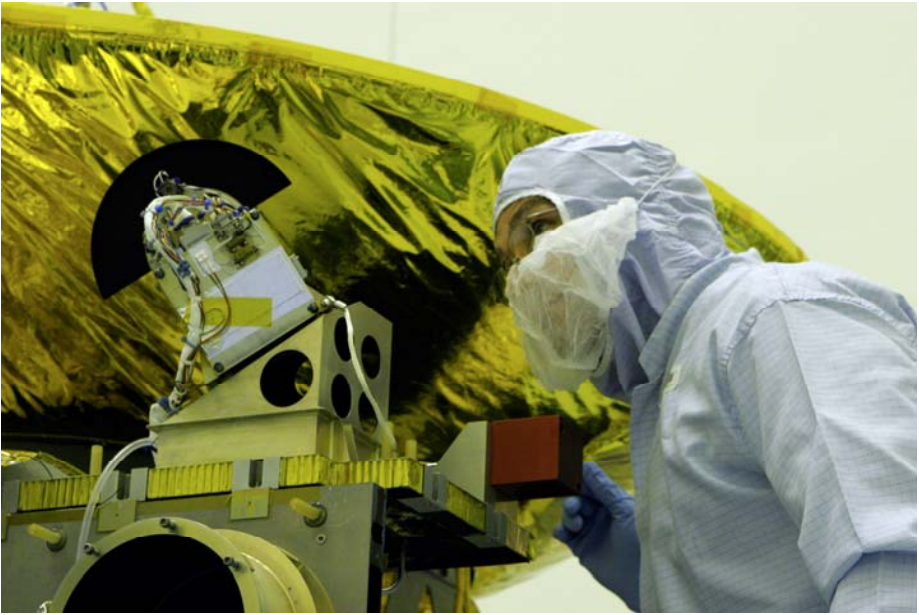

**Fig. 1.** PEPSSI mounted on the New Horizons spacecraft via its complex-angle bracket in the clean room at APL prior to being covered with its thermal blanket.

The first such instrument developed at APL was the Ion Composition Telescope (ICT) on the Firewheel satellite (1105 kg) that was launched aboard the second Ariane test flight by the European Space Agency (ESA) (De Amicis 1988). The ICT was a combined tine-of-flight (TOF) and total energy ion spectrometer to operate over the range > 15 keV/nuc to >500 keV/nuc (the satellite also carried four lithium canisters and eight barium canisters that would release their contents in space). The ICT mass was 8.5 kg, power 5 W, and nominal data rate 180 bps.

The satellite was launched 23 May 1980; but a fuel system fault resulted in failure of all four first stage engines, and the vehicle fell into the South Atlantic after launching from Kourou, French Guiana. The design was the basis for the Medium-Energy Particle analyzer (MEPA) instrument on the AMPTE CCE (Active Magnetosphere Particle Tracer Experiment Charge Composition Explorer) spacecraft (McEntire *et al.* 1985).

MEPA was designed to measure spectra and composition of magnetospheric particles from ~10 keV per nucleon (oxygen) to ~6 MeV. A similar time-of-flight scheme was incorporated into the Energetic Particles Detector (EPD) on the Galileo spacecraft (Williams *et al.* 1992) and into the Magnetospheric Imaging Instrument (MIMI) now operating at Saturn as part of the Cassini spacecraft payload (Krimigis *et al.* 2004). By including a "front-end" electrostatic analyzer and post-acceleration (after that analysis section), the technique can also be applied to lower-energy particles (Gloeckler *et al.* 1985).

Typically, sweeping magnets have been used to eliminate electron signals in configurations without a "front end" electrostatic analyzer. On the Ulysses Heliosphere Instrument for Spectra, Composition, and Anisotropy at Low Energies (HI-SCALE) experiment both a sweeping magnet and foil spectrometer technique are employed (Lanzerotti *et al.* 1992). In the latter, an aluminized parylene foil is used to accept electrons and reject ions at one of the detectors. This is the same approach used on PEPSSI as well as on the energetic Particle Spectrometer (EPS) on the MESSENGER spacecraft. This approach saves the mass and complexity of a separate electron detector, or of a magnet.

A similar approach has been selected for comparable instruments selected for inclusion on the Juno New Frontiers mission to Jupiter and the Radiation Belt Storm Probes mission at Earth.

The MESSENGER EPS and New Horizons PEPSSI instruments have been known as "the hockey puck" due to the cylindrical aperture/detector/TOF section that has about the same proportions and size as a regulation ice hockey puck. The original work on such a miniaturized sensor was begun at APL under internal research and development funds. Following the Space Physics SDT report, a proposal was submitted 7 Sep 1993 in response to NRA-93-OSS-01 (*Space Physics Supporting Research and Technology and Suborbital Program*) entitled "An Integrated Compact Particle Detector for a Mission to Pluto." Rated "Very Good" overall, the proposal was not funded (The proposal could not respond to NRA 93-OSSA-5 issued to explore the instrumentation for the strawman payload on the Pluto Fast Flyby mission because a particle instrument was not included in the strawman payload).

The next year, a revised proposal was submitted to the Planetary Instrument Definition and Development Program (PIDDP) under NASA NRA 94-OSS-11. The proposal "A Compact Particle Detector" was submitted 6 Oct 1994 and targeted to a PFF new start authorization in FY98. Requested funding was for 1 Jan 1995 through 31 Dec 1997. In the abstract of the proposal are found the words: "In accord with the recommendations of the Science Definition Team for Space Physics Objectives for the Pluto Fast Flyby Mission (SDTSPOPFFM – M. Neugebauer, chair) that met August 1993, the instrument concept to be developed is a Suprathermal Particles Sensor that would be one part of a Combined Particle Sensor. The CPS was recommended as the highest priority particles and fields instrument for PFF."

The same basic proposal as to the Space Physics Division was used; but with a block diagram and the idea of a fan of six solid state detectors (SSDs) added. The instrument was called the Pluto Ion Composition Analyzer or PICA. The technical specifications were <0.5 kg, <0.5 W; particles ~15 to 20 keV to ~3 MeV including ion spectra and composition, electrons, and neutrals. Possible applications presented were Pluto Express (PFF was now history), Solar Probe, Interstellar probe, and Jupiter MEASURE. Development work under this program led to the inclusion of the instrument in the first MESSENGER proposal in the discovery program under NASA AO-96-OSS-02.

The NASA proposal system transitioned over to the Research Opportunities in Space Science (ROSS) system as a broad agency announcement as defined in FAR 6.102 (d) (2) with the 1998 release on 5 February 1998. In this new scheme, the PIDDP program was element A.3.5 with a due date of 3 Aug 1998.

A new PIDDP proposal entitled "Miniaturized Energetic Particle Development" was submitted to NASA 31 Jul 1998. The basic design was to be carried forward from the previous PIDDP work – the energy chip design was characterized as "almost complete" for a single channel per chip – although this never materialized as flight hardware for a variety of reasons. Other electronics development topics were also pursued. Little from this effort was available for incorporation into the MESSENGER flight hardware given the timing of MESSENGER's selection for flight (McNutt *et al.* 2006), but some improvements were possible for PEPSSI due to work under this second grant prior to the selection of New Horizons for flight (Andrews *et al.* 1998; McNutt *et al.* 1996). As had been predicted in the initial community 1992 work, the most difficult aspect was the development of a scientifically useful energetic particle instrument with the small mass and power limits that had to be met to allow for inclusion on a Pluto mission.

## 2   Scientific Background and Objectives

### 2.1.   The Interaction of Pluto with the Solar Wind

The interaction of the various planets of the solar system with the solar wind goes back to the initial forays outside of the Earth's magnetosphere in the early 1960s. As part of the studies that eventually led to the Voyager program (Dryer *et al.* 1973) made an initial scoping study of what a solar wind interaction with Pluto might look like. They noted that for the anticipated scale lengths a kinetic approach was more proper for Pluto as for Mercury and that for a vanishingly small ionospheric scale height at Pluto, a long, induced magnetotail is, nonetheless, expected. Anticipating the then-future Voyager encounter with Neptune and its large moon Triton (McNutt 1982–4) wrote: "The outermost known planet in the solar system, Pluto, is similar in  size and spectral signature to Triton [77]. Pluto apparently has a tenuous methane atmosphere [78] and, like the other medium-sized moons in the solar system, probably has no intrinsic magnetization. As there are no plans for a spacecraft flyby of Pluto in this century,

Voyager observations of Triton and its interaction with the solar wind and/or a Neptunian magnetosphere, will, for the foreseeable future, provide our best guesses for the interaction of Pluto with its plasma environment." Here references [77] and [78] refer to (Morrison *et al.* 1982) and (Fink *et al.* 1980), respectively.

There is no more work in the published literature for several years. On 17 Jan 1989 F. Bagenal sent an E-mail message to R. McNutt: "Alan Stern is organizing an AGU session this spring on the Pluto-Charon system. He asked me if anyone had thought about Pluto's magnetosphere. He said he could not find anything in the published literature. He urged me to submit an abstract. Well, I guess I could do a simple standoff calculation using the latest atmospheric measurements (which Alan has made, I think). Have you thought about Pluto lately? I remember you kept going on about Pluto in the good old days. Perhaps we could through something together next week–unless you have already done so!!!"

This began a collaboration between Bagenal and McNutt on Pluto's escaping atmosphere and how it would interact with the solar wind. Current thinking is that the interaction of Pluto with the solar wind is something between that of an unmagetized planet, such as Venus and that of a comet, depending upon the strength of the atmospheric outflow. If Pluto were to have even a weak intrinsic magnetization, then the interaction would be more akin to that of the magnetized planets due to the weak solar wind ram pressure at ~30 AU and beyond. Estimates of the overall outgassing rate of the atmosphere $Q_0$ are between ~$10^{27}$ and $10^{28}$ molecules s$^{-1}$. Such rates have effects ranging from just shielding the surface from the solar wind to producing a well-formed magnetosphere encompassing the the orbit of Charon (Fig. 2). Detailed discussions of past and current thinking about the interaction is given by (Bagenal *et al.* 1997) and (McComas *et al.* 2007), respectively, and more detailed simulations have been carried out as well (Delamere & Bagenal 2004; Harnett *et al.* 2005).

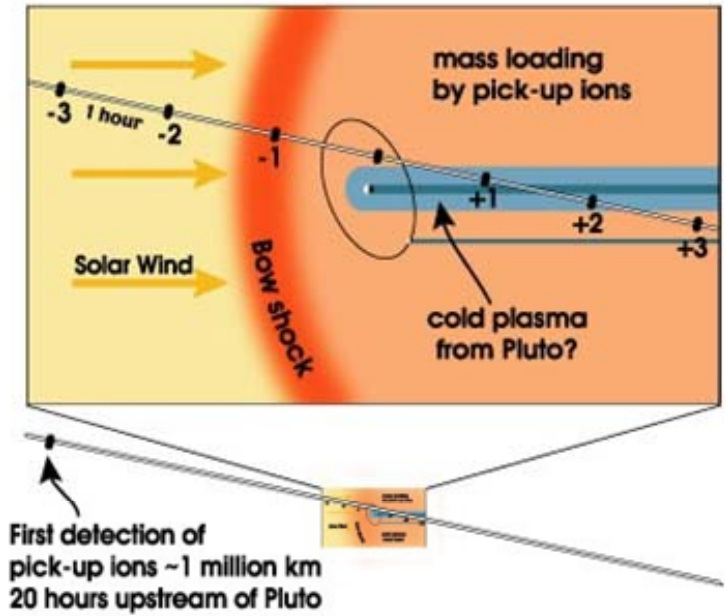

**Fig. 2.** Schematic of expected interaction of Pluto with the solar wind for the strong-interaction limit. Based upon scalings from cometary interactions, the first detection of pickup ions from Pluto are expected as early as ~20 hours prior to closest approach.

## 2.2. PEPSSI Science Objectives

The PEPSSI sensor is designed to perform *in situ* measurements of the mass, energy spectra, and distributions of moderately energetic particles in the near-Pluto environment and in the Pluto-interaction region. The instrument measures particle velocity and energy, derives particle mass, and discriminates between electrons, protons, alphas, and carbon-nitrogen-oxygen (CNO – taken as a closely-spaced group in atomic weight), and heavier ions. The direction of particles is also discerned. PEPSSI objectives, within the context of New Horizons science mission group objectives include:

*Group 1 Objective*. A group 1 objective is characterization of the neutral atmosphere of Pluto and its escape rate. To support this objective, PEPSSI will detect heavy ions and measure associated energy spectra and spatial variation along the trajectory. By analogy with cometary measurements, these measurements will be used to determine the neutral particle escape rate, which along with UV spectral measurements made in the upper atmosphere (Stern *et al.* 2007), will be used to put together a fully self-consistent model of Pluto's upper atmosphere to satisfy this group 1 objective.

*Group 2 Objectives*. A group 2 science objective is characterization of Pluto's ionosphere and interaction with the solar wind. This characterization will be aided by PEPSSI measurements of the spatial extent and composition of pickup ions; these measurements are complementary to those that will be made by SWAP (McComas *et al.* 2007).

*Group 3 Objectives*. A group 3 science objective is characterization of the energetic particle environment of Pluto and Charon. This will require measurement of the spatial extent and velocity-space distributions of energetic ions (e.g., $H^+$, $N^+$, and $N_2^+$). The PEPSSI instrument will make the required energetic ion measurements.

## 2.3. Measurement Requirements

2.3.1  Measurement Ranges

Energy thresholds and energy ranges depend upon the energy measurement mode, i.e, TOF-only, energy (SSD) only, or coincidence measurements through the entire system. The ranges as a function of species and the mode are given in Table I.

Table I.  Energy Measurement Ranges for PEPSSI

| Species | Energy measurement range | | |
| --- | --- | --- | --- |
| | Energy + TOF measure | Energy - only measure | TOF - only measure |
| Energetic electrons | Not applicable | 25 keV to 500 keV | Not applicable |
| Protons | 25 keV to 1 MeV | Not applicable | 700 eV |
| Atomic ions, e.g., CNO group, $Mg^+$, $Si^+$, $Ne^+$ | 60 keV to 1 MeV | Not applicable | 15 keV |
| Molecular ions e.g., $N_2$, $O_2$ | 100 keV to 1 MeV | Not applicable | 30 keV |

2.3.2 Derived Instrument Specifications

*2.3.2.1 Mass Resolution (Mass Uncertainty)*

Particle mass is derived from energy and TOF measurements. The uncertainty in the derived mass, i.e., the mass resolution, is determined by (a) energy measurement resolution, (b) TOF measurement resolution, (c) particle mass, and (d) calibration accuracy. For Energy-plus-TOF measurements, the mass resolution for three species of particles (spanning light, medium, and heavy mass) is specified as <2 atomic mass units (AMU) for H+ (25 keV to 1 MeV), < 5 AMU for C+/N+/O+ (60 keV to 1 MeV), and <15 AMU for Fe+ (60 keV to 1 MeV).

For TOF-Only measurements, the means to ascertain particle mass is less precise, and this requirement, with respect to mass resolution, is to distinguish between $H^+$ and CNO group particles. To support derivation of species mass, for particle energies in the 700 eV to 1 MeV range, PEPSSI was specified to be capable of measuring particle TOF over a range of 1 to 320 ns.

*2.3.2.2 Species Mass Range.*

The PEPSSI instrument is constrained in downlink capability from Pluto as well as in the mass and power available for the instrument. Hence, prudent choices had to be made to meet all of the constraints while still enabling the collection of appropriate data from the vicinity of Pluto and its transmission to Earth following the flyby. Species resolution for the various energy spectra is limited by the counting statistics and the physical size of the detector that limits the TOF drift space. To enable the discrimination of solar wind particles (primarily protons and alpha particles, i.e. doubly-ionized helium nuclei) from pickup particles from Pluto, including atomic "debris" as well as ionized molecules of nitrogen, methane, and carbon monoxide, and allow for discovery science within the confines of the requirements, energy spectra are output for proton events, electron events, CNO events, heavy particle events (> 24 AMU, typified by Fe).

*2.3.2.3 Sensitivity and Geometric Factor Requirements.*

With PEPSSI mounted on the spacecraft, including installation of the RTG power source, and with the PEPSSI covers closed, the background ion particle count rate was specified not to exceed one particle per second. Expected fluxes at Pluto are relatively low (~100 events per second), so to stay within low power limits of operation, the PEPSSI instrument was specified to be capable of processing at least $10^3$ particle events per second, where this event rate is applicable to the total of all classes of measurements, i.e., Energy-plus-TOF, Energy-Only, and TOF-Only. The PEPSSI instrument has the potential to measure particle events at a much higher rate; this rate should be established once the analysis of data from the Jupiter flyby is fully analyzed.

*2.3.2.4 Geometric Factor*

Expected count rates at Pluto are unknown but expected to be low. Hence, the geometric factor was required to be as large as possible, consistent with the targeted low mass of the instrument of ~1 kg.

On the basis of these trades, The PEPSSI geometric factors, for electron and ion detection, were specified to meet or exceed the values given in Table II. The geometric factors for electron and ion detection are different because of the difference in numbers of ion and electron detectors. The values that follow apply to the entire aperture acceptance angle of 160° by 12°, i.e. the geometric factor per "pixel" is less.

**Table II. PEPSSI Sensor Specifications**

| | |
|---|---|
| Ion detection geometric factor | $\geq 0.1$ cm$^2$ steradian |
| Electron detection geometric factor | $\geq 0.033$ cm$^2$–steradian |
| Acceptance angle | 160° by 12°, 6 sectors of 25° by 12° each, 2° gaps between sectors |
| Aperture area | 0.6 cm × 1.2 cm per sector |
| TOF length | 6 cm nominal between entry, exit foils |
| Number of Detectors Per Sector | 2 |
| Detector Area | > 0.4 cm$^2$, ion and electron detectors |
| Number of Ion Detectors | 9 |
| Number of Electron Detectors | 3 (located in sectors 1,3,6) |

*2.3.2.5 Integration Interval.*

Nominally, energy-plus-TOF measurements, used to determine particle species and associated energy spectra, are integrated over a 10-second interval (based upon consideration of spacecraft speed, and hence spatial resolution, telemetry rates and data volume playback during the Pluto encounter). TOF measurements, used to determine particle velocity distribution, are integrated over the identical time interval. By command, the integration interval may be adjusted from 1 to 7200 seconds.

### 2.3.3 Measurement Resolution Requirements

These specifications flow, in turn, to the next level of implementation requirements that drove the system design and implementation.

#### *2.3.3.1 Energy Resolution*

As a goal, instrument energy measurement resolution, which includes the effects of all noise sources including analog-to digital converter (ADC) quantization noise, was 5 keV full-width at half maximum (FWHM) or less. As a requirement, instrument energy measurement resolution for ions was ~7 keV or better, for electrons 8 keV or better; ADC quantization is equivalent to 1 keV. The Energy resolution is determined to a large degree by the performance of the energy-peak detector chips for the given SSD detector capacitance and leakage currents. The energy resolution of the energy-peak detector chips is a trade-off between power dissipation, mainly in the charge sensitive amplifier, and integration time of the shaper. Both parameters are fixed on the energy board with two resistors (common for all 12 channels) as described in more detail in the electronics section.

The overall energy resolution is also determined by the digital noise (mainly the ADC noise) on the energy board. The clock frequency of the temperature remote input/output (TRIO) application specific integrated circuit (ASIC) chip for the ADC can be set comparatively low, to 150KHz, and, given the very low digital power dissipation of the chip (1 mw digital power), the system noise is within the specification levels.

#### *2.3.3.2 TOF Resolution*

Time measurement resolution, which includes the effects of time jitter, time walk, path dispersion, and time quantization, are specified as 1 ns FWHM or less.

The time resolution is dependent on the constant fraction discriminator (CFD) ASIC chip time walk and time jitter and the time jitter of the time to digital converter (TDC) ASIC chip. The CFD time walk is ~200–300ps for a 1:100 input energy dynamic range; the time jitter is in the range 50–100ps depending on the power dissipation. The time jitter of the TDC chip is ~50ps. Thus the overall electronics time resolution is < 500ps. The optimum is achieved with proper delay line selection and optimization of the power dissipation within the power budget limits.

### 2.3.4  Platform Requirements

For proper interpretation of the PEPSSI data, the instrument alignment with respect to the spacecraft and the knowledge of the spacecraft attitude are both required to 1.5°. For deciphering the measurements to be made at Pluto during the Pluto flyby, knowledge of spacecraft distance to Pluto is required to an accuracy of 225 km. This knowledge of position figure is equal to roughly 1/10 the diameter of Pluto, and is achieved with the spacecraft timing system and navigation of the spacecraft at the system level.

### 2.3.5  Time Resolution Requirement

#### *2.3.5.1  Timetagging Requirements*

For proper interpretation and processing of science data, knowledge of the data intervals over which PEPSSI science data is collected must be known to within to ±1 second of spacecraft mission elapsed time (MET). To this end, data packets sent from PEPSSI to the command and data handling (C&DH) system are time tagged with MET time to 1-second resolution (Fountain et al. 2007). Based on spacecraft time-keeping requirements, knowledge of MET time relative to UTC time to ~10 ms accuracy is known after the fact.

#### *2.3.5.2  Science Data Synchronization*

PEPSSI science data (species energy spectra, velocity, and pulse-height analysis (PHA)) are collected over fixed time intervals. These data intervals are synchronized by, and time aligned with one pulse-per-second (1PPS) timing epochs input from the spacecraft that are coincident with the MET one-second time increments.

2.3.6   Calibration Requirements

2.3.6.1  *Required Ground Calibration*

Calibration of the PEPSSI instrument is required in order to meet instrument performance specifications. Ground calibration tests were planned using both linear accelerator facilities at APL and Van De Graaff facilities located at GSFC. These facilities provide a calibrated source of electrons and a variety of ions (e.g., protons, helium, CNO, iron) at a variety of energies. Time and availability issues reduced the calibration plan to using the APL facility only. Calibration was performed with the PEPSSI instrument in vacuum at a temperature of 25°C for each sector. Calibration was performed prior to installation on the spacecraft. A calibrated alpha-particle source is installed in the collimate assembly and described below. This source was used to during thermal vacuum tests to verify instrument performance has not changed.

2.3.6.2  *In-flight Calibration Characterization*

Following launch, the PEPSSI instrument calibration must be characterized during initial instrument test and prior to all major data collecting operations. To characterize the instrument, background particle data was collected and downlinked to the PEPSSI payload operations center. Archived energy and TOF measurement data in the PHA data packet, as well as particle species data, have been processed to evaluate instrument calibration. To correct for some drift in measurements, new 'look-up' tables have been configured and uplinked. Further characterization is occurring during the close approach to Jupiter, using that planet's magnetospheric plasma as a calibration source. The characterization will be monitored during the yearly checkouts during the cruise to Pluto. The installed alpha-particle calibration source will also be monitored to look for any changes in instrument performance prior to arrival in Pluto-space in 2015.

2.3.6.3  *Energy Board Temperature Monitor*

Energy measurements vary slightly as a function of energy board component temperature. A temperature sensor is installed on the energy board, and this temperature data is telemetered in instrument housekeeping telemetry. Energy board temperature information allow the instrument ground processing software to account for any slight errors in energy measurement due to energy-board temperature variation.

# 3 Technical Description

## 3.1. Instrument Overview

Given the requirements, as well as ongoing lessons from the near-concurrent development of the EPS detector for the MESSENGER mission (Andrews et al. 2007), the required functionality was achieved within the prescribed constraints. These properties are given in Table III and Table IV. Some of the numbers are still being refined based upon the final analysis of instrument performance during the New Horizons flyby of Jupiter (28 Feb 2007).

PEPSSI consists of a collimator and sensor assembly, referred to as the sensor module, mounted atop an electronic board stack (Figs. 3 and 4). The electronic stack consists of six metal-framed electronic boards. The stack is a cube measuring approximately 10 cm on a side. The sensor module is approximately 2 cm high and protrudes out from the stack an additional 6 cm. A simplified block diagram of the PEPSSI instrument is shown in Fig. 5. The corresponding simplified schematic for operation is shown in Fig. 6.

As shown in the block diagram, the sensor module includes a time-of-flight (TOF) section about 6 cm long feeding a solid-state Si detector (SSD) array. The SSD array, connected to the energy board, measures particle energy. Secondary electrons, generated by ions passing through the entry and exit foils, are detected to measure ion TOF. Event energy and TOF measurements are combined to derive mass and to identify particle species.

## Table III. PEPSSI Performance Parameters

| | |
|---|---|
| Geometric Factor (Start Foil) | 0.15 cm$^2$ sr |
| Geometric Factor (each Electron Pixel) | 0.008 cm$^2$ sr |
| Geometric Factor (each small Ion Pixel) | 0.008 cm$^2$ sr |
| Geometric Factor (each large Ion Pixel) | 0.016 cm$^2$ sr |
| Electron Efficiency vs. Energy | Est. roughly at 100% at mid energies |
| Low Energy Ion Efficiency vs. En × M | > 4% details in Section 5 |
| High Energy Ion Efficiency vs En × M | > 4 % details in Section 5 |
| Energy Coverage | See Table Below |
| ΔE / E capability at 50 keV | 14 % |
| ΔE / E capability at > 100 keV | 10 % |
| Energy / TOF Channel resolution | 250 eV/167 ps Granularity |
| TOF dispersive spread | Expected 2.5 to 3.5 ns |
| Mass Species Separation | See Table Below |
| Angular Coverage (total) | 160° × 12° |
| Angular Coverage (Electrons) | 147° × 12° |
| Angular Coverage (Low Energy Ions) | 160° × 12° |
| Angular Coverage (High Energy Ions) | 147° × 12° |
| Angular Coverage (Diagnostic Ions) | 147° × 12° |
| Angular Pixels (Electrons) | 12° × 27° (13° with 13.6° gap) |
| Angular Pixels (Low Energy Ions) | 12° × 30° (full ~27° coverage) |
| Angular Pixels (High Energy Ions) | 12° × 27° (13° with 13.6° gap) |
| Angular Pixels (Diagnostic Ions) | 12° × 27° (13° with 13.6° gap) |
| Electron Scattering Angle Contrast | Several percent |//

| **Energy Coverage:** | | **Technique** |
|---|---|---|
| Electrons | 25 keV – 500 keV | Singles / Foil Technique |
| Low Energy Ions | 700 eV – 1 MeV | TOF Only |
| Low Energy CNO | 15 keV – 1 MeV | PH-based mass discrimination |
| High Energy Protons | 40 keV – 1 MeV | TOF vs. Energy |
| High Energy CNO | 100 keV – 1.2 MeV | TOF vs. Energy |

While PEPSSI uses the same type of energy and time-of-flight measurement scheme as the previous such instruments mentioned above, it also employs multiple (six) angular apertures in a swath in order to provide angular information without relying upon either a mechanical scanning mechanism or the rotation of the spacecraft. The PEPSSI acceptance angle is fan-like and measures 160° by 12° with six 25° segments. Each segment is separated by a 2° gap. Its total ion geometric figure is greater than 0.1 cm$^2$ sr. Particle direction is determined by the particular 25° sector in which it is detected. The angular resolution of the instrument was required to determine the incoming direction to better than 40° for the ions in each of the six sectors, i.e., within a 25° by 12° window. Electron particle direction is limited to sector 1, sector 3, and sector 6.

Table IV.  Final PEPSSI flight parameters and assessment against requirements

| Parameter | Requirement | Measured or derived from measurements | Comments |
|---|---|---|---|
| **Engineering Parameters** | | | |
| Mass | < 1.5 kg | 1.475 kg | Measured at delivery |
| Power Average | 2.55 W | 2.49 W | Not including heater power |
| Power Peak | same | 2.51 W | Additional power needed one time to deploy cover |
| Data Volume | < 135 bps | 91 bps | With FAST Compression |
| Size | < 1200 cm$^3$ | 352 cm$^3$ | Includes sunshade and mounting tabs / structures |
| Performance Parameters | | | |
| Geometric Factor | >0.1 cm$^2$ sr | 0.15 cm$^2$ sr | Calculated: PDR |
| Ion Energy Range | 15 keV/n to 1 MeV/n | 700 eV/n to 1 MeV/n | Lowest energies with TOF-only, as expected |
| Electron Energies | 25 to 500 keV | 25 to 500 keV | Not verified in calibration |
| Energy Resolution | 10 keV | <5 keV | |
| Species | H, He, CNO, e$^-$ | H, He, CNO, Fe, e$^-$ | |
| Field-of-View | 160° × 12° | 160° × 12° | |
| Angular Resolution | 25° × 12° | 25° × 12° | |
| TOF Range | 1-320 ns | 1-320 ns | Verified electronically |
| TOF Resolution | < 5 ns | < 4 ns (FWHM) | |
| Integration period | 10 s adjustable | 10 s default | Programmable from 1 to 7200 s |

The entrance apertures for the axially symmetric time-of-flight (TOF) section are 6 mm wide. Each aperature is covered by a 50Å aluminum/350Å polyimide/50Å aluminum foil. These foils reduce the TOF UV Lyman-alpha photon background. The exit apertures are covered by a 50Å palladium/500Å polyimide/50Å palladium foil. Both entrance and exit aperture foils are mounted on a high-transmittance stainless steel grid supported on a stainless steel frame.

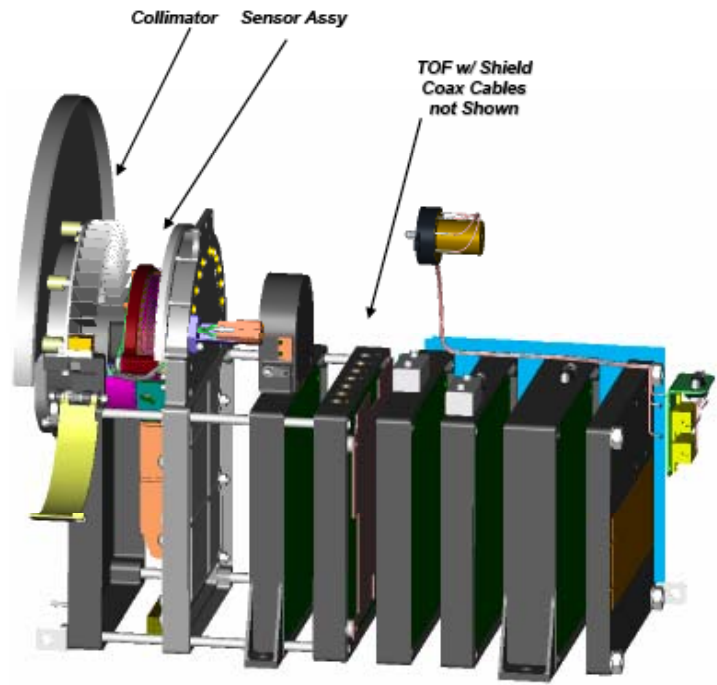

**Fig. 3.** Expanded view of the PEPSSI instrument showing, left to right: sunshade, collimator assembly (with acoustic doors open - flight configuration), sensor assembly, and electronics boards stack.

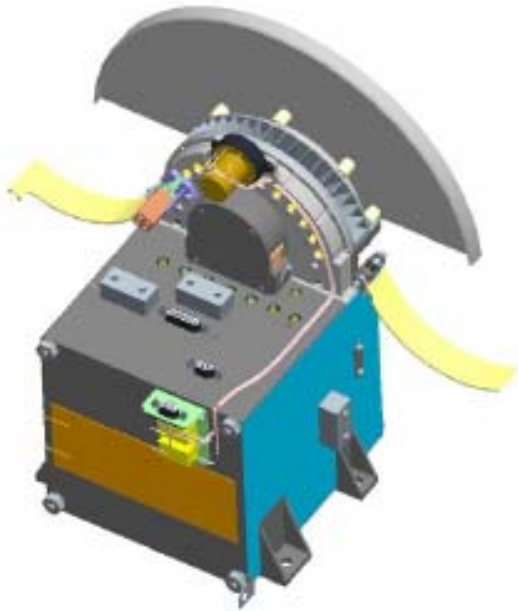

**Fig. 4.** Assembled view of the sensor, again in flight configuration.

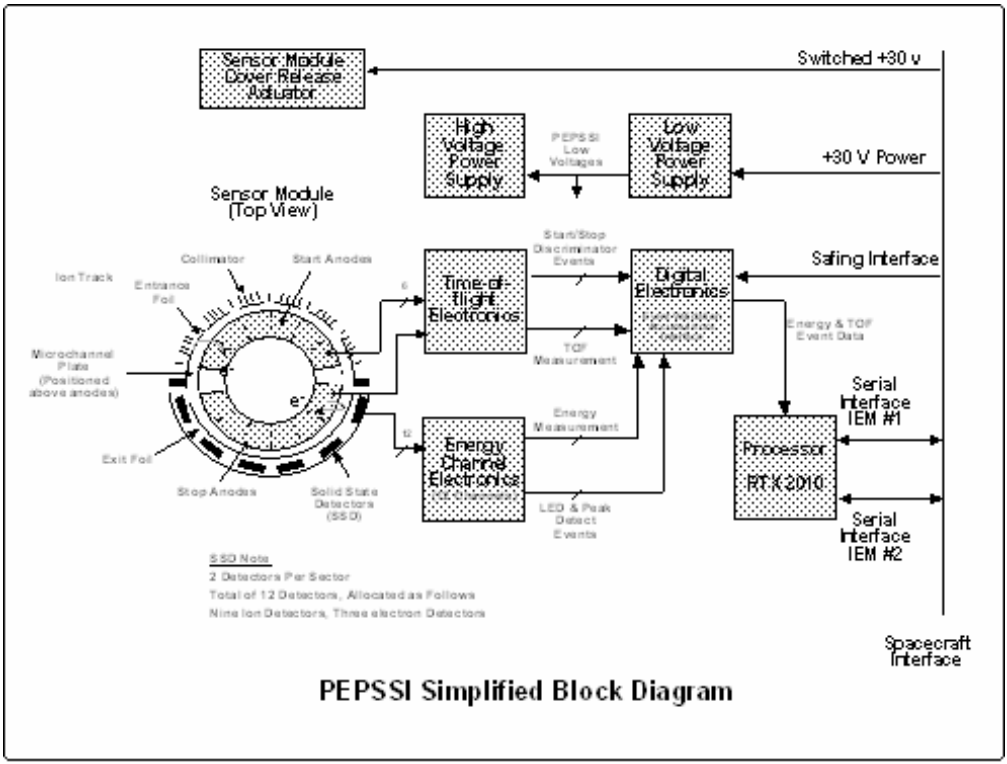

**Fig. 5.** Simplified block diagram of the PEPSSI instrument. Flow of information from the sensor to the spacecraft interface is shown.

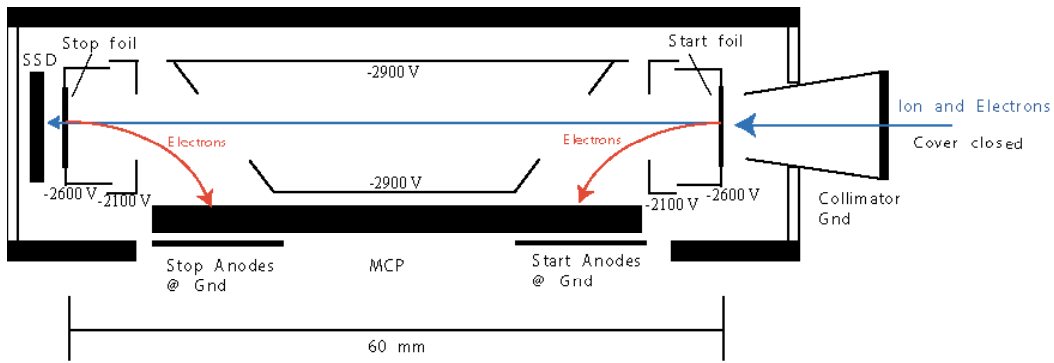

**Fig. 6.** Simplified operation of the PEPSSI sensor (1 of 6 shown). Incoming ions and electrons pass through the collimator assembly hitting the "start" foil and ejecting secondary electrons. The start electrons hit the start anode flagging the time-of-flight start. The incoming particles cross the 6-cm drift space and then hit the stop foil, again ejecting secondary electrons that hit the stop anode. The incoming particle then hits the SSD where its total energy is recorded. The electronics detect and classify both nominal events and valid "triples" (that have valid start, stop, and energy signals within appropriate electronic windows such that the incoming particle can be fully categorized.

### 3.1.1 Differences from EPS on MESSENGER

The MESSENGER EPS unit was the first flight unit of this basic design. PEPSSI presented different requirements with respect to available power and maximum count rates. While larger rates were anticipated for the encounter with Jupiter and could be used to good advantage for some calibration activities, the requirements were always focused at the conditions at Pluto. In addition, the launch of New Horizons over two years after that of MESSENGER allowed for some upgrades to be included in the PEPSSI unit.

The combination of lower expected counting rates and the need to save power led to the implementation of (1) lower-current MCPs, (2) a new low-voltage power supply (LVPS) with a complete custom converter design to save power, (3) a complete redesign of the energy board using new application-specific integrated circuit (ASIC) chips to save power, and (4) a redesign of the time-of-flight (TOF) board with different amplifiers and delay lines to achieve higher gain.

With the need for long-term reliability during the cruise to Pluto, PEPSSI added a high – voltage (HV) safing system, which is always active. This system samples the HV current 1000 times per second, and shuts down the supply if the HV current exceeds a threshold. A different HVPS driver is used on PEPPSI that provides improved performance (not incorporated on PEPSSI due to the MESSENGER launch schedule). The PEPSSI unit also incorporates a buffer amplifier and a filter, neither of which are present in the EPS unit.

The EPS instrument has 24 detectors (six large for ions, six large for electrons and the same counts with small ones), whereas PEPSSI uses only 12, and on its detectors ties the inside one to the outside one, rather than all being independent, as on EPS which has six processing chains for ions and six for electrons, with large or small detectors selected by ground command.

The front – end collimator is more open on PEPSSI than on EPS to increase the geometric factor and PEPSSI also has an added sunshade to be able to look back near the solar direction (EPS is mounted well away from the Sun-looking direction on the MESSENGER spacecraft to keep the temperature down and avoid spurious signals from the eleven Sun solar luminosity at Mercury's perihelion.

EPS is thermally isolated from the MESSENGER spacecraft with a separate thermal radiator to reject internal heat to space, while PEPSSI has no radiator, but is quasi-thermally coupled to the spacecraft mounting bracket (conductively isolated but radiatively coupled). Furthermore, the EPS thermal design uses heaters and mechanical themostats to keep the instrument from getting too cold, while PEPSSI relies on the spacecraft to keep it just warm enough to not need heater power.

**3.2. Mechanical Design**

Top-level mechanical design requirements and design factors of safety are in accord with those used throughout the New Horizons project. In particular, the PEPSSI instrument was designed to withstand a quasi-static load limit of 30 g along 3 orthogonal axes and (applicable to the primary structure design and as multiplied by the appropriate factor of safety). In addition, first mode structural frequencies were specified to be above 70 Hz in the spacecraft thrust direction and above 50 Hz in the lateral directions and the instrument designed to withstand maximum pressure rate change of 1.0 psi/sec. All of these specifications were verified during environmental tests.

3.2.1 Dimensions and Mounting

The PEPSSI envelope is 19.7 cm × 14.7 cm × 21.6 cm as installed and 25.1 cm 14.7 cm × 21.6 cm following the one-time opening of the acoustic doors. The instrument itself is mounted on a bracket that provides for a complex angular offset of the field of view (FOV) of the viewing fan from any of the spacecraft decks (Fig.1; Fig. 7; Fig. 10).

The complex angular offset was employed to optimize the viewing of freshly ionized pick-up ions in the vicinity of Pluto due to charge exchange with neutrals from Pluto's atmosphere. This allows for the instrument to be mounted to the spacecraft deck while looking past the high gain antenna (HGA) while not being obscured by it (Fig. 8**Error! Reference source not found.**).

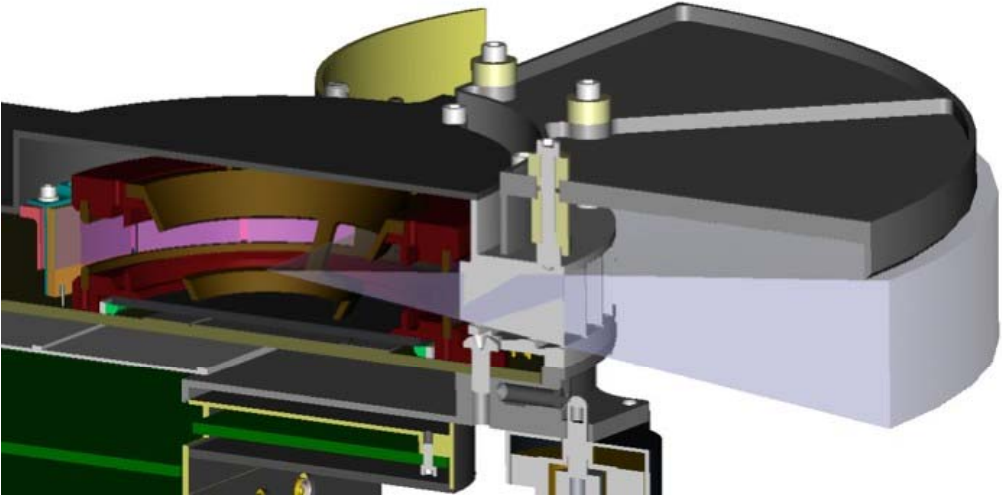

**Fig. 7.** Cutaway view of the PEPSSI sensor showing the FOV defined by the internal sensor structure, collimator assembly, and Sun shade (top).

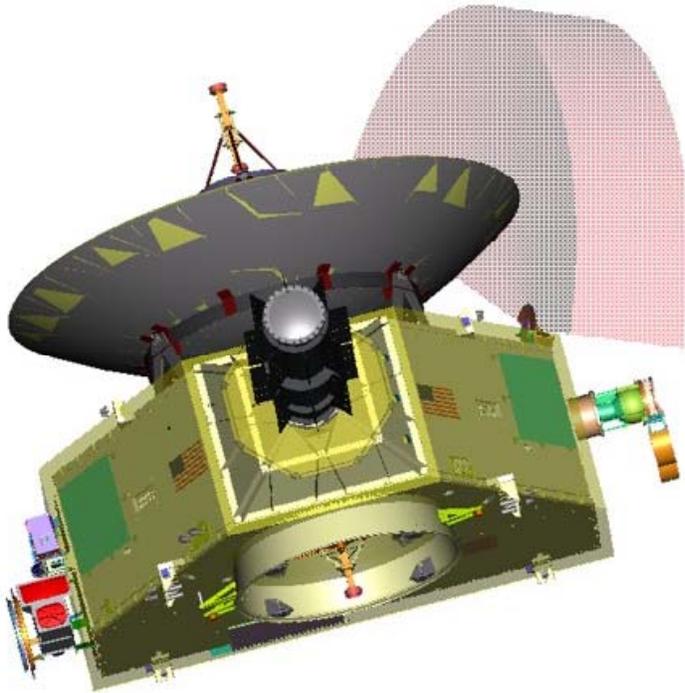

**Fig. 8.** Location of PEPSSI on the New Horizons spacecraft. The lightly shaded area denotes PEPSSI Field-of-View (FOV).

Alignment for PEPSSI's FOV on the observatory's top deck was determined by the design and location of its mounting bracket. Alignment control was to be kept within 1.5° of the Euler angle rotations defining the bracket's mounting surface. Knowledge of PEPSSI's FOV was specified to be within 1.5° of the observatory's coordinate system, as referenced to the orthogonal plane surfaces of the PEPSSI instrument. An error in interpretation of the one of the interface control documents (ICDs) led to a misalignment in the bracket surface as manufactured. Figure 9 shows the planned and actual fields of view in the coordinate system of the observatory, i.e., the New Horizons spacecraft.

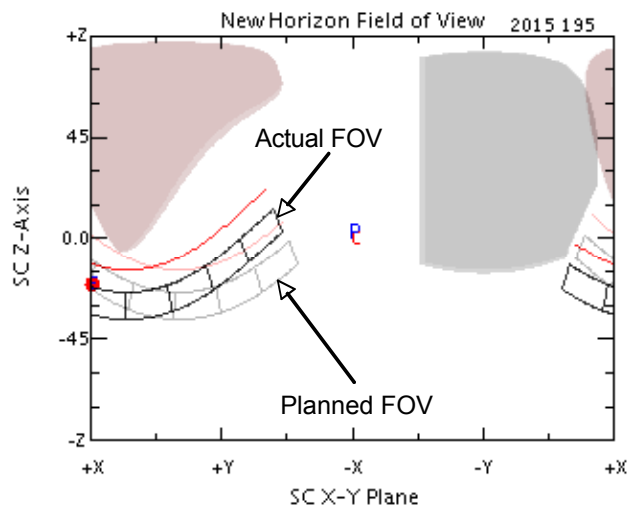

**Fig. 9.** The PEPSSI FOV in angular coordinates referenced to the main deck of the spacecraft. The shaded area above the FOV traces indicates the HGA and the shaded area to the right indicates the body of the spacecraft. The red line indicates the Sun shield, the red dot the location of the Sun and the P and C (near the center) the locations of Pluto and Charon near the beginning of the mission.

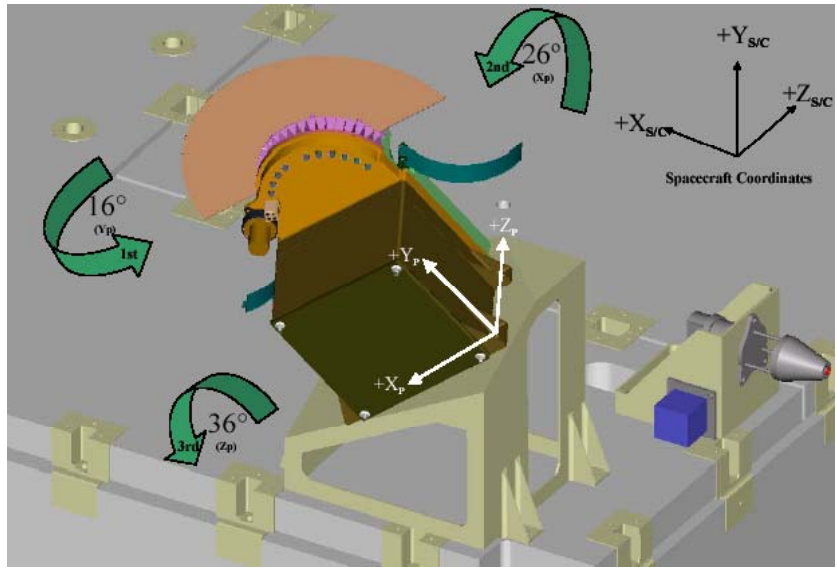

**Fig. 10.** PEPSSI mounting orientation with respect to the New Horizons deck.

With this FOV PEPSSI points at an off-angle to the Sun during spacecraft downlink operations and most of Pluto flyby operations. (The HGA points towards the earth, along the +Y axis, during downlink operations). This allows PEPSSI to detect pickup ions, coming from the general direction of the Sun, while keeping the Sun out of the FOV.

There are no rigid requirements to keep the Sun out of the instrument FOV; however, for normal operation the Sun should be ~ 10° from the instrument boresight. In the event the Sun is in the instrument FOV, no damage to the instrument will occur. However, the flood of energetic photons may cause a degradation of both energy and TOF measurements. To mitigate these affects, in the event the Sun is in the FOV, PEPSSI includes include the capability to disable by command the start and stop anode(s) of the sector(s) pointed in the sunward direction. In fact, this phenomenon with the Sun in the FOV was what led to the mapping of the actual FOV.

PEPSSI is mounted to the spacecraft's top panel with four (4) #8-32 stainless-steel screws into threaded inserts on a dedicated spacecraft-designed mounting bracket. The instrument is thermally isolated from the bracket by thermal washers between the bracket and PEPSSI's mounting tabs. PEPSSI's alignment and repeatability of mounting with respect to the observatory's coordinate system was satisfied by dimensional tolerancing and position of the mounting holes in the bracket and on the spacecraft.

### 3.2.2  Mass Properties

The mass budget for the PEPSSI instrument was 1500 grams, including 120 grams of margin. This mass included all sensor pieces as well as the mounting bolts and was worked to on a board-by-board basis (including the metal framing for the boards). The measured sensor mass at delivery to the spacecraft was 1475 g.

### 3.2.3  Deployable Cover

The PEPSSI design includes two cover doors installed over the instrument aperture. One purpose of the instrument covers is to prevent acoustic (air pressure) damage to the thin entry/exit foils within the sensor module during launch. Another is to keep out air-born dust and contaminants during instrument ground test. During all ground operations and storage, except functional test of cover release, the aperture covers were maintained in place to prevent damage to the entry/exit foils and to prevent contamination of the MCP.

The covers are crescent shaped and each covers one-half of the 160° aperture angle. Each cover is mounted at one end on a hinge bracket assembly that includes a Mandrel with torsion spring. In the closed position an actuator pin holds the covers in place. To open the covers, the actuator is fired to retract the retaining pin, allowing the covers to spring open. A spacecraft command must be sent to fire the actuator. Once open, the covers are maintained in the open position by torsion-spring action. To allow for ground tests (e.g., EMI tests) with the covers open, the actuator used to pin the doors closed shall be capable of at least 100 cycles of operation.

Cover opening was tested in a vacuum chamber prior to flight for correct actuator function. The covers were open on the flight unit on 3 May 2006.

### 3.2.4  Instrument Purge

The PEPSSI instrument interface included a purge manifold. Once PEPSSI is assembled with the MCP, it requires constant purge with at least MIL-P-27401D Grade C (99.995% pure) nitrogen, at a flow rate of 0.1 liters/minute.

The purge system was capable of being disabled for a short period of time (up to 30 minutes) while in a class 10,000 or better clean room. Records were maintained of all periods when purge has either been removed or failed.

### 3.2.5 Handling Requirements

The PEPSSI instrument was kept under purge at all times, except during test in vacuum chambers. The PEPSSI door remained closed to protect the entry and exit foils. When open, all airflow toward the foils, including breath, was prevented. As long as the cover is closed and the unit is being purged, it could be in any normal room environment.

Proper procedures and equipment to eliminate the risk of electrostatic discharge were followed and used. The collimator was not cleaned or handled in any manner, and the cleaning of the outside cover was minimized.

A high-voltage power supply (HVPS) safing plug was included and removed prior to launch. In addition, a cover release arming plug was installed prior to launch to enable cover release commands in space.

### 3.2.6 Transportation and Storage

The PEPSSI instrument required a shipping container to provide protection during transportation and storage. The instrument was double bagged and filled with dry nitrogen for transport and constrained and padded within the container, to minimize vibration and shock during transport.

### 3.2.7 Vacuum and Outgassing Requirement

The PEPSSI HVPS was required to be turned on to full voltage only in a vacuum environment $< 3 \times 10^{-6}$ torr. As a consequence, the HVPS must not be activated until sufficient time in vacuum has elapsed for spacecraft and instrument outgassing. A high concentration of outgassing products and particles could contaminate the MCP under high-voltage conditions, so outgas products must be at a sufficiently low level for safe turn-on of the HVPS.

Prior to launch, during both instrument-level and spacecraft-level thermal-vacuum test, appropriate detectors were placed inside the test chamber to monitor the level of outgas products. During spacecraft-level testing, 24 hours of time in vacuum (at < $3\times10^{-6}$ torr) was required before the HVPS may be safely turned on. During instrument-level test, the test conductor determined when it was safe to turn on the HVPS. Prior to HVPS activation, the PEPSSI instrument could be (and was) powered on for low-level engineering level checkout. At full atmosphere, during instrument and spacecraft level testing, the HVPS voltage (applied to the MCP) was kept less than 700 volts. After the New Horizons launch, the requirement was for at least two weeks to pass prior to first activation of the HVPS. The PEPSSI instrument was powered earlier (with HVPS off) to allow for initial checkout (20–22 Feb 2006) and engineering-level checks (1–2 Mar, 27 Apr, and 2 May 2006).

### 3.3. Detectors and Electronics

Power as initially allocated to the instrument took into account the limited resources on the spacecraft. Referenced to the +30V input the peak power specifications were <2000 mW in science mode and <1750 mW for the high voltage power supply (HVPS) disabled. During development, these numbers were found to not be realizable without extensive additional development. To remain within financial resources a request was granted to raise the allowable power required by 550 mW.

#### 3.3.1 Energy Measure

Each SSD sector in the sensor module has 2 detectors. In sectors 1, 3, 6, one detects ions, the other electrons. The electron detectors are covered with a 1-µm Al layer, to block low-energy protons and heavier ions. The detector pairs in sectors 2, 4, 5 are just ion detectors. A key to the electronics and functional layout is given in Fig. 11.

Energetic electrons from 25 keV to 500 keV are measured by the electron detectors. The Al layer blocks protons and ion particles with energies less than 100 keV, and particle energy levels above 100 keV are expected to be rare in the near-Pluto environment. For those rare events, where ion energy levels exceed 100 keV, coincident TOF measurements will be used to discriminate between ions and energetic electrons.

Ion energy measurements using the ion detectors are combined with coincident TOF measurements to derive particle mass and identify particle species. Particle energy for protons greater than 40 keV and heavy ions (such as the CNO group) greater than 150 keV are measured, up to a maximum of 1 MeV. Lower-energy ion fluxes are measured using TOF-Only measurements; detection of micro-channel plate (MCP) pulse height provides a coarse indication of low-energy particle mass.

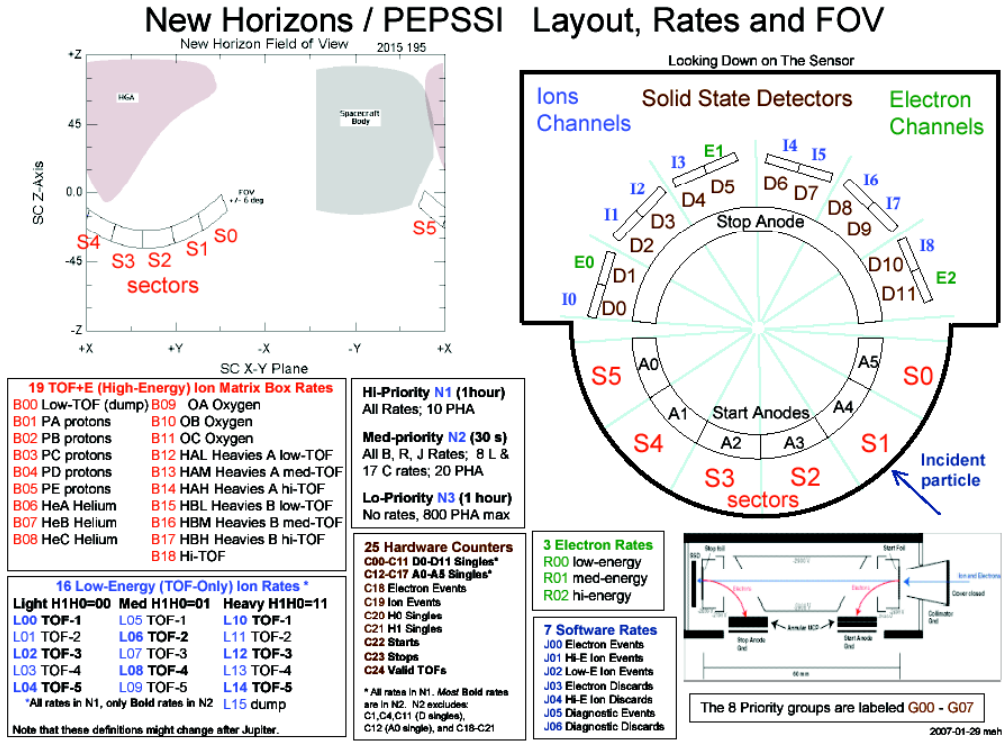

**Fig. 11.** Schematic layout of PEPSSI showing the nominal FOV and detectors and rates for the various modes and channels.

3.3.2   Time-of-Flight Measure

Before an ion passes through the TOF head, it is first accelerated by a 2.6-kV potential. Secondary electrons from the foils are electrostatically separated on the MCP, providing start and stop signals for TOF measurements. The segmented MCP anode, with one start anode for each of the six angular segments, determines the direction of travel, particularly for lower-energy ions that do not yield an SSD signal above threshold. A nominal 500-volt accelerating potential between the foil and the MCP surface controls the electrostatic steering of secondary electrons. The dispersion in transit time is less than 400 picoseconds (ps).

### 3.3.3 Electronics

The six electronics boards in the electronics stack (cf. Fig. 3) each provide a specific function for the instrument (Table V). The common board size has also been used on other APL spacecraft subsystems, notable the electronics for the (redundant) digital processing units and several of the instruments on the MESSENGER spacecraft (Gold *et al.* 2001).

**Table V. Electronics Boards in the PEPSSI Instrument**

| PEPSSI board | Board function(s) |
| --- | --- |
| Energy Board | Amplify SSD event signals (12 channels – signal proportional to particle energy) and digitize (10 bits). The measurement is achieved with low-power, low-noise Energy (CSA + Shaper) – Peak Detector ASICs |
| TOF Board | Amplify start (six) and stop anode signals, compute TOF (time duration between start and stop anode signals). The measurement is achieved with the low-power, low-jitter, and low-walk TOF chip (CFD + TDC) |
| HVPS Board | Provide high voltage to entry/exit foils (–2600 V), MCP (–2100 V and –100 V), deflector plates (–2900 V), and SSD bias voltage (–100 V) |
| Digital (Interface) Board | Event validation logic; energy and TOF event counters; interface functions |
| Events Processor Board | Command and Telemetry communications to spacecraft; events processing |
| LVPS Board | Supply ±15 V, ±5 V to electronic boards |

### 3.3.4 Operation

#### 3.3.4.1 *Science Mode*

In the science mode, three basic classes of measurements are concurrently made. The first class, called TOF-plus-Energy, uses SSD measurements in coincidence with TOF measurements to determine energetic ion particle mass (species) and associated energy spectra. Measurements are collected from six different sectors and particle direction to a particular 25° by 12° sector is determined by the start sector.

The second class of measurement is referred to as TOF-Only. It is a measure of ion particle velocity based on the TOF measurement and is made if no coincident energy event is detected at the SSDs. Particle composition is categorized as light (protons), medium (mass < CNO), or heavy (mass ≥ CNO). These measurements occur if the ion particle energy is below the energy threshold of the SSD detector.

The third class of measurements consists of Energy-Only measurements of particles where the SSD measurement event is not coincident with any TOF measurement event. These measurements include electron energy measurements, where transit time of electrons through the TOF chamber is, for practical purposes, zero, as well as ion energy measurements made when the ion fails to generate any secondary electrons (for TOF measurement).

*3.3.4.2 HVPS Activation*

The high voltage (HV) supply of the HVPS can only be turned on in vacuum ($< 3 \times 10^{-6}$ torr). If the HV supply is accidentally turned on during ground test in air, the MCP may be degraded. Therefore, when the PEPSSI instrument is powered up on the ground, the HVPS (consisting of HV and bias power supply) is disabled and no science data is collected. A sequence of instrument commands is required to enable the HVPS, fine-tune voltages, and tweak HVPS clock frequency for optimum efficiency. Valid science data is output after the HVPS command sequence has been executed and science data enabled.

*3.3.4.3 Cover Door Open*

At first power-up on orbit, the PEPSSI covers are in the closed position. A spacecraft command must be sent to fire the actuator and open the front covers. Once open, the covers remain open indefinitely, and no cover release commands are required at any subsequent turn-on of the instrument.

*3.3.4.4 Test Mode*

The instrument can implement a test mode that allows for data input and output through a test port instead of the C&DH communication ports. This is strictly for use during instrument software development pre-flight and is not for spacecraft level test or for use in space operations.

*3.3.4.5 Calibration*

Ground-based calibration is performed with the instrument in a dedicated calibration mode. In this mode, energy and TOF measurements are telemetered at a high rate. Based on an evaluation of telemetered energy and TOF data, instrument calibration tables are configured and subsequently loaded into the instrument.

3.3.5   Electrical Interface

There are five electrical interfaces from PEPSSI back to the spacecraft. Only the power and command and telemetry interfaces are still of use. The others were used for ground testing, safing, and the deployment of the cover that occurred as part of the overall commissioning.

The PEPSSI instrument does not require a survival heater for operation. However, PEPSSI does require a spacecraft-provided and controlled heater that can be can be powered during cruise phase when observatory power is available. The heater provides thermal margin, if required, and dissipates 1 watt when switched on.

*3.3.5.1   PEPSSI Power Interface.*

Switched power is provided to PEPSSI at 30 volts nominal and is regulated to within ±1 volt of nominal. Primary power is supplied to the PEPSSI support electronics through a separate dedicated connector that has no signal or secondary lines. Total power consumed by PEPSSI was initially specified to be 2 watts or less, but was later increased during the design process.

There are specifications regarding input voltage characteristics (operating and survival) regarding regulation, input voltage ripple, source voltage transients, source impedance, and voltage turn-on/turn-off rates. Inrush current transients at power turn-on and power turn-off imposed on the instrument are also specified. The instrument was designed and tested to operate over and survive respective voltage operating conditions as required.

*3.3.5.2   PEPSSI Command and Telemetry Interface*

PEPSSI implements a low-speed command and telemetry interface circuit that conforms to the EIA RS-422 standard for serial data transmission. The low-speed buses for PEPSSI each consist of three circuits: a 1 pulse/second (1PPS) sync signal, a command circuit, and a telemetry circuit.

PEPSSI implements two RS-422 bi-directional universal asynchronous receive and transmit (UART) ports to the spacecraft. The first port connects to the spacecraft C&DH system located in the integrated electronics module (IEM) #1, and the second port connects to the spacecraft C&DH system located in IEM #2. Normally, only one of the two IEMs is active, and PEPSSI responds and interacts with the active IEM (Fountain *et al.* 2007).

The serial ports are bi-directional and PEPSSI is capable of simultaneously sending and receiving at 38.4 k baud rate with 8 data bits, no parity, 1 start bit, and 1 stop bit. The least significant bit of each byte is sent first. For multi-byte values, "big endian" format is used, where the most significant byte is sent first; the least significant bit is referred to as "b0". PEPSSI receives instrument commands and MET time from the spacecraft over this port, and sends instrument science and housekeeping data to the spacecraft. Data sent to the spacecraft is normally recorded in the SSR of the respective IEM for latter downlink to the ground station.

The active IEM and PEPSSI exchange messages over the UART using the standard UART protocol to govern the lower-level aspects of the transfer. A higher-level construct, an Instrument Transfer Frame (ITF) protocol, is used for higher-level synchronization and error control. The IEM provides PEPSSI with command messages and spacecraft time messages. PEPSSI provides the IEM with TLM messages that contain instrument state data as well as science data. PEPSSI does not exchange messages with any other instrument and every message to or from PEPSSI is contained within an ITF.

In addition to Transmit (to spacecraft) and Receive (from spacecraft) signal lines, each port includes the 1PPS signal input to PEPSSI. This signal provides nominal one-second timing information; command and telemetry transfer is synchronized to 1PPS epochs as illustrated in Fig. 12.

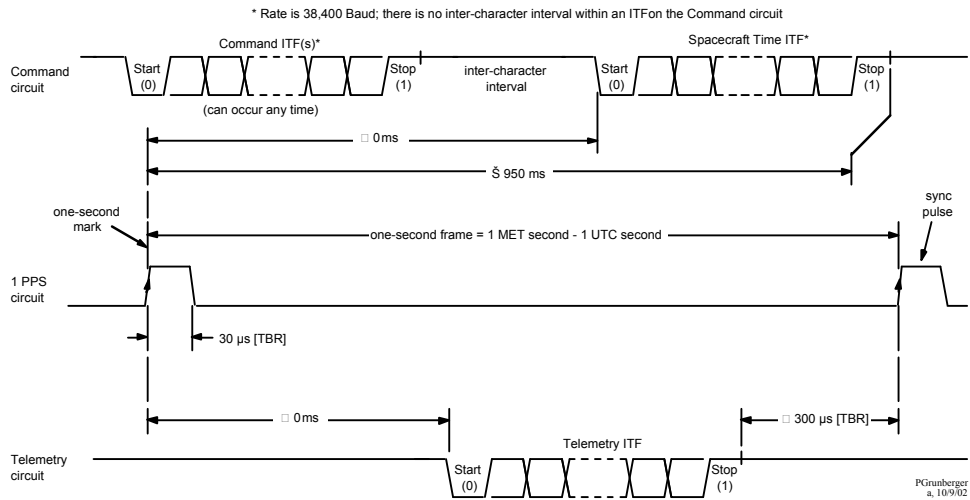

**Fig. 12.** PEPSSI command and telemetry timing.

Commands, instrument telemetry, and MET time are packed in the ITF format that is described below. This frame is defined by the time between the rising edge of any two adjacent 1PPS signals. Command and telemetry data frames are transferred as a serial 8-bit byte stream.

MET time is transferred from the spacecraft to PEPSSI over each one-second interval. MET time is transferred to 1-second precision and is valid when the 1PPS signal is asserted. MET time is transferred within 0 to 950 ms following each 1PPS.

PEPSSI telemetry data may be transferred anytime between 1PPS epochs except when the 1PPS rising edge occurs. All telemetry transmissions terminate 300 µs prior to the rising edge of the 1PPS. All three signals (Transmit, Receive, 1PPS) are implemented as complementary RS-422 electrical standard interfaces.

PEPSSI sends a number of different science and housekeeping data products to the spacecraft. These products are packed in CCSDS telemetry packet format. A number of CCSDS telemetry packets may be included in each telemetry ITF sent to the spacecraft.

*3.3.5.3 Test Port Interface*

For early hardware and software development, PEPSSI implemented an RS-232 protocol test interface to the instrument processor. However this was intended only for early development and test and was not accessed at the spacecraft level of integration and test.

*3.3.5.4 HVPS Safing*

The PEPSSI HVPS could be safely turned on to full level only when the instrument is in vacuum. Damage to the MCP may result if high voltage power is applied at atmospheric or partial atmospheric pressures. To prevent accidental turn-on of the HVPS, the PEPSSI instrument was designed to accept a safing plug connection– installation of the safing plug in the spacecraft harness was required to disable HVPS operation. The safing plug was removed before launch to enable on-orbit HVPS operation.

*3.3.5.5 PEPSSI Cover Release*

The PEPSSI design includes a cover for the instrument aperture. One purpose of this cover is to protect the instrument foils during ground vibration and acoustic shock tests. Prior to launch, the cover is opened only during T-V, calibration, and special cover tests; the cover remained closed at all other times. The instrument cover is opened by spacecraft command. The spacecraft command system and associated spacecraft ground support equipment (GSE) included implementation of an arming plug. The arming plug was removed during ground system test to prevent accidental opening of the cover and then installed before launch to enable on-orbit cover release operation.

The spacecraft provided a one-time activation signal to open the cover. An electrical non-explosive shape memory alloy (SMA) pin-puller is incorporated in the instrument and was used for the release. The SMA pin-puller employs redundant circuitry. Each circuit requires a current load of between 0.5 to 2.0 amps at 30 VDC. Depending upon the operational load selected, the spacecraft power distribution unit could apply this power for a minimum of 200 msec. The pin-puller employs an auto shut-off switch removing power after it has activated, approximately 100 msec after power is applied. The cover has been successfully opened, and this feature will no longer be used.

## 3.4. Telemetered Data Products

3.4.1 Proton and Electron Energy Spectra

In the energy spectra data for detected electron, proton, and heavy ion particles each energy bin corresponds to a particular particle energy range (defined in the software data base look-up Table VI), and the telemetered value represents the number of particles detected in the respective energy bin. The initial look-up table that defines the energy bin coordinates for each particle was determined and loaded during pre-launch instrument calibration.

Ion particle data is output for each of six sectors; electron data is output for each of three sectors. The integration interval, i.e., the time interval over which the data is collected, is nominally 10 seconds, but data integration interval may be changed by command from 1 second to 7200 seconds.

For those sectors that include an electron detector (sectors 1, 3, 6), the number of ion events counted in each energy bin represents the number of ion particles impinging on the respective single ion detector. For those sectors that include ion detector pairs (sectors 2, 4, 5), the number of ion events counted in each energy bin represents the sum of ion particles impinging on the respective ion detector pair.

Table VI. Proton/Electron/Heavy Ion Energy Spectra Data Products

| Species | Number of energy bins | Bits per bin | Scaling | Number of sectors | Number of bits per integration interval |
|---|---|---|---|---|---|
| Electrons | 3 | 10 | Log compressed, 5-bit mantissa, 5-bit exponent, base 2 | 3 | 90 |
| Protons | 6 | 10 | | 6 | 360 |
| Heavy Ions ( > 1 AMU) | 4 | 10 | | 6 | 240 |
| Total bits per integration interval–> | | | | | 690 |

3.4.2 Heavy Ion Energy Spectra

The energy spectra data for heavy ions, output from each of six sectors, is conditional; it is output only if the particle event rate exceeds a predefined threshold (Table VII). The event rate threshold includes hysteresis, so that once heavy ion spectra data is output, the event rate must drop to a somewhat lower value for output to cease. The particle event rate threshold is programmable by instrument command and is typically set at 100 particle events per second. The data integration interval, nominally 10 seconds, is identical to the integration time interval for all other spectra type data. It may be adjusted by instrument command over a range of 1 to 7200 seconds.

For those sectors that include an electron detector (sectors 1,3,6), the number of heavy ion events counted in each energy bin represents the number of ion particles impinging on the respective single ion detector. For those sectors that include ion detector pairs (sectors 2,4,5), the number of heavy ion events counted in each energy bin represents the number of ion particles impinging on the respective pair of SSD ion detectors.

Table VII. Heavy Ion Energy Spectra

| Species | Number of energy bins | Bits per bin | Scaling | Number of sectors | Number of bits per integration interval |
|---|---|---|---|---|---|
| CNO | 6 | 10 | Log compressed, 5-bit mantissa, 5-bit exponent, Base 2 | 6 | 360 |
| $N_2$ | 3 | 10 | | 6 | 180 |
| Total bits per integration interval–> | | | | | 540 |

3.4.3  TOF-Only Velocity Spectra

Particle velocity spectra data, derived from TOF-only measurements are comprised of three sets of particle velocity data corresponding to particle mass categorized as light, medium, or heavy (**Error! Reference source not found.**). The two stop-anode discriminators are employed to differentiate between light, medium, and heavy particles. The discriminator threshold settings normally are set so that light particles correspond to protons, medium particles correspond to the CNO group, and heavy particles correspond to particles with mass greater than the CNO group. Similar to heavy ion spectra, particle velocity spectra data is output only if the particle event rate exceeds a predefined threshold. As with the other products just mentioned, the data integration interval, nominally 10 seconds, is identical to the integration time interval for all other spectra type data. It may be adjusted by instrument command over a range of 1 to 7200 seconds.

Table VIII. TOF-Only Velocity Spectra

| Mass category | Number of spectral bins | Bits per bin | Scaling | Range | Number of sectors | Number of bits per integration interval |
|---|---|---|---|---|---|---|
| Light | 5 | 10 | Units = km/sec Log compressed, 5-bit mantissa, 5-bit exponent, base 2 | 200 km/sec to 2000 km/sec spanning 5 bins | 6 | 300 |
| Medium | 5 | 10 | | | 6 | 300 |
| Heavy | 5 | 10 | | | 6 | 300 |
| Total bits per integration interval–> | | | | | | 900 |

3.4.4   Singles-Event Data (for Event Validity Check)

To allow validation of instrument science data, PEPSSI counts and telemeters a number of instrument single events (Table IX). Single-event data is collected over the same identical data integration intervals as ion-energy spectra and velocity spectra data. It is nominally 10 seconds, but may be adjusted by instrument command over a range of 1 to 7200 seconds.

Table IX. Singles Event Data (Event Validity Check)

| Event description | Number of event counters | Bits per counter | Number of count bits | Scaling |
|---|---|---|---|---|
| SSD Leading Edge Discriminator –Ion Event | 9 | 10 | 90 | Log compressed, 5-bit mantissa, 5-bit exponent, base 2 |
| SSD Leading Edge Discriminator –Electron Event | 3 | 10 | 30 | |
| SSD Peak Detector – Ion Event | 9 | 10 | 90 | |
| SSD Peak Detector – Electron Event | 3 | 10 | 30 | |
| Start Anode Discriminator Event | 6 | 10 | 60 | |
| CFD Start Event | 1 | 10 | 10 | |
| CFD Stop Event | 1 | 10 | 10 | Log compressed, 5-bit mantissa, 5-bit exponent, base 2 |
| Valid TOF Events | 1 | 10 | 10 | |
| Ion Events | 1 | 10 | 10 | |
| Electron Events | 1 | 10 | 10 | |
| Software Counters | 3 | 10 | 30 | |
| Total data bit count per integration interval –> | | | 380 | |

3.4.5   PHA (Pulse Height Analysis) Event Data

Particles detected by PEPSSI are prioritized and ranked according to 'interest'. A table, called the PHA rotating priority structure, loaded prior to launch, defines the order in which priority is assigned. Energy and TOF measurements, collectively referred to as PHA data, are telemetered for particles with highest priority. The specific data telemetered for each PHA event are specified in Table X.

PHA event data are collected over fixed ten second intervals and output as a single block. The number of PHA events collected in any single second is limited to a maximum number that is programmed by instrument command. Typically, the limit is set to 40 PHA events per second (400 PHA events per 10 second collection interval). When low particle event rates are encountered, the PHA event rate will most likely average less than the maximum limit.

Table X.  PHA Event Data

| Data item | Description | Number of bits |
|---|---|---|
| TOF [units = ns] | Particle TOF Measurement. All zero fill for energy-only events. Log Compressed Scaling | 8 |
| Energy [units = keV] | Particle energy measurement. All zero fill for TOF-only events. Log Compressed Scaling | 8 |
| Ion sector | SSD channel/sector identifier, 9 channels | 4 |
| Start sector | Identifies sector where the TOF start anode signal is detected | 3 |
| Heavy Discriminator | TOF Stop Anode Discriminator States (Flags). Identifies event particle as light, medium, or heavy | 2 |
| Quality flag | Set if multiple-ion hits detected | 1 |
| Reserved | Reserved | 6 |
| Bits / PHA event–> | | 32 |
| Bits / 40 PHA events –> | | 1280 |

3.4.6    Non-Packetized Housekeeping Data

PEPSSI telemeters non-packetized housekeeping data. These include voltage and current measurements that are monitored during activation of the instrument energy measurement system and TOF measurement system. PEPSSI non-packetized housekeeping data are placed in the instrument-defined status field of Instrument Transfer Frames sent from PEPSSI to the spacecraft C&DH system. Housekeeping data is updated at a regular 1-second rate. The total number of bits in each housekeeping frame is 56 bits, resulting in an average output data rate of 56 bps (Table XI).

Table XI. PEPSSI Non-Packetized Housekeep Data

| Parameter | Name | No. of bits | Units |
|---|---|---|---|
| HV Input Current Mntr. | CDH_PEPSSI_HV_CURR | 8 | MICROAMPS |
| HV Output Voltage Mntr. | CDH_PEPSSI_HV_VOLT | 8 | VOLTS |
| Bias Input Current Mntr. | CDH_PEPSSI_BIAS_CURR | 8 | MICROAMPS |
| Bias Output Voltage Mntr. | CDH_PEPSSI_BIAS_VOLT | 8 | NONE |
| Energy Board Tmp. | CDH_PEPSSI_ENE_BD_TEMP | 8 | DEGREES C |
| HVPS Board Tmp. | CDH_PEPSSI_HVPS_BD_TEMP | 8 | DEGREES C |
| LVPS Board Tmp. | CDH_PEPSSI_LVPS_BD_TEMP | 8 | DEGREES C |
| Total bits | | 56 | |

### 3.4.7 Quick look Diagnostic Data

For a quick look analysis of instrument status and measurement results, PEPSSI can output a diagnostic set of data at a fixed 300-second rate. It is intended that PEPSSI diagnostic data be downlinked as soon as possible following the near Pluto encounter, i.e., on a first-look basis. This will allow for examination of instrument performance in the quickest and most timely manner. It is estimated, over a 24-hour timeline, roughly the time duration of the near-Pluto encounter, about 480 kbits of PEPSSI quick-look data will be available for downlink.

The diagnostic data is identical to the electron/proton/heavy-ion energy spectra and rate-counter data previously defined, except that it is collected over 300-second integration time intervals. In addition, 600 bits of extended housekeeping data are included. The diagnostic integration time interval is constant at 300 seconds, and the subsequent average output data rate is less than 6 bits per second.

### 3.4.8 Instrument Data Rate Summary

The telemetry system of the PEPSSI instrument may be configured in a number of different ways by instrument command. The data integration interval for a number of products may be adjusted, heavy ion and TOF velocity data products may be enabled or disabled, and the maximum PHA event rate may be changed. The volume of data telemetered from PEPSSI, and the resulting average data rates, will vary accordingly.

Table XII lists PEPSSI data products and computes telemetry data rates for a number of representative telemetry system configurations. The first four products listed (proton/electron energy spectra, heavy-ion spectra, TOF velocity spectra, and rate-counter data) represents data collected over a data integration time interval that is changeable and may be set by instrument command. The remaining data products (PHA data, non-packetized housekeeping, and quick look diagnostic data) are collected and output at fixed rates.

Table XII. Example PEPSSI Telemetry Data Rates As Function of Telemetry Configuration

| Telemetry data product | Number of data bits | Data integration interval | Example bit rate (bps) | | | | |
|---|---|---|---|---|---|---|---|
| | | | In space low rate configurations | | In space typical configurations | | Ground test only |
| | | | 300 s data interval | 10 s data interval | 300 s data interval | 10 s data interval | 1 s data interval |
| Proton/Electron Energy Spectra | 690 | 10 seconds typical<br><br>Programmable 1 to 7200 seconds | 2.3 | 69 | 2.3 | 69 | 690 |
| Heavy Ion (CNO, N2) Energy Spectra [1] | 540 | | 1.8 | Disabled | 1.8 | 54 | 540 |
| TOF-Only Velocity Spectra [1] | 900 | | 3.0 | Disabled | 3.0 | 90 | 900 |
| Singles Events (Rate Counters) | 380 | | 1.3 | 38 | 1.3 | 38 | 380 |
| PHA<br>Event Rate ≤ 40 events/Sec<br>Event Rate ≤ 10 events/Sec | 32 per PHA event | 10 second (fixed) | ≤ 320 | ≤ 320 | ≤ 1280 | ≤ 1280 | ≤ 1280 |
| "Non-Packetized" Housekeep Data [2] | 64 | 1 seconds (fixed) | 64 | 64 | 64 | 64 | 64 |
| Quicklook Diagnostic<br>   Proton/Electron Spectra<br>   Rate Counters<br>   Extended housekeep | 1670 | 300 seconds (fixed) | 5.6 | 5.6 | 5.6 | 5.6 | 5.6 |
| CCSDS Frame Overhead | | | 8.5 | 16.5 | 8.5 | 16.5 | 88.3 |
| Average total data bit rate, bps –> | | | ≤ 407 | ≤ 513 | ≤ 1367 | ≤ 1617 | ≤ 3948 |

Note [1]: Heavy ion spectra and TOF velocity spectra outputs are conditional. They are output only if the particle event rate exceeds a predefined threshold. The particle event rate threshold is programmable by instrument command. The output may also be disabled by instrument command regardless of particle event rate.

Note [2]: Non-Packetized housekeeping data is packed in the Instrument Transfer Frame instrument status field and is intended for inclusion in spacecraft telemetry packets.

The data integration interval, for the first four products listed, is nominally set to 10 seconds and is never set any lower during in-space operation. For anticipated in-space operations, where the heavy-ion spectra and TOF velocity spectra are enabled, and the PHA limit is set to 40 PHA events per second, the maximum output data rate is about 1620 bits per second. For certain ground test operations, prior to spacecraft integration, where data is integrated over short, one-second time intervals, the output data rate averages about 3950 bits per second. If required by mission operation or spacecraft constraints, as illustrated, the instrument telemetry system may be configured to telemeter less data.

The PEPSSI instrument will output four types of telemetry packets on regular time intervals:

–High Priority
–Medium Priority
–Low Priority, and
–Common Status

The time intervals may be modified by instrument command, but it is expected that the default telemetry packet rates are sufficient to adequately support the Pluto encounter as well as cruise phase operations. Table XIII lists the packet types and provides current best estimates of average data output rates and 24-hour data storage volume for the Pluto encounter. The data rate and volume figures are raw values and do not include reductions due to use of the Fast data compression algorithm. The amount of reduction, in data volume and rate, depends upon the variableness of the raw data stream and therefore, is not fixed. It is anticipated the reduction of data volumes and rate will be 50% or more; therefore, the actual output data rates at the PEPSSI command and telemetry interface are expected to be on the order of 300 to 500 bps. The data rates and volumes given in the table should be considered upper bounds.

The table also indicates that the 24-hour volume for High Priority data is on the order of about 1.55 Mbits. However, the Fast compression algorithm can reduce this volume by up to 50% leaving the remaining volume less than 2% of the spacecraft 24-hour 'first-look', day one downlink capability following the Pluto encounter.

**Table XIII. Average Data Rates and Volumes**

Estimates of nominal PEPSSI telemetry data rates and volumes (prior to compression by fast algorithm). (The 152 bit ITF overhead is not included)

| Packet | Bit length (including CCSDS header) | Nominal output rate (seconds) | Average data rate (bits / second) | 24-Hour data volume (MegaBits) |
|---|---|---|---|---|
| High priority packet | 5,392 | 300 | ≈ 18 | ≈1.55 |
| Medium priority packet | 4,800 | 10 | ≈ 480 | ≈ 41.5 |
| Low priority packet (first) | 32,848[1] | 300 | ≈ 110 | ≈ 9.5 |
| Low priority packet (second) | 32,848 [1] | 300 | ≈ 110 | ≈ 9.5 |
| Common status packet | 488 | 300 | ≈ 1.5 | ≈ 0.12 |
| Average bit rate | | | ≈ 720 | |

Note [1]: Assumes a maximal length CCSDS packet of 4096 data bytes.

The instrument is quite flexible; for example, the actual bit rate at Jupiter was ~60 bps.

### 3.4.9 Calibration

For instrument calibration prior to launch, PEPSSI test software included a capability to read out raw instrument measurements. Raw measurement data was evaluated in order to configure the look-up calibration table that was subsequently loaded into the instrument software database.

### 3.4.10 Memory Image Dump

PEPSSI implements a capability to load selected portions of processor memory, e.g., look-up tables for instrument calibration. The PEPSSI instrument includes a capability to telemeter an image of the newly loaded segment of memory, to allow load verification in the spacecraft command center.

## 3.5. Commands

PEPSSI uses a variety of commands to provide functionality to the instrument and set various parameters that affect how the data is collected and processed before being sent to the ground.

### 3.5.1 Energy Commands

Commands related to energy measurement are given in Table XIV. These include commands to adjust energy detection thresholds of leading edge discriminators. The leading edge discriminator thresholds operate in current mode so that a threshold current essentially corresponds to an energy threshold.

Table XIV. PEPSSI Energy Measurement System Commands

| Command description | LED threshold | |
|---|---|---|
| | Granularity | Range |
| Energy discriminator threshold commands | | |
| Set leading edge discriminator (LED) threshold [Chan 1:Chan12] | ≈ 1 keV | ≈ 15 to 255 keV |

### 3.5.2 HVPS Commands

High-voltage PEPSSI operation requires a vacuum environment. At power turn-on, the HVPS is disabled by default. PEPSSI includes a set of HVPS and bias voltage supply commands to enable high-voltage operation, adjust high-voltage power supply clocks for optimum efficiency, and fine-tune the high-voltage outputs. A summary of these commands is given in Table XV.

Table XV. PEPSSI HVPS commands

| Description | Scaling | |
|---|---|---|
| | Resolution | Range |
| **High voltage power commands** | | |
| Enable / disable HVPS commands (safety feature) | NA | NA |
| Enable / disable HVPS & bias supply clocks | NA | NA |
| Adjust high voltage clock rate | ≈ 750 Hz | ≈ 75 kHz ±20% |
| Set high voltage command limit | ≈ –15 v | ≈ –3500 v |
| Set high voltage level | ≈ –15 v | ≈ –3500 v |
| Set high voltage alarm level | ≈ –15 v | ≈ –3500 v |
| Enable/disable high voltage alarm | NA | NA |
| | | |
| **Bias voltage power commands** | | |
| Set bias voltage level | ≈ –0.5 v | ≈ –125 v |
| Adjust bias voltage clock rate | ≈ 350 Hz | ≈ 47.25 kHz ±20% |

### 3.5.3 TOF Commands

In concert with energy threshold commands, PEPSSI implements a set of commands to adjust the thresholds of the start and stop constant fraction discriminators (CFD). With the threshold set to zero, about $5 \times 10^5$ electrons are required to generate a start or stop pulse respectively (and thus initiate a TOF measurement). The threshold can be adjusted upward, requiring more electrons and proportionally higher energy levels to initiate a TOF measurement event. A summary TOF command list is given in Table XVI.

Table XVI. PEPSSI TOF Measurement System Commands

| Description | Units scaling | |
|---|---|---|
| | Resolution | Range |
| Start anode leading edge discriminator thresholds | | |
| Set TOF sector threshold [Sector 1:Sector 6] | | |
| Stop anode leading edge discriminator thresholds | | |
| Set heavy 0 threshold | ≈ 10 mv | ≈ 0 to 2500 mv |
| Set heavy 1 threshold | | |
| TOF constant fractional discriminators | | |
| Set CFD start threshold | | |
| Set CFD stop threshold | | |
| Anode enable/disable | | |
| Enable/disable start anode[anode 1:anode 6] | NA | NA |
| Enable/disable stop anode [anode1:anode 6] | NA | NA |
| Miscellaneous | | |
| Set TOF chip mode | NA | NA |
| Reset TOF chip | NA | NA |

3.5.4 Process Control Commands

PEPSSI implements a number of commands to manage instrument operations, including energy measurement integration times, enable/disable event pile-up checks, etc. A summary list follows in Table XVII.

Table XVII. PEPSSI Process Control Commands

| Description | Scaling | |
|---|---|---|
| | Resolution | Range |
| **Measurement process control** | | |
| Load look-up tables | NA | NA |
| Load PHA rotating priority structure | NA | NA |
| Set integration time for energy, TOF spectra (seconds) | 1 | 7200 |
| Set maximum telemetered PHA event rate (events per second) | 1 | 1 to 100 |
| Set event rate threshold to enable telemetry of heavy ion & velocity spectra (events per second) | 1 | 0 to 1000 ( 0 => enable unconditionally) |
| Enable/disable telemetry of heavy ion & velocity spectra | NA | NA |
| Set coincidence window delay | Determined during design phase | |
| Set coincidence window width | | |
| Mask SSD discriminator firings for selected channel | NA | NA |
| Enable/disable electron multiple-hit check | NA | NA |
| Enable/disable ion multiple-hit check | NA | NA |
| *Miscellaneous* | | |
| Reset event FIFO | NA | NA |
| NO-Op command | NA | NA |
| Memory load command | NA | NA |
| Memory dump command | NA | NA |

## 3.6. Telemetry and Command Format

The C&DH system receives all telemetry from PEPSSI as non-packetized critical housekeeping data and as CCSDS packets. The CCSDS formatted packets can be common packets with standard formats, common packets with PEPSSI-specific formats, and PEPSSI-specific packets.

### 3.6.1 Data Rate and Volume

As just noted, the raw data rate, with the instrument telemetry system configured in the preferred manner for space operations, is a maximum of about 1620 bps, representing a total raw data volume collected over 24 hours of about 140 Mega-bits. These are raw figures that apply prior to application of the Fast data compression, which may reduce data volume by up to one-half. If necessary, the instrument telemetry system may be configured for lower data rates.

Telemetry data sent from the PEPSSI instrument is compressed using the Fast algorithm. Fast is an APL-developed data-compression algorithm that is imbedded in instrument common code used by a number of MESSENGER and New Horizon instruments. Fast is a lossless compression algorithm; the reduction of data volume is variable, data volume may be reduced by up to one-half.

### 3.6.2 Telemetry Formatting

#### 3.6.2.1 *Instrument Transfer Frame (ITF)*

All messages exchanged between the spacecraft C&DH and PEPSSI are transported using the ITF format. The format applies to all telemetry data sent from the instrument and to commands and MET time received from the spacecraft. The 48-bit frame header associated with this format is outlined in Table XVII.

**Table XVIII. Instrument Transfer Frame Format**

| Major field | Field | Byte # | Description |
|---|---|---|---|
| ITF Header | sync pattern 1 | 0 | 0xfe |
| | sync pattern 2 | 1 | 0xfa |
| | sync pattern 3 | 2 | 0x30 |
| | Message type | 3 | S/C TIME, COMMAND, or TELEMETRY |
| | Checksum | 4 | Byte-wise exclusive-OR of all bytes following the checksum |
| | Message length | 5-6 | Number of bytes following the length field |
| Message | Message data | 7 to N-1 | *New Horizons* formatted S/C time, command, or instrument state/telemetry message. N = length of ITF in bytes. max N= 12 for S/C time message max N= 279 for Command, max N= 1250 for Telemetry |

Byte 0 is first byte transmitted/received by PEPSSI. The instrument resynchronizes to the incoming sync pattern on each ITF to limit error propagation in case of bit errors. An ITF contains a single message, (S/C time, Command, or Telemetry) and will have variable length depending on the length of the contained message. The maximum length for command messages is selected to accommodate memory load commands. The maximum length for telemetry messages is selected to provide adequate margin for the one-second timing requirements for telemetry transfer. S/C time messages are fixed length. The time messages contain information in addition to time that vary from instrument to instrument. The ITF header includes a checksum (byte 4) on the data that follows. In the case of Command and S/C Time messages transferred to PEPSSI, the C&DH computes and adds the checksum to the ITF when the telecommand packet from the spacecraft is parsed into PEPSSI commands. PEPSSI computes and adds the checksum to all telemetry messages sent to the C&DH. PEPSSI does not receive or send messages to any other instrument or subsystem other than the C&DH.

*3.6.2.2 Telemetry Interface*

Once per second, the PEPSSI instrument formats a telemetry message containing data in an ITF for transfer to the C&DH. This frame is defined by the time between the rising edge of any two adjacent 1PPS signals. The message includes one or more CCSDS-formatted telemetry packets. The format is shown in Table XIX.

Transfer of the ITF may begin anytime during the one-second frame but must be completed prior to the rising edge of the 1PPS. If there are more packets than can fit in the ITF, then the maximum number of bytes are put into the ITF, even if this means a partial packet is used. The remaining (unsent) portion of the packet is held in memory until the next ITF is formatted, whereupon it is inserted at the start of that ITF's telemetry data. If PEPSSI does not have enough telemetry to fill an ITF during that second, the size of the ITF will be smaller than the maximum allowed.

*3.6.2.3 CCSDS Packetization*

The header of the CCSDS-formatted packet is defined in Table XX. The length of the packet is variable and is specified in the CCSDS header. The data field of each CCSDS telemetry packet is compressed using the lossless FAST algorithm.

The application process id (ApID) in the primary header identifies the type of packet. The eleven-bit id is divided into a four-bit source and a seven-bit data. This gives an instrument up to 128 possible packet types that can be produced. The four-bit source for the PEPSSI instrument is shown in the table below. For memory dump packets, an instrument should set the lower seven bits of the ApID to a value of 1 (0000001). All other ApID assignments are at the discretion of the instrument. The mapping of PEPSSI instrument data products to specific application process identification (ApID) packets are specified in the PEPSSI software specifications document.

**Table XIX. Standard Telemetry Instrument Transfer Frame Format**

| Name | Length (bits) | Value | Description |
|---|---|---|---|
| Sync pattern 1 | 8 | $0 \times fe$ | |
| Sync pattern 2 | 8 | $0 \times fa$ | |
| Sync pattern 3 | 8 | $0 \times 30$ | |
| Message type | 8 | $0 \times 04$ | Telemetry |
| Checksum | 8 | --> | Byte-wise exclusive-OR of all bytes following the checksum |
| Message length remaining | 16 | --> | Number of bytes following the message length field |
| Heartbeat | 1 | --> | Alternating bit |
| Boot / application (optional) | 1 | 0 = Running boot code<br>1 = Running appl. code | Flag describes if boot or application code is running |
| Turn-off request (optional) | 1 | 0 = No turn-off requested<br>1 = Turn-off requested | Flags describes if instrument requests spacecraft to turn off power or reset |
| Reserved standardized instrument status | 5 | --> | Other bits not yet assigned. |
| Valid command count | 8 | --> | Module $2^8$ count of commands accepted |
| Invalid command count | 8 | --> | Module $2^8$ count of commands rejected |
| Instrument defined status | 8*N | N/A | Instrument non-packetized status or housekeeping. The C&DH processor will place this data in spacecraft-produced telemetry packets, and will use it in the evaluation of autonomy rules. N is TBD |
| Offset to first packet header | 16 | $0 \times 0000$ = A packet begins in first byte of telemetry packets field<br>$0 \times FFFE$ = There is no telemetry data in this frame<br>$0 \times FFFF$ = There is telemetry data in this ITF, but no packet begins in this frame | Offset within telemetry data field to start of first data packet header in frame. If an instrument never has more telemetry than can fit in a single ITF, this field will always be 0. |
| Telemetry data | 8*M | --> | New Horizons formatted CCSDS telemetry packets (maximum value for M is TBD) |

**Table XX. CCSDS Telemetry Packet Header Format**

| Item | Bit Field | Length (Bits) | Value (Binary) | Description |
|---|---|---|---|---|
| Primary header | Version number | 3 | 000 | Designates a source packet |
| | Type indicator | 1 | 0 | Designates a telemetry packet |
| | Secondary header flag | 1 | 1 | Secondary header flag present |
| | Application process identifier | 11 | 1101 xxxxxxx | PEPPSI packet Identification |
| | Grouping flags | 2 | 01 | First packet in group |
| | | | 00 | Intermediate packet |
| | | | 10 | Last packet in group |
| | | | 11 | Not part of group |
| | Source count | 14 | | Continuous sequence count (mod 16384) of packets for specific ApID |
| | Packet length | 16 | | Number of bytes in [packet secondary header + packet data field] – 1 |
| Secondary header | MET | 32 | | Spacecraft MET at time the source Packet is constructed |
| Data | | 8*M | | Additional data bytes, M, can range from 1 to TBD |

### 3.6.2.4  *Memory Dump Packet*

New Horizons uses a single memory dump telemetry format for all onboard processors. The format of the telemetry packet is specified in Table XXI.

**Table XXI. Memory Dump Telemetry Packet Format**

| Item | Length (bits) | Value | Description |
|---|---|---|---|
| Packet primary header | 48 | | For memory dump packets, the lower seven bits of the 11-bit ApID should be binary '000 0001'. |
| Packet secondary header | 32 | | MET |
| Address | 32 | | Start Address |
| Byte count | 16 | 1-256 | Number of bytes in this dump packet (not including pad) New Horizons has chosen a maximum dump size of 256 bytes. |
| Memory type | 8 | TBD | This field is available for memory architectures that cannot determine the memory type from the address. This field is spare (0x00) for architectures that can. |
| Spare | 8 | 0 | Unused (0x00) |
| Data | N*8 | | $1 \leq N \leq 256$ (From 1 to 256 8-bit bytes) |
| Pad | M*8 | 0 | $0 \leq M \leq 3$ Pad up to 32-Bit boundary. |

### 3.6.3 Command Formatting

#### 3.6.3.1 *Command Message*

The format of a command ITF is defined in Table XXII. The command message may be variable in length and length is specified in the header. One command is included in each ITF; however multiple command ITFs may be received from the C&DH in any one-second frame. For command verification, command acceptance and command rejection counts are included in the PEPSSI telemetry ITF.

**Table XXII. Command Message in Instrument Transfer Frame**

| Name | Byte # | Value | Description |
|---|---|---|---|
| sync pattern 1 | 0 | 0×fe | 0×fe |
| sync pattern 2 | 1 | 0×fa | 0×fa |
| sync pattern 3 | 2 | 0×30 | 0×30 |
| Message type | 3 | 0×02 | *COMMAND* |
| Checksum | 4 | | *byte-wise exclusive-OR of all bytes following the checksum* |
| Message length | 5–6 | | *number of bytes following the length field (field sized for length <= TBD bytes; messages do NOT span this data structure)* |
| Message data | 7 to N | | *New Horizons formatted instrument command max N is TBD* |

*3.6.3.2 Command Format.*

The format of the message data within a command ITF is specified in Table XXIII.

**Table XXIII. Message Data**

| Name | Length (bits) | Value | Description |
|---|---|---|---|
| Opcode | 16 | | Instrument splits this into 2 8-bit fields (dest, opcode). |
| Macro field | 2 | 00 = Execute real-time<br>01 = Invalid<br>10 = Add to Inst. Macro<br>11 = Add to DPU Macro | Instrument use only.<br><br>Unused or spare (0x00) for non-instrument commands. |
| Command word count | 14 | N+2 | Number of 32-bit command words (N) + 2 |
| Command words | 32*N | | |
| Checksum | 32 | | 32-bit XOR of the fields defined above |

This format is identical to the NH Spacecraft formatted commands and allows for multiple commands to be packed into a single CCSDS packet. However, only a single instrument command will be sent via ITF at a time.

*3.6.3.3 Memory Load Command*

New Horizons has adopted a single memory load command format for all onboard processors. The format of the command is specified in Table XXIV.

**Table XXIV. Memory Load Command Message**

| Item | Length (bits) | Value | Description |
|---|---|---|---|
| Opcode | 16 | | |
| Macro Field | 2 | → | 00 = Execute real-time<br>01 = Invalid<br>10 = Add to Inst. Macro<br>11 = Add to DPU Macro<br>Unused (00) for non-APL instruments |
| Length | 14 | 5-68 | Command length in 32-bit words (4 + (N+M)/4). |
| Address | 32 | → | Start address |
| Byte Count | 16 | 1-256 | Number of bytes to be loaded (not including pad).<br>*New Horizons* has chosen a maximum load size of 256 bytes. |
| Memory Type | 8 | TBD | This field is available for memory architectures that cannot determine the memory type from the address. This field is spare (0x00) for architectures that can. |
| Spare | 8 | 0 | Unused (0x00) |
| Data | N*8 | | 1≤ N ≤ 256 (From 1 to 256 8-bit bytes) |
| Pad | M*8 | 0 | 0≤ M ≤ 3<br>Pad up to 32-bit boundary. This data will not be loaded but will be included in XOR calculation. |
| Checksum | 32 | → | 32-bit XOR |

### 3.6.3.4 Memory Dump Command

The format for the New Horizons memory dump command is specified in Table XXV.

**Table XXV. Memory Dump Command Message**

| Item | Length (bits) | Value | Description |
|---|---|---|---|
| Opcode | 16 | → | |
| Macro Field | 2 | → | 00 = Execute real-time<br>01 = Invalid<br>10 = Add to Inst. Macro<br>11 = Add to DPU Macro<br>Unused (00) for non-APL instruments |
| Length | 14 | 5 | Command length in 32-bit words |
| Address | 32 | → | Start Address |
| Byte Count | 32 | → | Number of bytes to be dumped. |
| Memory Type | 8 | TBD | This field is available for memory architectures that cannot determine the memory type from the address. This field is spare (0x00) for architectures that can. |
| Spare | 24 | 0 | Unused (0x00) |
| Checksum | 32 | → | 32-bit XOR |

### 3.6.4 MET Time Message

An MET time message is sent from the C&DH to the instrument once per second. The MET time is the predicted time of the next 1PPS. The MET ITF is defined in Table XXVI. The MET ITF is transferred within 950 ms following each 1PPS epoch.

**Table XXVI. C & DH-to-Instrument Spacecraft Time Message in Instrument Transfer Frame**

| Name | Length (Bits) | Value | Description |
|---|---|---|---|
| sync pattern 1 | 8 | $0 \times fe$ | |
| sync pattern 2 | 8 | $0 \times fa$ | |
| sync pattern 3 | 8 | $0 \times 30$ | |
| Message type | 8 | $0 \times 01$ | S/C-Time |
| Checksum | 8 | | Byte-wise exclusive-OR of all bytes following the checksum |
| Message length | 16 | $0 \times 0004 + N$ | Expressed in number of bytes of time information plus any additional information sent to that instrument. |
| S/C-Time | 32 | MET | Mission elapsed time corresponding to the predicted S/C time at the next SYNC pulse |
| Additional information sent to the instrument | 8*N | N/A | Additional information sent to an instrument can include a closest approach countdown, whether that instrument is allowed to produce a memory dump packet in the next second, scan rate, etc. This field will be customized for PEPSSI, and will be fixed from one second to the next. |

## 3.7. Instrument Environmental Design Requirements

### 3.7.1 Thermal Interface

The PEPSSI-to-spacecraft thermal interface includes a conduction component via the spacecraft-mounting surface and a radiation component derived by a ratio between blanketed and painted surfaces Two spacecraft-monitored temperature sensors are continuously monitored via the spacecraft Command and Data Handling (C&DH) system. These sensors are used to ascertain the temperature of the instrument. The spacecraft thermal control system maintains the temperature on the spacecraft side of the PEPSSI mounting plate, and the support electronics box mounting points. The spacecraft has no specific requirement for control of the temperature gradient across the mounting interface from any one point to another. The spacecraft also has no specific requirement for control of the interface temperature rate of change. Internal temperature measurements for PEPSSI are included in PEPSSI's housekeeping telemetry data.

### 3.7.2 PEPSSI Thermal Design Requirements

Design and test limits give the temperature ranges over which PEPSSI was cycled during thermal vacuum testing. The S/C Interface Temperatures Operating/Test temperatures are 0°C to +40°C, the S/C Interface Temperatures Non-Operating Survival are –20°C to +40°C, and the Internal Operating Limits are with Baseplate Control for the instrument itself and –45ºC to +45ºC for the electronics boards and –35ºC to +35ºC for the solid state detectors. The non-operating survival limits assume the PEPSSI cruise heater is not on. The heater may be turned on during the cruise phase when the PEPSSI is off and when observatory power is available, to provide margin with respect to the non-operating −20°C limit.

3.7.3  Radiation Shielding Requirements

The New Horizons radiation environment includes exposure to solar protons, Jovian electrons, galactic cosmic rays, and neutrons and gamma-rays emitted from the radioisotope thermoelectric generator (RTG) power source. PEPSSI is designed to survive a total ionizing dose of 15 kilorads without failure and to be immune to latch-up. Additional spot shielding to parts was added as required to bring the instrument to this overall shielding level.

3.7.4  Electrostatic Requirements

The electrostatic requirement for PEPSSI is that the observatory's total charge be less than 1000 volts. Analysis by the New Horizons project for the Jupiter flyby showed show a worst case (35 $R_J$ closest approach) potential of ~ +5 to +10 volts on isolated solar-illuminated surfaces, and ~ −200 volts on isolated surfaces in darkness. There is no hard information of the plasma environment of Pluto, but it is not expected to possess a strong intrinsic magnetic field and, therefore, is not expected to have a mechanism to produce a population of particles sufficiently energetic to charge the observatory to a level of concern. The worst Jupiter case is less than the maximum allowable for PEPSSI by a margin of five; and the electrostatic charge at Pluto is expected to be substantially less. The requirement is expected to be easily met.

**3.8.  Test Requirements**

3.8.1  Vibration

Instrument test vibration levels are specified in project documentation for levels appropriate to the launch vehicle stack used for New Horizons.  Testing philosophy was based on a protoflight approach, where flight hardware is tested to qualification levels for flight acceptance durations. Tests included sine vibrations test at levels applicable at the spacecraft-mounting interface, with vibration test of three orthogonal axes, including the thrust axis, performed, as well as random vibration tests with the same geometry.

3.8.2 Thermal

The PEPSSI instrument is mounted on a bracket on a side deck of the spacecraft and is thermally conductive through the mounting interface. The instrument temperature is controlled by a tailored thermal coupling at the mounting interface and use of instrument radiating surfaces. No heaters are required. The ranges of mounting interface temperatures, at the bracket-to-spacecraft deck interface, are given in Table XXVII.

**Table XXVII. PEPSSI Thermal Test Limits**

| Condition | Limits |
|---|---|
| Typical | +30°C to +40°C |
| Operating | 0°C to +40°C |
| Operating test | −10°C to +50°C |
| Non-operating survival | −30°C to +50°C (Includes 10°C margin) |

3.8.3 Acoustics and Shock

All instruments, including PEPSSI, were mounted on the spacecraft during separation shock and acoustic test. In addition, acoustic tests were performed on the PEPSSI engineering model instrument, using protoflight acoustic levels, to verify the integrity of the design with respect to acoustic noise. No instrument level acoustic test was carried out for the flight PEPSSI hardware.

3.8.4 EMI / EMC

PEPSSI electromagnetic interference (EMI) and electromagnetic contamination (EMC) level requirements are set forth in the New Horizons project documents. Full conducted emissions and conducted susceptibility as well as radiated emissions, radiated susceptibility, and transients measure tests were run prior to integration with the spacecraft.

## 4 Performance

### 4.1. Data Conversion to Physical Units

PEPSSI measurements are intended to generate the information needed to derive the charged ion and electron differential intensities (**I** [cm$^{-2}$ sr$^{-1}$ s$^{-1}$ keV$^{-1}$]) and phase space densities (**PSD**[s$^3$ cm$^{-6}$]). The significance of **I** and **PSD** are that, for many particle transport processes within planetary magnetospheres and interplanetary space, they satisfy the collision-less Boltzmann equation, derived from Liouville's Theorem for collision-less transport, for energy preserving, and for non-energy preserving processes, respectively. Our purpose here is to develop the quantitative procedures for converting the count rates (**R**[counts s$^{-1}$]) reported by PEPSSI in the many predefined channels described in preceding sections, or in channels defined on the ground using the pulse-height-analysis (PHA) data, into estimates of **I** and **PSD** for the various defined ranges of energies, particle species, and arrival angles. We begin by defining terms.

4.1.1 Flux, Differential Intensity and Phase Space Density

The rate **R**[particles/s] that traverse an area A can be given by the Flux **F** [cm$^{-2}$ s$^{-1}$] with:

$$\mathbf{R} = A * \mathbf{F} \tag{1}$$

or by the intensity **i** [cm$^2$ s$^{-1}$ sr$^{-1}$]

$$\mathbf{R} = A * \int \mathbf{i} \cos(\hat{u}) \, d\Omega \tag{2}$$

where $\Omega$ is solid angle and $\hat{u}$ is the angle to the area normal.

Often used is the quantity differential intensity **I** [cm$^2$ s$^{-1}$ sr$^{-1}$ keV$^{-1}$], defined as the number of particles of a given species S with energy between E and E+$\Delta$E that traverse the area A during the time t, where

$$\mathbf{R}(S, E) = \mathbf{I}(S, E) * A * \Delta\Omega * \Delta E \tag{3}$$

In three dimensions, with $\theta$ being the polar angle and $\phi$ the azimuthal angle of a polar reference system:

$$d^3\mathbf{R}(S, E, \theta, \phi) = \mathbf{I}(S, E, \theta, \phi) * A(\theta, \phi) \cos(\hat{u}) * t * dE \cos\theta\, d\theta\, d\phi \qquad (4)$$

where we note that the area A and the unit vector û pointing normal to A are both functions of $\theta$ and $\phi$.

Note that for non-relativistic energies, $\mathbf{I}(S, E, \theta, \phi)$ is related to the phase space density **PSD** (number of particles in the configuration space element $d^3R$ and with velocity between v and v+ $d^3v$) by the simple relationship

$$\mathbf{PSD}[s^3\ cm^{-6}] = \mathbf{I}[cm^{-2}\ s^{-1}\ sr^{-1}\ keV^{-1}] * m/v^2 \qquad (5)$$

For relativistic energies, the phase space density (identified with lower case letters here: **psd**) is defined in terms of particle momentum **p** ($\mathbf{psd}[s^3\ cm^{-6}\ kgm^{-3}]$), and the relationship to differential intensity **I** is:

$$\mathbf{psd}[s^3\ cm^{-6}\ kgm^{-3}] = \mathbf{I}[cm^{-2}\ s^{-1}\ sr^{-1}\ keV^{-1}] / \mathbf{p}^2 \qquad (6)$$

4.1.2 Definition of Sensor Transfer Function and Geometric Factor

Using (4), the rate **R**[particles/s] of particles of species S, in the energy band $\Delta E$ around mean energy E, angular band $\Delta\theta$ around mean polar direction $\theta$, and angular band $\Delta\phi$ around the mean azimuthal direction $\phi$, measured by the instrument can be expressed as:

$$\mathbf{R}(S, E) = \int_{\Delta E} \int_{d\theta} \int_{d\phi} \mathbf{I}(S, E, \theta, \phi) * A(\theta, \phi) \cos(\hat{u}) * dE \cos(\theta)\, d\theta\, d\phi \qquad (7)$$

One may take the kernel approach to relating **R** to **I** by allowing $\mathbf{I}(S, E, \theta, \phi)$ to be a Dirac $\delta$-function (mono-energetic, infinitely narrow beam). Then:

$$d\mathbf{R}(S, E, \theta, \phi) = d\mathbf{I}(S, E, \theta, \phi) * \mathbf{T}(S, E, \theta, \phi) \qquad (8)$$

where $\mathbf{G}(S, E, \theta, \phi)$ [$cm^2$ sr keV] is the kernel, also known as the "transfer function" of the instrument. Then:

$$\mathbf{R}(S,E) = \int_{\Delta E} \int_{d\theta} \int_{d\phi} \mathbf{I}(S, E, \theta, \phi) * \mathbf{T}(S, E, \theta, \phi) \cos(\theta) d\theta\, d\phi \qquad (9)$$

It is typical for the first estimate to assume that the particle intensities are constant over the energy-species-angle bandwidths of any one channel. The estimated intensity $I_{kj}$ associated with a specific detector **k**, which at a specific time views a range of angles centered in a specific direction ($\theta_k$, $\phi_k$), and which is associated with a specific energy channel **j**, which measures particle species $S_j$ over an energy range that stretches from $E_j$ to $E_j + \Delta E_j$, is given by equation (9) as:

$$\mathbf{R}_{kj} = \int_{\Delta E_j} \mathbf{R}_k(S_j, E_j) \, dE = \int_{\Delta E_j} \mathbf{I}_k(S_j, E) * \mathbf{T}(S_j, E)) \, dE \sim <\mathbf{T}_{kj}> \mathbf{I}_k(S_j, E_j) \Delta E_{kj} \quad (10)$$

where $\mathbf{R}_{kj}$ is the channel count rate and where $<\mathbf{T}_{jk}>$ is called here the Transfer Factor $\mathbf{H}_{kj}$. It is convenient to separate $\mathbf{H}_{kj}$ into a counting efficiency "$\varepsilon_{kj}$" and a factor that relates strictly to geometry, the so-called geometric factor $\mathbf{G}_{kj}$. That is:

$$\mathbf{I}_k(S_j, E_j) \sim \mathbf{R}_{kj} / (\varepsilon_{kj} \, \mathbf{G}_{kj} \, \Delta E_{kj}) \quad (11)$$

Therefore, the first goal of calibration is to obtain detector-channel-averaged transfer factors $\mathbf{H}_{kj}$ (or the equivalent efficiency factor $\varepsilon_{kj}$ since $\mathbf{G}_{kj}$ is easy to calculate), the detector-channel energies $E_{kj}$ and the detector-channel energy band pass $\Delta E_{kj}$.

More quantitative work may require a full inversion of the kernel equation given in Equation (9). Here we are usually aided by the fact that to some level of approximation the Transfer Function **T** is separable in the form:

$$\mathbf{T}(S, E, \theta, \phi) = \mathbf{K}(S, E) * \mathbf{P}(\theta, \phi) \quad (12a)$$

and hopefully even

$$\mathbf{T}(S, E, \theta, \phi) = \mathbf{K}(S, E) * \mathbf{C}(\theta) * \mathbf{D}(\phi) \quad (12b)$$

Note that even with these separations, an exact inversion of the integral is rarely possible, and we can compute only the coefficients of some tailored expansion of **K**, **C**, and **D**, as for example in spherical harmonics (using Legendre polynomials). The accuracy of these coefficients depends on both the raster coverage of the measurements and on the calibration.

### 4.1.3 Goals of the PEPSSI Characterization and Calibration Efforts

The goal of PEPSSI characterization and calibration efforts, then, is to relate the rates, $\mathbf{R}_{kj}$, of the on-board or ground-defined channels to the *in situ* particle intensities $\mathbf{I}(S, E, \theta, \phi)$ by developing a hierarchy of information about the transfer function $\mathbf{T}(S, E, \theta, \phi)$. Specifically, we wish to derive for each channel:

$E_{kj}$, $\Delta E_{kj}$, $\varepsilon_{kj}$ $G_{kj\,j}$, $\mathbf{C}_{kj}(\theta)$, $\mathbf{D}_{kj}(\phi)$, $\mathbf{K}_{kj}(S, E)$, $\mathbf{P}_{kj}(\theta, \phi)$, and finally, $\mathbf{T}_{kj}(S, E, \theta, \phi)$

Clearly, the further to the right we move into this list, the higher the fidelity the characterization and calibration efforts must be.

Typical calibration procedure consists of stimulating the instrument with energetic particles, first from radiation sources, and then from accelerator beams (fair approximation of a delta function) and recording the response of the different channels. The ideal calibration for PEPSSI meant a scan of the following variables:

Species and Mass: e, H, He, O, Ar (proxy for heavier species, e.g., Fe)

Energy: 1 keV – 1 MeV (~30 energies for 10 points/decade)

Polar angle: +/- 10 deg (12 degrees nominal FOV)

Azimuthal angle: +/- 90 deg (160 degrees nominal FOV)

Note that even given this level of discreetness, complete characterization to the level of establishing $T_{kj}(S, E, \theta, \phi)$ for all channels would have required: 5 masses * 30 energies * 21 polar angles * 180 azimuthal angles = 567,000 calibration points! Clearly it was (and is) not possible to run this many beam values without significant infrastructure and automation. Therefore, we are dependent to a substantial degree on the separability of the transfer function as described in Equation (12). An important goal of the characterization and calibration efforts was, in fact, to establish that degree of separability. Simulations play an important role in establishing expectations for the instrument, as well.

## 4.2. Simulations

Described here are the calculated expectations regarding the performance of the PEPSSI instrument. We describe various calculations and simulation runs that provide a higher level of fidelity in predicting the performance of the PEPSSI instrument than those based on experience with previous instruments. These simulations constitute the benchmarks against which subsequent ground and flight calibrations are compared

Numerical simulation provides an efficient and effective mean of verifying and quantifying the flight instrument response. Hence, a range of simulation tools are developed in order to characterize PEPSSI. The results from these simulation tools allow us to calculate the end-to-end expected performance of the instrument which one can use to compare with in-flight and calibration data. Different sections of PEPSSI are simulated using different tools to gain better understanding on each part of the instrument. The following sections described

### 4.2.1 Geometric Factor

The number of data points needed to measure the geometric factor is very large, and the time needed to perform these measurements was well beyond the available time. Instead, estimates of the geometric factor rely on computer modeling and analysis. The very same analysis formed the base for the selection of the collimator best configuration: 23 blades, each 25 mm long. This geometry for the collimator defines the maximum acceptance angles through the collimator and the geometric factor GF = 0.14 cm$^2$ sr, which is the total geometric factor for the start system.

### 4.2.2 Ion Measurements

#### *4.2.2.1 Electron optics*

As discussed previously, secondary electrons generated by primary ions in a start and a stop foil are used to generate timing signals to measure the ion's time-of-flight (TOF) within PEPSSI. They are also used to determine the directionality of the low energy ions that do not stimulate the SSDs. Figure 13 shows sample simulated secondary trajectories within PEPSSI from the foil to MCP, derived using the commercial COULOMB software (see web site: "**www.integratedsoft.com/products/coulomb/**").

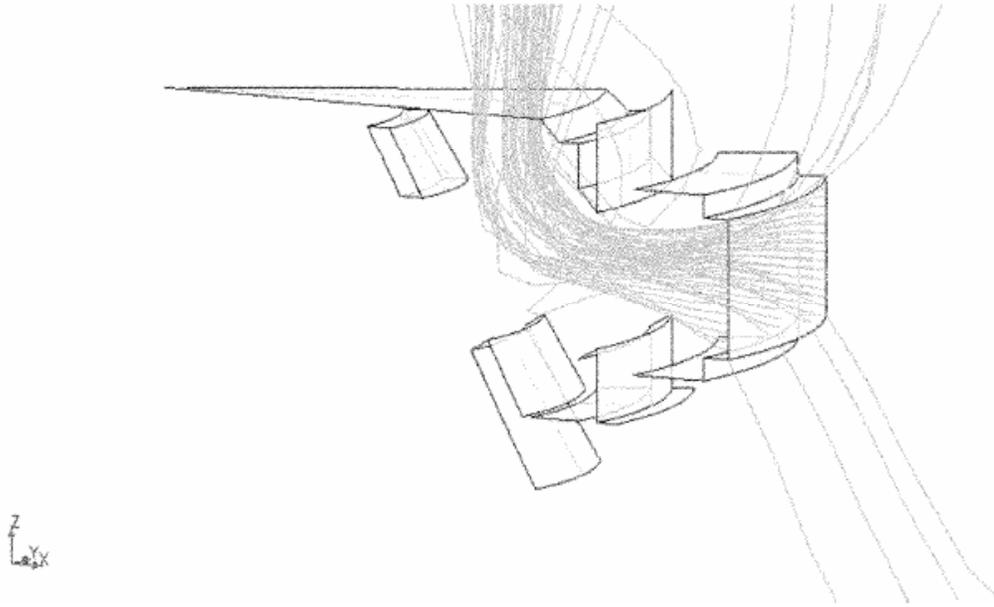

**Fig. 13.** Simulated secondary electron trajectories within the PEPSSI sensor.

Here, a pie-shaped cutout of the cylindrical PEPSSI sensor is shown (cylindrical symmetry is assumed for the electrostatic fields). The electron optics for both the start foil/MCP and stop foil/MCP are identical, hence the figure shows a good representation for both start and stop electron trajectories. Electrons are generated at the foil (right of the figure) when ions (not shown) transverse through the foil. These low energy electrons (typically <10 eV) are immediately affected by the accelerating voltages within PEPSSI (the high energy ions are not significantly affected by the accelerating voltages) and are steered towards the MCP (the flat surface at the top) as evident in the figure. An important issue is the dispersion in the arrival times of the secondary electrons as a function of foil position and the angle of emittance of the secondary electrons from the foil. Such dispersion adds to the error in the measurement of time-of-flight. Figure 14 (top) shows the time dispersion associated with varying angles for electrons emitted from the center of the foil. In Fig. 14 (bottom) the dispersion shown for any one column are associated with varying the emission position over the ~6 mm vertical extent of the foil. The combination of position and angle yields a time dispersion of roughly 1.5 – 2 ns. Combining the dispersions associated with the start and stop detections in a root-mean-square sense yields a total time dispersion error of roughly 2.1 – 2.8 ns. Combining that value in a root-mean-square sense with the ~1.5 ns electronic dispersion yields a total time dispersion that resides between 2.5 and 3.2 ns.

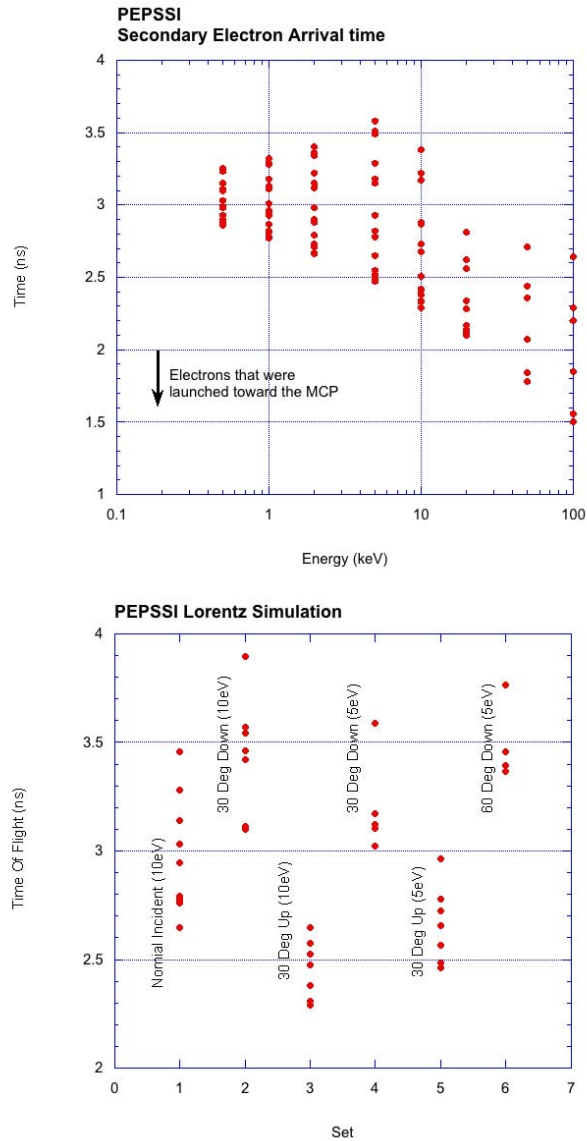

**Fig. 14.** Time dispersion associated with varying angles for electrons emitted from the center of the foil.

*4.2.2.2   Ion energy losses*

As ions transverse the TOF section of PEPSSI through the front foil (~10 μg/cm2), back foil (~19 μg/cm2), dead layer of the solid state detector (~550 Å), and eventually stop and deposit their total energy in the SSD, they lose energy and scatter depending upon the ion's initial energy, mass, and the medium through which they pass. We have incorporated data from the SRIM particle-interaction-with-matter code **(Biersack & Haggmark 1980)** in our simulation to simulate realistically the energy loss and scattering as ions transverse through different materials in the PEPSSI sensor.

*4.2.2.3 TOF measurements*

When ions penetrate through the front foil, a distribution of ion velocities is created, as calculated using SRIM. This distribution of ions is then used to calculate the distributions of ion TOFs. The uncertainty due to secondary electron dispersion and electron noise together with SRIM data are all incorporated in our simulation. Figure 15 the simulated response of PEPSSI TOF spectra as a function of particle initial energy for four species (H, 3He, 16O, and 56Fe). At low energy (~10s of keV), ions lose significant amounts of energy and scatter significantly when going through the front foil. These effects explain the spread in TOF measurement at low energy. However, at higher energy (~100s of keV), the TOF spreads are mostly consequences of the uncertainties in the TOF measurement from both the electronics jitteri (~1.5 ns), and secondary-electron dispersion in the TOF optics as discussed in previous section (except for very heavy ions, e.g., Fe).

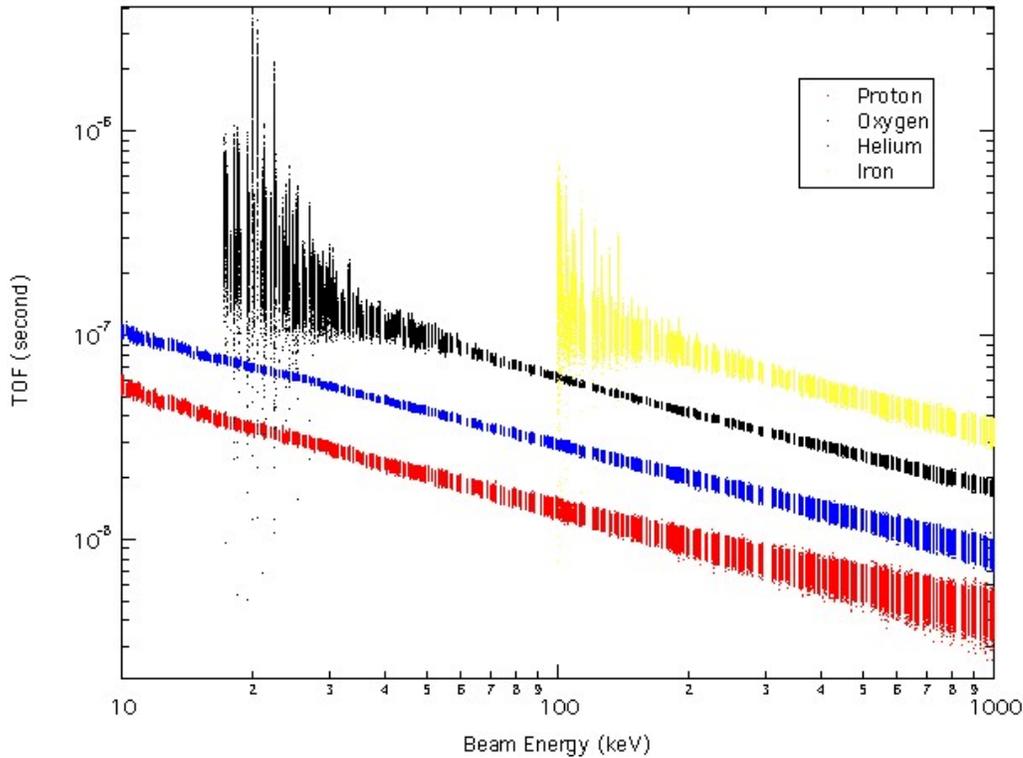

**Fig. 15.** Simulated TOF spectra as a function of accelerator beam energy for proton, oxygen, helium and iron.

*4.2.2.4   Total energy measurement*

If an ion has sufficient energy left once it transits the front and back foil, it ends up in the SSD. Depending on the ion's final energy and mass when it reaches the SSD, it can penetrate through the dead layer of the SSD and produce an electronic signal to be measured. Figure 16 the simulated measured energy by PEPSSI as a function of the ion's initial energy before entry into PEPSSI. The measured energy shows the cumulative effect of the energy spread due to energy lost throughout the PEPSSI (start foil, stop foil, dead layer, and electron hole pair production). The energy loss is greatest for ions with the greatest nuclear charge, here Fe.

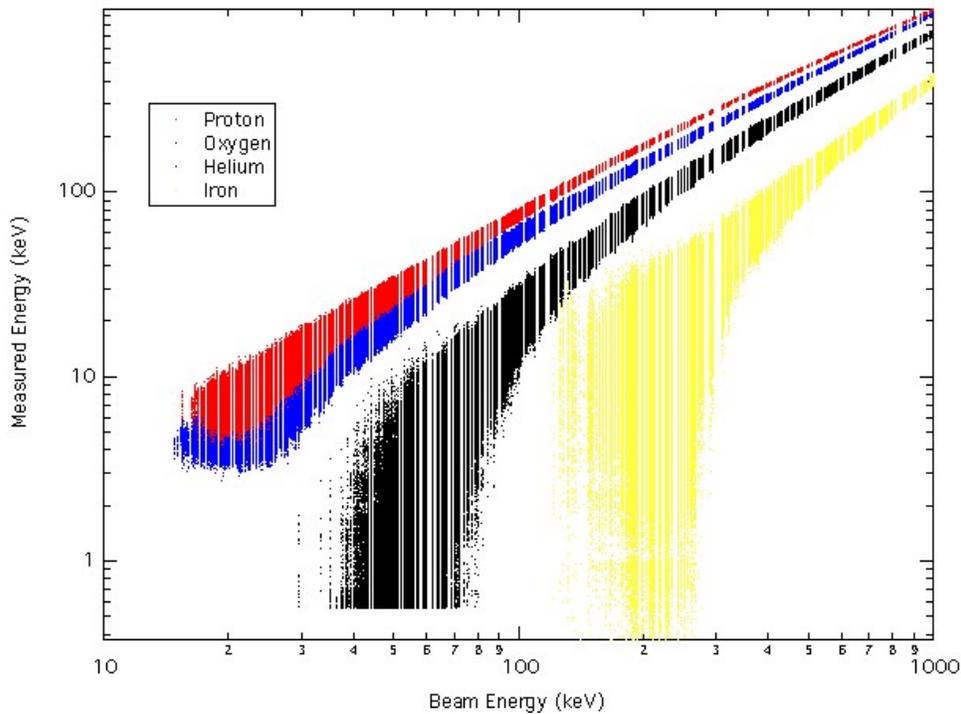

**Fig. 16.** Simulated measured energy versus incident energy for four different ions.

*4.2.2.5 TOF versus energy*

Finally, **Error! Reference source not found.** 17 shows the result of combining the energy measurement with the TOF measurement on a particle-by-particle basis. Notice that even with a ~3 ns time dispersion, the major elemental species of interest (H, He, CNO, Fe) are discriminated except at the lowest energies for CNO and Fe, where the error in the energy measurements dominates.

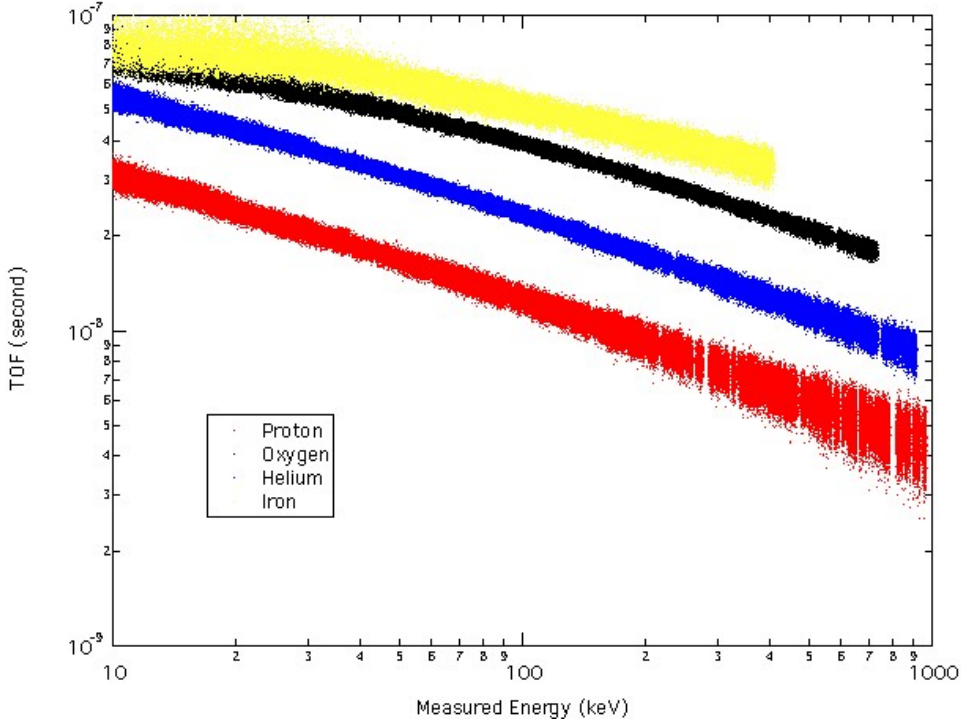

**Fig. 17.** Simulated TOF and measured energy for PEPSSI.

*4.2.2.6 Efficiencies*

The efficiency for detection of an ion within the SSD's is roughly ~100% (except at the very lowest energies where energy straggling can position the energy below the low energy threshold). The efficiency for obtaining a TOF measurement is estimated using the efficiency of generating secondary electrons in both the front and the rear foil. To emit a secondary electron, such electrons must be generated close enough (distance "ρ") to the surface of the foil so that the electron can escape before it is re-assimilated. Thus, very roughly, it is expected that the efficiency for the generation of a secondary electron is proportional to the amount of energy per unit distance (dE/dX [keV/micron]) that an ion deposits as it goes through this outer thin layer of the foil. The canonical number of secondary electrons generated out of each surface as a proton with 10's of keV energy encounters the foil is between 0.5 and 1 (**Frischkorn & al. 1983**); R. W. McEntire, Private communication, 2004; estimated here as 0.75). Since we require a simultaneous start and stop electron, the efficiency of proton detection at 10's of keV is $\left(1-e^{-0.75}\right)^2$, where $\left(1-e^{-0.75}\right)$ is the Poisson probability of having at least 1 or more electrons emitted when the mean emission number is 0.75. Thus, the efficiency is roughly 28%. Other energies and species may be roughly scaled with this number using tabulated dE/dX values. For example, at 50 keV total energy the dE/dX values for protons and oxygen ions are 120 and 250 keV/micron, respectively (we ignore for now the energy losses suffered by the ion in getting to the position in either foil where the secondary electron is generated). Thus, since the average number of secondary electrons for oxygen will by ~0.75 x (250/120), or 1.56. Poisson statistics tells us that the probability of detecting an oxygen ion with 50 keV energy is $\left(1-e^{-1.56}\right)^2$ or 62%.

### 4.2.3 Electron Measurements

The PEPSSI electron measurement strategy depends on the use of the aluminum flashing on the electron SSD. We therefore need to understand the effect of that flashing on both the ion and electron measurements within the electron SSDs. Based on simulations with GEANT-4 (Agostinelli & al. 2003), Fig. 18 shows the effect of Al flashing on proton measurements, and **Error! Reference source not found.** shows the effect on electron measurements, for a varying thickness of Al flashing. The baseline spectrum assumed for both protons and electrons is a power-law spectrum with a spectral index of 3; that is, Intensity = A E-3, where "A" is a constant. With the flashing thickness utilized with PEPSSI, 1 micron Al, the proton intensities are severely depleted for energies < 250 keV. With that flashing thickness, the electron intensities are maintained to energies down to about 20 keV. The drop off of electron intensities at the highest energies occurs because of the electron penetration of the 500 micron detectors. That drop-off is rounded because of electron scattering within the SSD, as demonstrated with an examination of the detailed trajectories of individual electrons with the GEANT-4 simulations. There may be periods of time when the ion counts overwhelm the electron measurements. The most important issue is not that the electrons may be occasionally contaminated, but whether or not the electrons are contaminated in a way that the ion counts are confused with what are really electron counts. This problem is intrinsic to the technique used here to measure the electrons. Extensive experience with similar measurement made by the EPIC instrument on Geotail (Williams & al. 1994) demonstrates that, because of the clean proton measurements (below and above 250 keV) made in roughly the same direction, it is unlikely that ions and electrons will be confused.

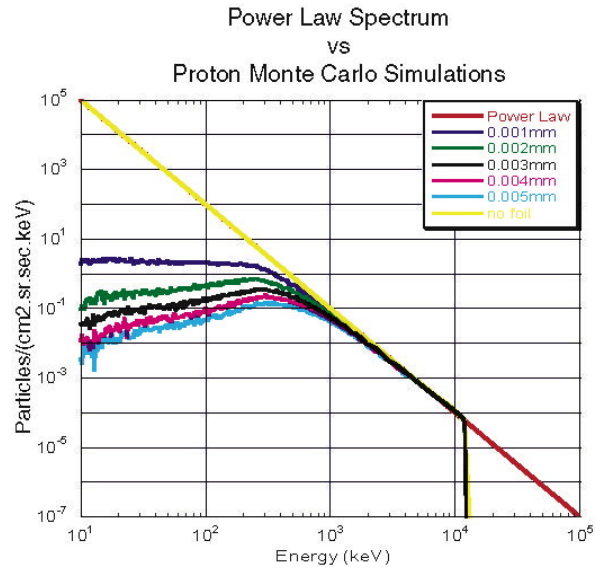

**Fig. 18.** Simulated measured proton spectra for different Al flashing thickness from a known incident spectra (red) on a 500 μm thick silicon detector.

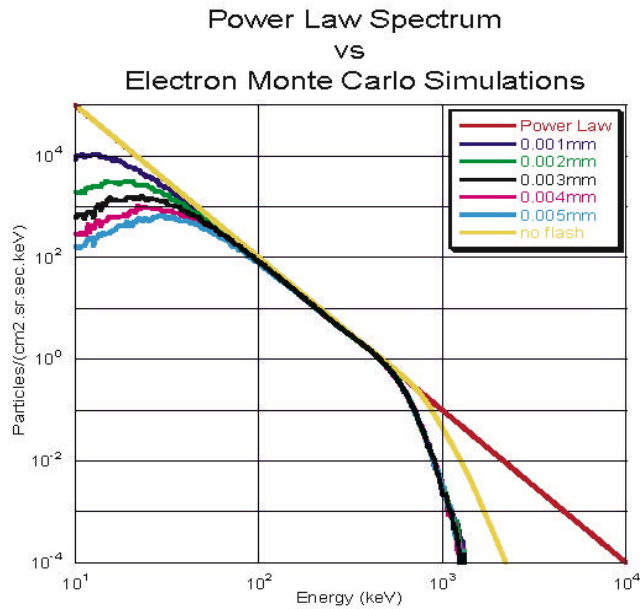

**Fig. 19.** Simulated measured electron spectra for different Al flashing thickness from a known incident spectra (red) on a 500 μm thick silicon detector.

## 4.3. PEPSSI Flight Unit Calibration

4.3.1 The JHU/APL Calibration Facility

Calibration of the PEPSSI instrument was conducted at the JHU/APL instrument testing facility through a combination of long-duration, radiation-source exposures (Am-241 degraded alpha source) and discrete exposures from the 170-kV particle accelerator beam. The Air Insulated Accelerator, obtained from Peabody Scientific, is a versatile system capable of producing a broad range of ion species to energies of 170 keV. The system produces beams of H, O, and noble gas ions with intensities over the range of $\sim 10^2$ to $10^6$ particles/cm$^2$/sec at the target position in an energy range of 3 to 170 keV. Figure 20 shows the picture of the accelerator HV terminal.

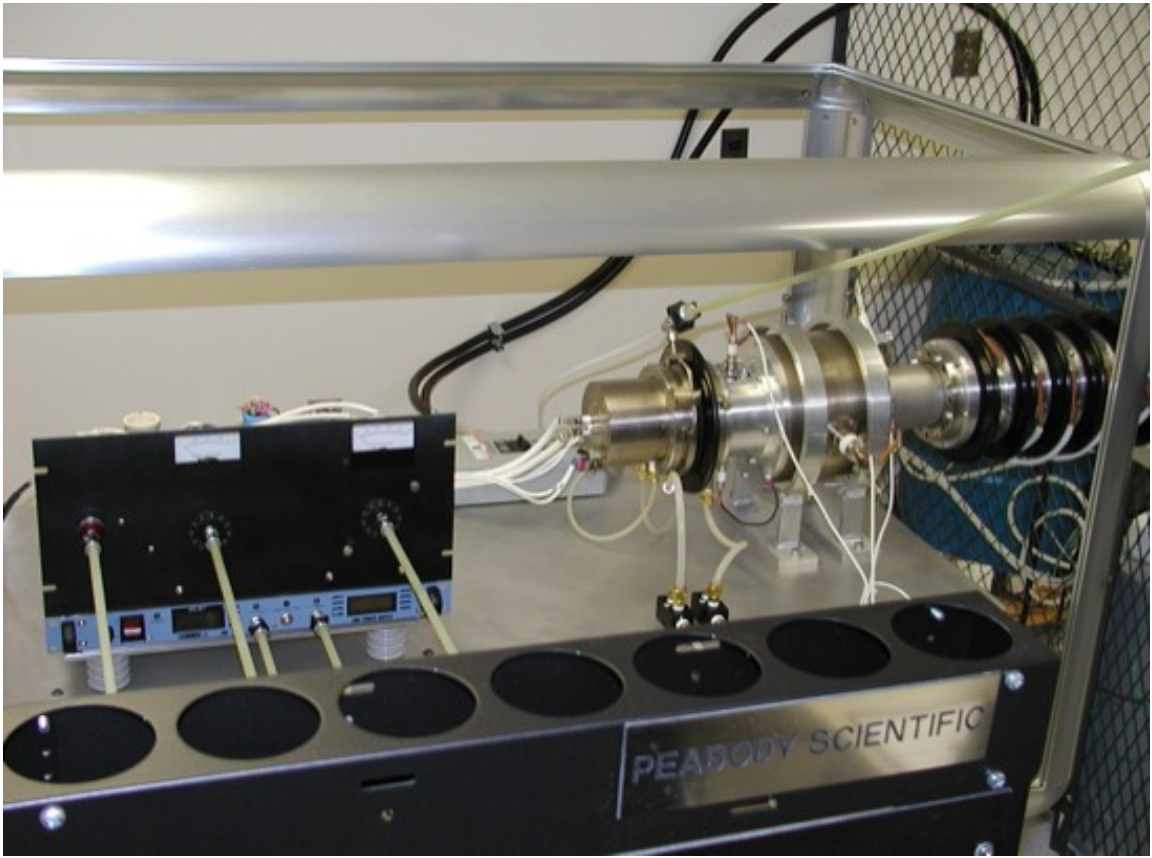

**Fig. 20.** The JHU/APL particle accelerator showing the high-voltage terminal.

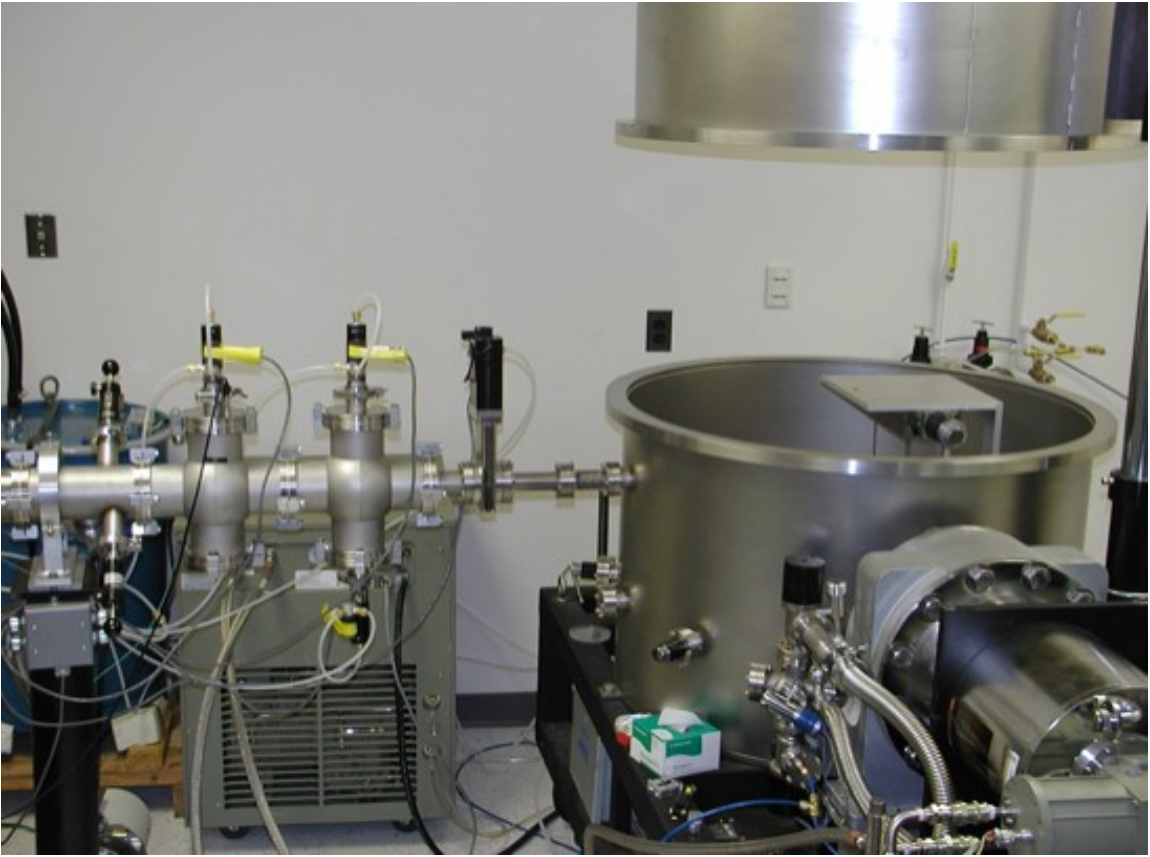

**Fig. 21.** The vacuum chamber used in calibrating the PEPSSI unit. The beam line enters the chamber horizontally from left to right in the picture.

A one-meter diameter and one meter tall cylindrical vacuum chamber is attached to the end of the accelerator beam line with a small gate valve to serve as the shutter (shown in Fig. 21). The chamber has two rotation stages ($\pm 0.1°$) and two translation stages ($\pm 2\mu m$) that are controlled by stepper motors with digital read-out capability. Within the beam line and in the chamber itself there are various mechanical shutters that are used to control the ion beam size from 1 mm$^2$ to 1 cm$^2$. Three different beam monitors are used in the accelerator and chamber for diagnostics. A Faraday cup in the beam line before the beam enters the chamber wall is used for initial diagnostics of the beam intensity. In addition, within the chamber itself there is a stepper-motor-controlled moveable fixture that is mounted with both a channeltron and an SSD to serve as the primary beam monitor. The moveable fixture allows measurement of the species, energy and intensity of the beam inside the chamber itself.

### 4.3.2 Test Set Up

For both the alpha particle and beam calibration test, the PEPSSI unit was mounted inside the vacuum chamber. An emulator box with was connected to PEPSSI to act as the New Horizons spacecraft in providing commands and power to the instrument. The emulator box was controlled by a Window-PC running GSEOS to operate and collect data from PEPSSI. All the calibration data were recorded under GSEOS as a recorded file inbinary format.

PEPSSI was in full flight configuration during calibration, including the collimator, flight SSDs and foils. The PEPPSI instrument was mounted to a rotating turntable that allowed the beam to be scanned in full 160° in the azimuth direction. The response of the instrument to the vertical angle was not investigated with the accelerator beam.

### 4.3.3 Ground Calibration

Numerous proton, helium, and oxygen beams, along with radioactive sources, were used to calibrate and characterize PEPSSI performance over the course of two weeks during the ground calibration period. Protons at energies from 30 to 170 keV and oxygen ions at energies from 50 to 170 keV were used to stimulate PEPSSI.

#### *4.3.3.1 Directionality*

Calibration runs like those depicted in Error! No bookmark name given. document the ability of the PEPSSI sensor determine the directionality of the incoming charged particles. Figure 23 shows the ability of the discrete "start" anodes to determine the directionality of the incoming particles for the low-energy ion events (no SSD signals). During this test, a constant, collimated oxygen beam is used, and PEPSSI rotated in the full 160° in the azimuth direction. All twelve internal collimator fins can be seen clearly in the figure overlaying the discrete anodes count rate. The discrete anodes perform the imaging function for low-energy ions. The proton beams used have angular spreads on the order of ~1°. The angular resolution achieved is adequate for the science to be performed. Figure 24 shows the normalized response of the SSDs to the same angular scan. All six pairs of SSD are accounted for except sector 0, 2, and 5 where the electron SSDs are located.

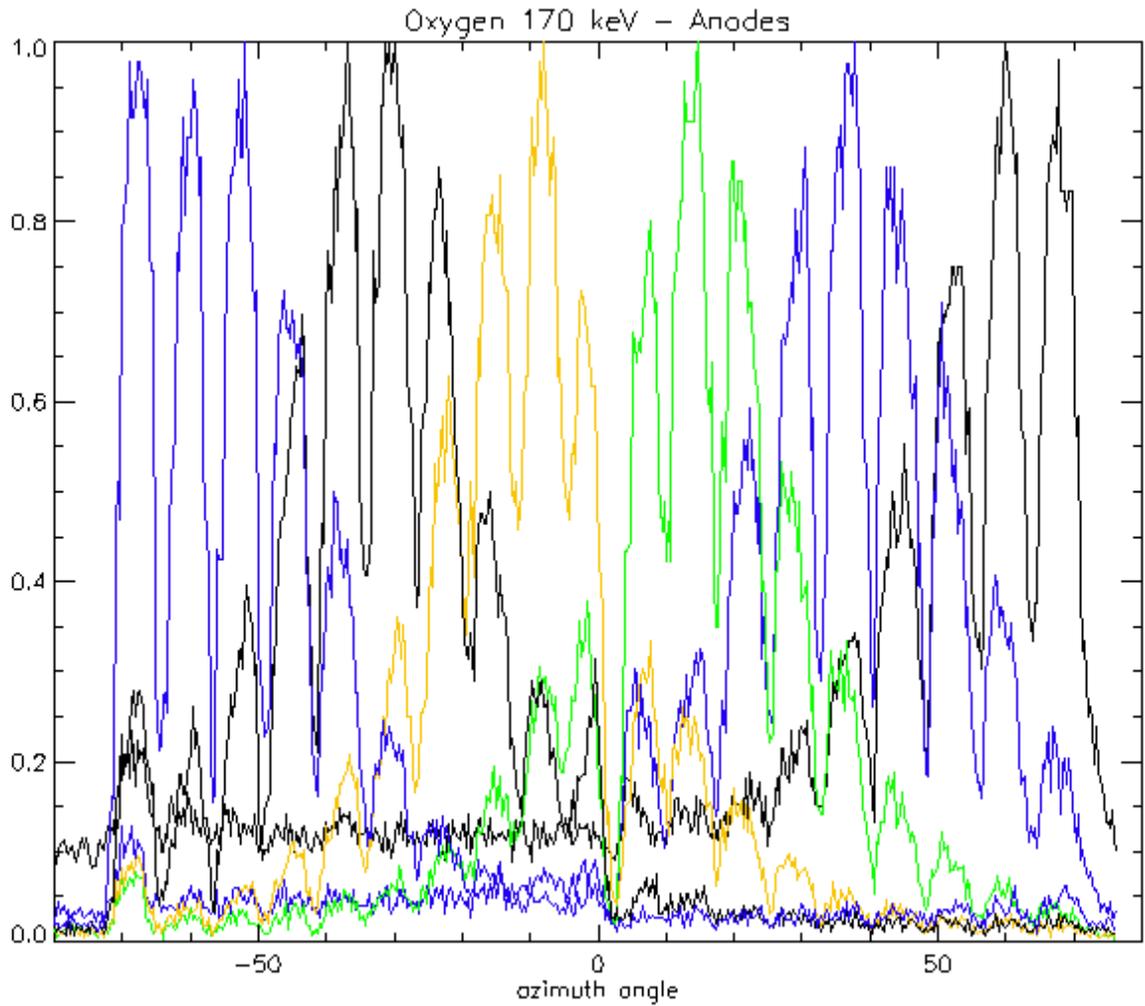

**Fig. 22.** Angular scan in the azimuth angle direction showing response of discrete start anode (colors) and the internal fins of the collimator to a narrowly collimated proton beam (170 keV). All 20 internal fins are seen in the scan. The proton beams used have angular spreads on the order of ~1 degree.

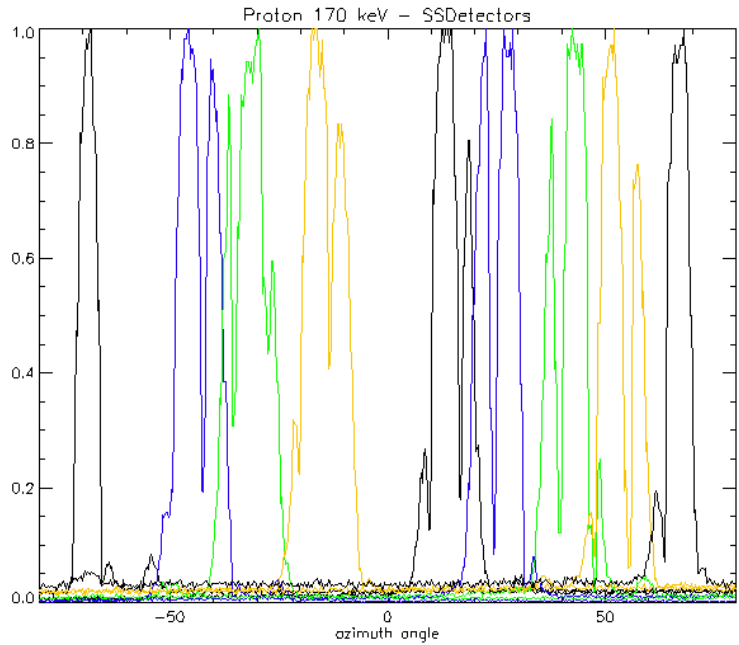

**Fig. 23.** Normalized response of all SSDs to the same angular scan. All detectors showed response from the beam, except sector 1, 5, and 11 where the electron SSDs are located.

### 4.3.3.2 Time-of-flight Resolution

The TOF spectrum of a 80 keV proton beam is shown in Fig. 24 with a 2.0 ns FWHM of the TOF peak. Intrinsically the TOF spread ranges between ~2 and 4 ns for the PEPSSI instrument. There are two causes that contribute to the TOF spread: 1) the dispersive spread in the secondary electrons (~2 ns); and 2) the performance of the TOF electronics (most notably the CFDs which is < 1 ns). The measured TOF spread agrees with the expected PEPSSI performance (2.5 – 3.5 ns). Figure 25 shows the conversion coefficients taken from simulation to relate the measured TOF bin number into actually TOF in nanoseconds. Figure 26 summarize the TOF performance of PEPSSI during beam calibration, the lowest energy that has a resolvable TOF spectrum for protons (oxygen ions) is 30 keV (50 keV).

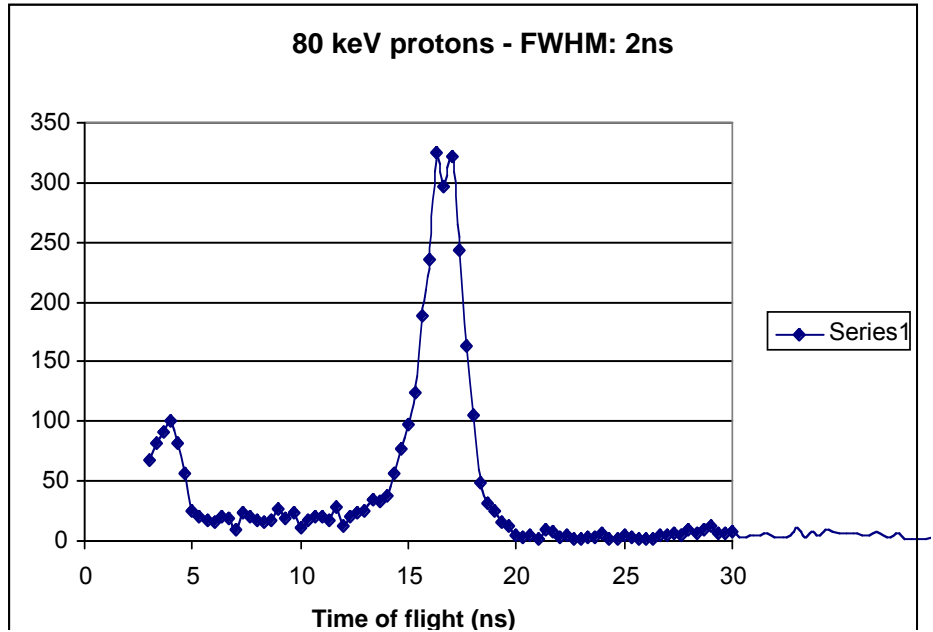

**Fig. 24.** The TOF spread of a 80 keV proton beam is shown in this figure. The fitted TOF FWHM is ~2 ns, better than the expected 2.5 ns TOF resolution.

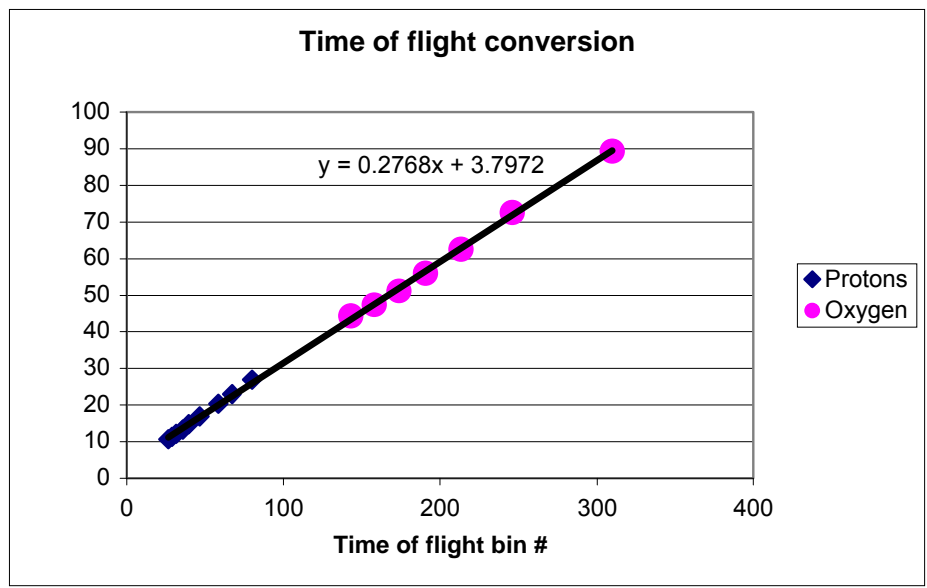

**Fig. 25.** The conversion coefficients relating the measured bin number and the actual TOF in nanoseconds. Conversion coefficients are calculated from the sensor simulation, taking into account pre-acceleration as well as losses through start foil.

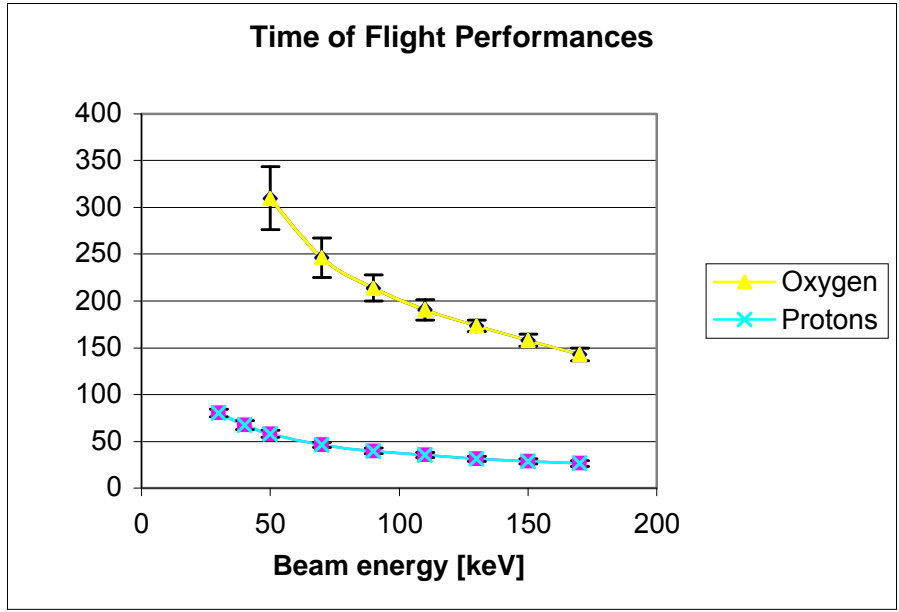

**Fig. 26.** TOF performance of all beam runs that were performed during the PEPSSI flight unit ground calibration. The lowest energy TOF spectrum that was observed for proton (oxygen) is 30 keV (50 keV).

*4.3.3.3 Energy Resolution*

Figure 27 shows the energy spectrum of a 130 keV proton beam measured by PEPSSI. The upper panel shows the enter spectra from 0 to 200 keV, and the lower panel is a close-up of the spectrum. The measured energy peak is at 120 keV and the fitted FWHM is 8.9 keV in excellent agreement with the simulation. The conversion coefficients for relating energy bin number into actual energy in keV are shown in Figs. 28 and 29 summarizes the energy performance of PEPSSI during beam calibration, the lowest energy that has a resolvable energy spectrum for protons (oxygen ions) is 40 keV (150 keV).

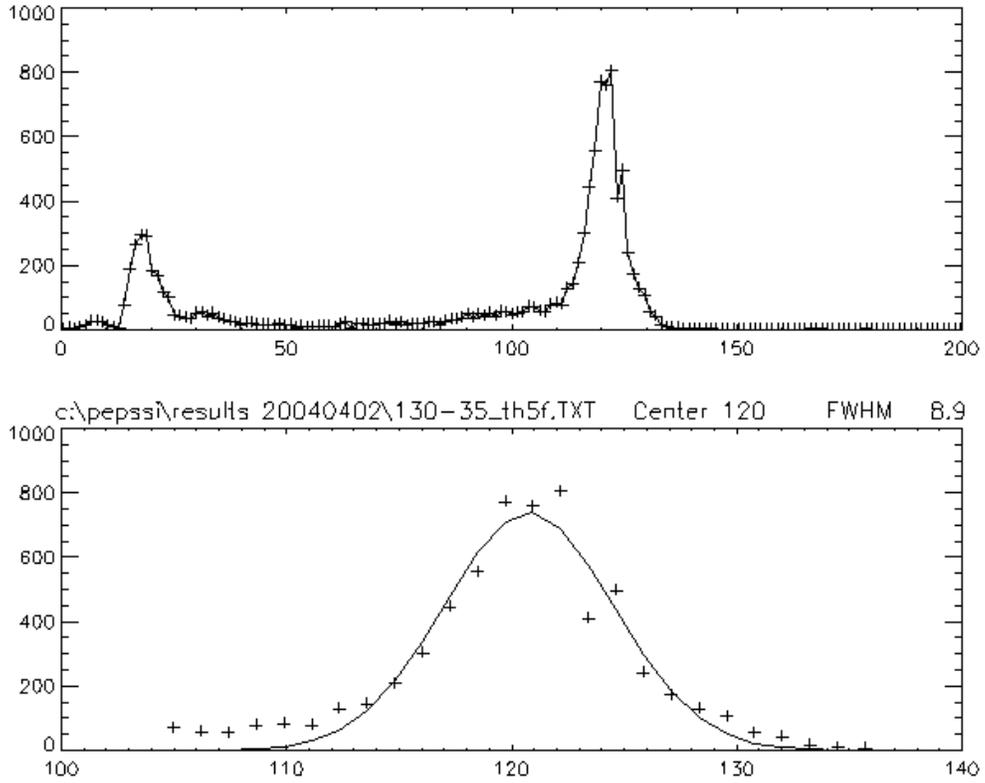

**Fig. 27.** The figure shows the energy spectrum of a 130 keV-proton beam measured by PEPSSI. The upper panel shows the energy spectra from 0 to 200 keV, and the lower panel is a close-up of the spectrum. The measured energy peak is at 120 keV and the fitted FWHM is 8.9 keV in excellent agreement with the simulation.

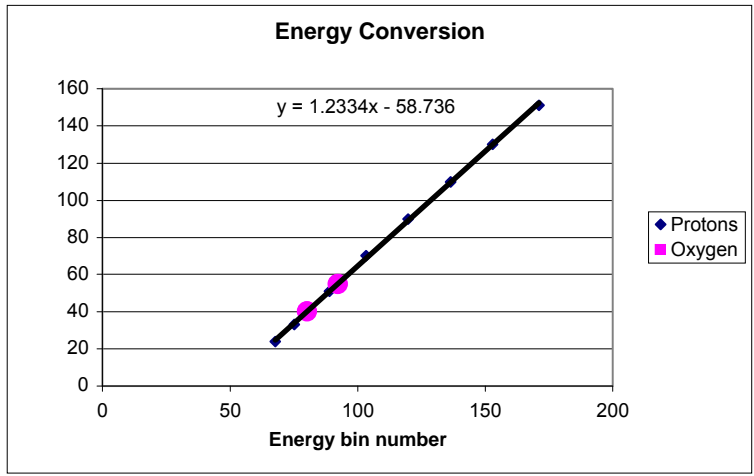

**Fig. 28.** Conversion coefficients for relating energy bin number into actual energy in keV. The coefficients are slightly detector chain dependent, however, the variation are ~ ±2%.

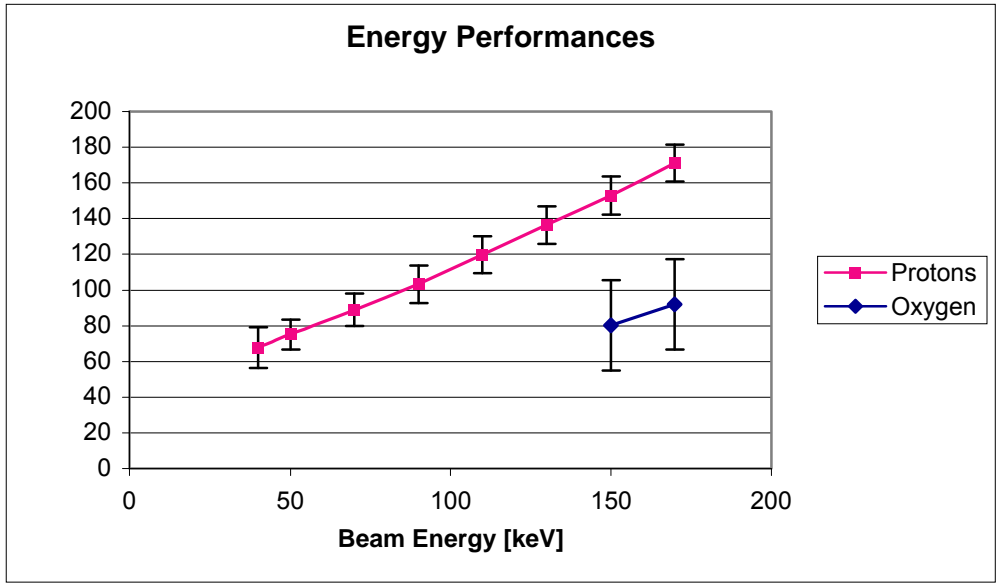

**Fig. 29.** Energy performance of all beam runs that were performed during the PEPSSI flight unit ground calibration. The lowest energy spectrum that was observed for protons (oxygen ions) is 40 keV (150 keV).

*4.3.3.4 Time-of-Flight versus Energy*

Events that have both TOF and energy measurement are called triple events. Depending on the ion mass, each species will have distinct track in a TOF vs. E plot. Figure 30 shows the 100 keV and 170 keV proton beam and the expected location of the track through simulation. Both energies fall exactly where we expected them to be, and Fig. 31 shows similar results from oxygen beams. Multiple energies and species were calibrated, and their combined results are shown in Fig. 32. Typical TOF (Energy) FWHMs are 2, 3.5, and 4.7 ns (10, 15, and 25 keV) for H, He, and O respectively.

Degraded radioactivity source provides a broad energy spectrum that is difficult to reproduce using a mono-energetic accelerator beam. A Mylar-degraded $^{241}$Am radioactivity source was employed to extend the ~5 MeV alpha line into a broad energy spectrum all the way down to ~100 keV. Figure 33 shows the color spectrogram of the source spectrum as measured by PEPSSI during ground calibration. The y-axis shows the TOF in nanoseconds, while the x-axis shows the measured energy in keV. The spectrum falls exactly on top of the red curve, which is the expected TOF by E curve for helium calculated from the simulation.

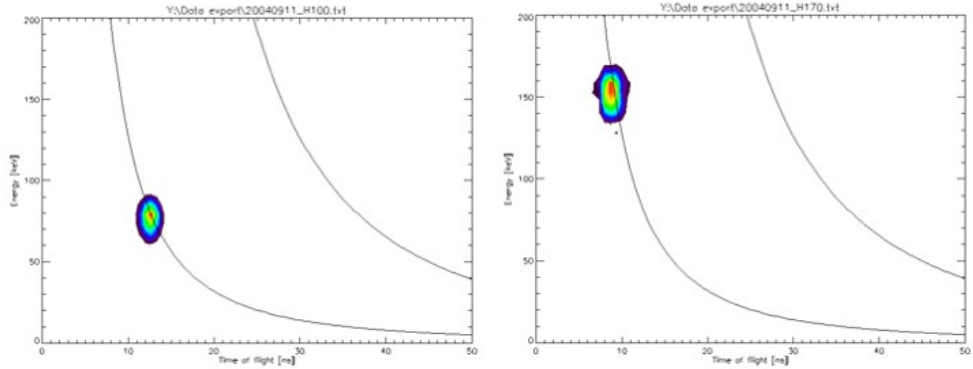

**Fig. 30.** Two proton beam results are shown in this figure in TOFxE plot. The estimated proton and oxygen tracks are shown as a thick line in the figure. The measured TOFs and energies of both proton beams fall exactly on the expected values.

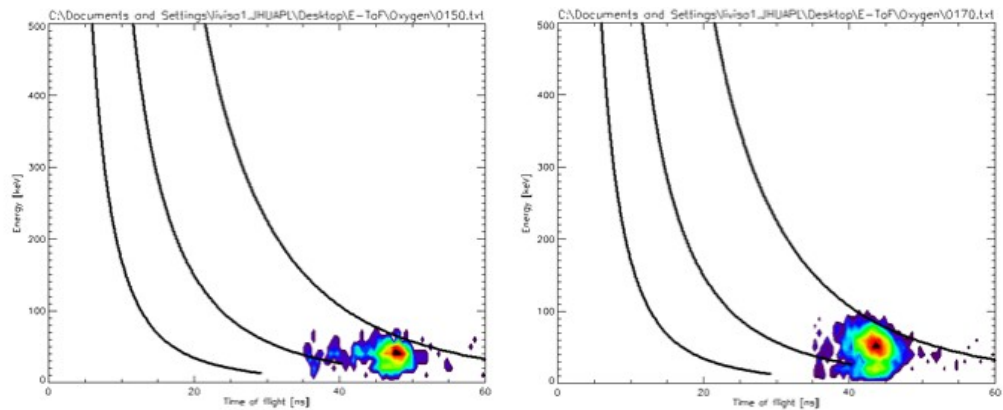

**Fig. 31.** Two oxygen beam results are shown in this figure in TOF x E plot. The estimated proton, helium, and oxygen tracks are shown as thick lines in the figure. The measured TOFs and energies of both oxygen beams fall lower than on the expected values because the expected tracks were calculated based on a thinner stop foil.

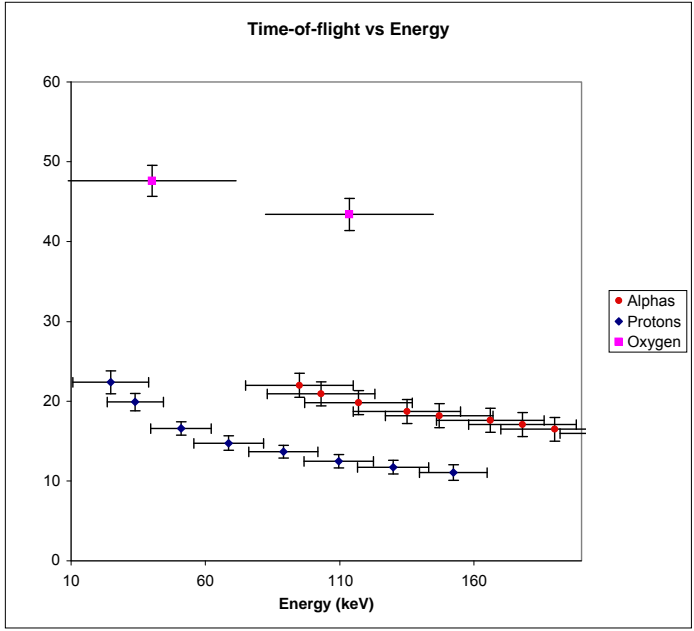

**Fig. 32.** This figure shows combined results from all the beam runs performed to characterize PEPSSI performance. All parameters met the flight requirements.

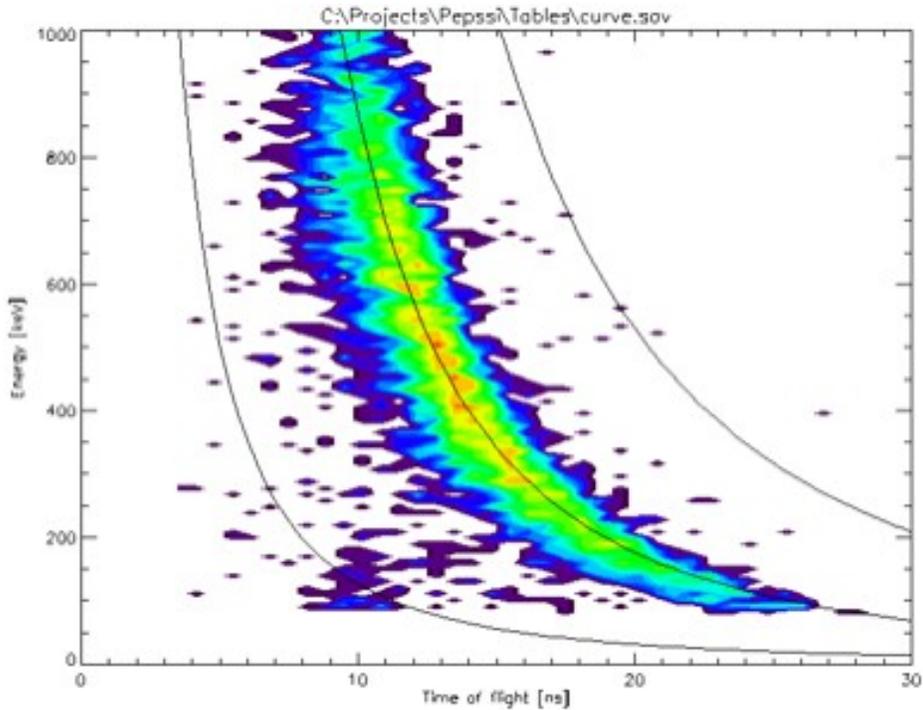

**Fig. 33.** Alpha-particle spectrum from the stimulation of the PEPSSI flight unit by a degraded $^{241}$Am source. The produces a flat energy spectrum in the range of ~100 keV to ~5 MeV.

*4.3.3.5 Pulse Height Analysis*

Due to the limited time available for calibration on the ground, the dependency of pulse height on the particle mass was not calibrated. Characterization of the heavy ion discriminators is being done during the commissioning phase in flight.

*4.3.3.6 Efficiencies*

Figures 34 and 35 show the counting efficiency for both proton and oxygen beams as a function of the voltage across the MCP plates. The expected efficiency for 170 keV protons is 30–40% for the singles rate, and the measured stop foil efficiency is in agreement with that value. However, the start foil efficiency is clearly lower than expected, and it dominates the overall measured efficiency. The expected single efficiency for oxygen is in the 80–90% range. Here the efficiency is dominated by the start foil and does not reach a plateau even for relatively high values of the MCP voltage.

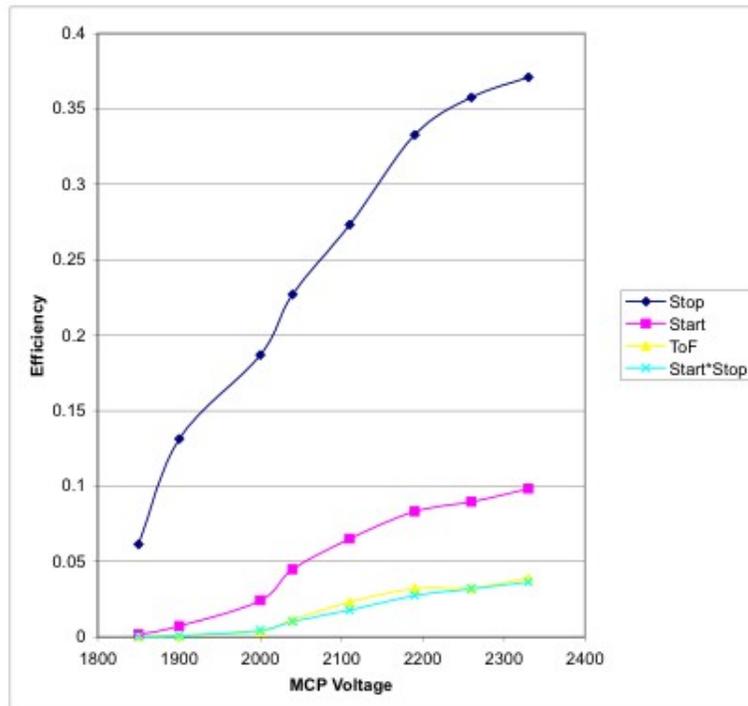

**Fig. 34.** The 170 keV proton efficiency plot as a function of the MCP bias voltage (across the plate). Expected efficiency for proton is 30–40%. Start foil efficiency is clearly lower than expected, and dominates the overall efficiency.

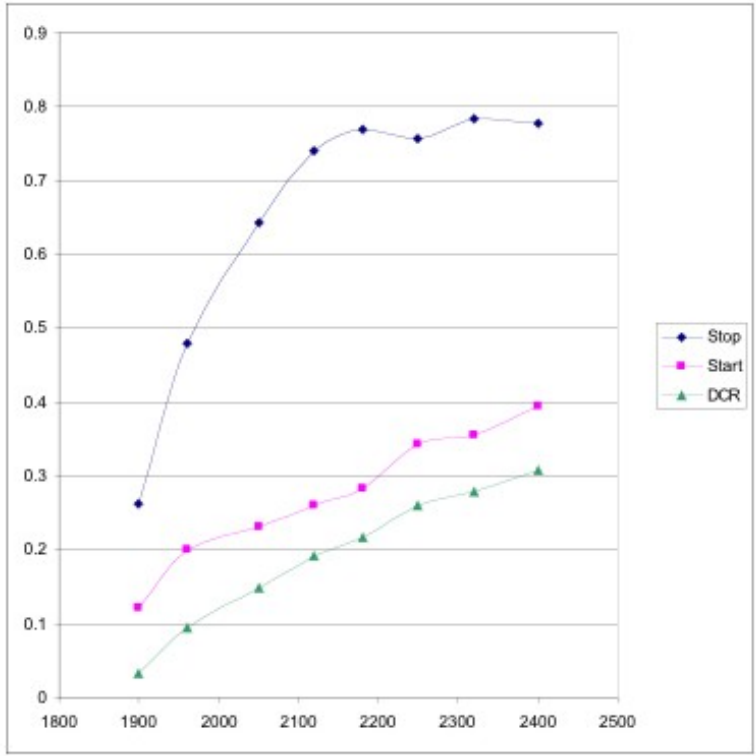

**Fig. 35.** The 170 keV oxygen efficiency as a function of the MCP bias voltage (across the plate). Expected efficiency for oxygen is 80–90%. Here the efficiency is dominated by the start foil efficiency, and does not reach a plateau even for relatively high values of the MCP voltage.

4.3.4   Spacecraft Integration and Test Calibrations

Additional testing with degraded alpha sources was performed during the instrument thermal vacuum tests. The goal of the test was to characterize PEPSSI performances over a wide range of temperatures. Figure 36 shows the internal radioactive calibration source during the thermal test.

4.3.5   Summary of PEPSSI Flight Model Calibrations

Tables III and IV summarize the results of the PEPSSI flight model ground calibration. Ion-beam calibrations of the flight PEPSSI instrument demonstrate it met most of the flight requirements. Most notably, the measured energy resolutions are in agreement with the expected simulated performance. The measured 2.5 to 3.5 ns TOF (FWHM) resolution is better than anticipated (probably through better than expected electronics performance).

The major outstanding issues following calibration of the PEPSSI flight model include: 1) lower than expected ion counting efficiencies, and 2) higher energy threshold of heavy ions. Major progress has been made in addressing in both issues with the engineering model.

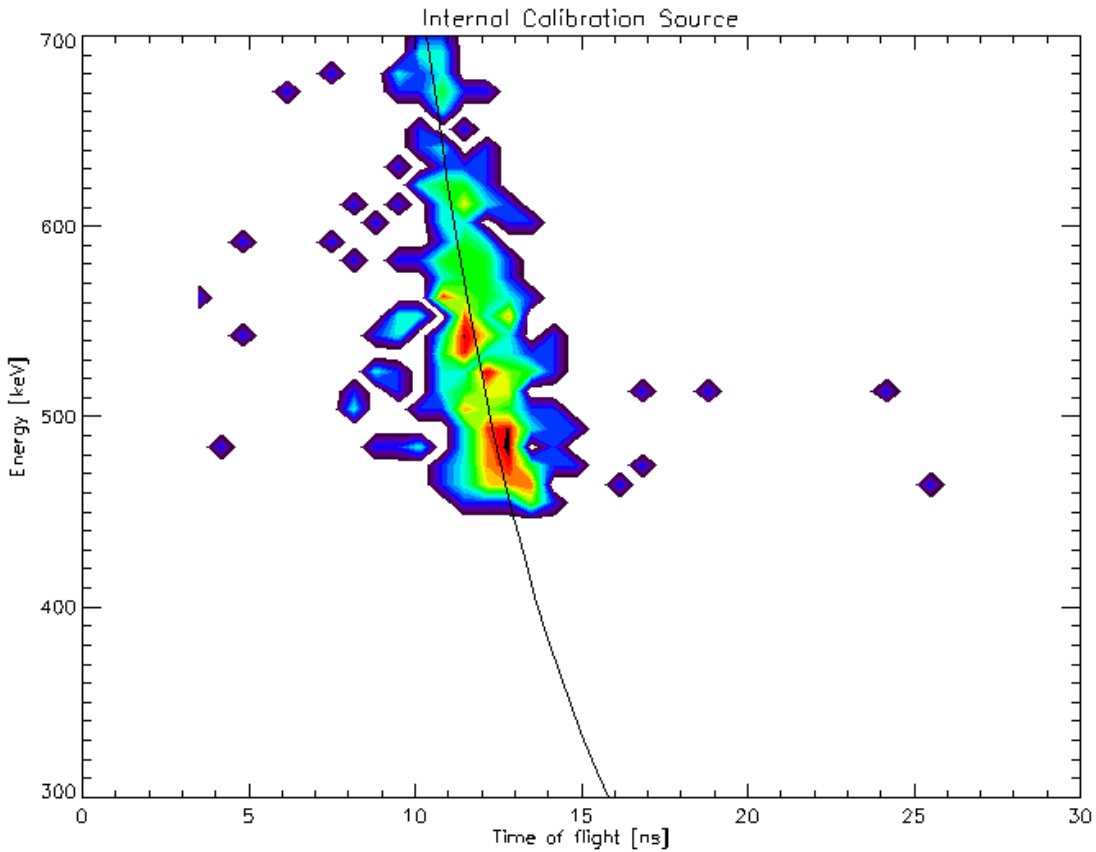

**Fig. 36.** PEPSSI flight unit response to the internal radioactive calibration source during the instrument thermal vacuum test.

### 4.4. PEPSSI Engineering Model Calibrations

After the ground calibration and completion of all environmental and acoustic tests, PEPSSI was mounted on the spacecraft in December 2004. As noted previously, PEPSSI uses thin foils to generate timing signals to gate the time needed for an ion to transverse a known path. The flight unit has a thicker back foil than the original design. The change was due to the fact that during the initial ground calibration on the PEPSSI engineering unit, we observed a lower than expected efficiency on the instrument TOF system. Subsequently, the problem was traced to both front and back foils being used on PEPSSI.

Replacement of both front and back foils before instrument delivery to the spacecraft was not possible; hence, a stopgap measured was taken. A flight-qualified MESSENGER Energetic Particle Spectrometer (EPS) back foil (Andrews *et al.* 2007) was located and installed in the PEPSSI flight unit as the new back foil. It is this combination of PEPSSI original front foil and EPS back foil that has gone through ground calibration and is in the PEPSSI flight unit.

### 4.4.1 Original PEPSSI Foils

The original PEPSSI design uses aluminum-coated polyimide as the material for both the front and back foils. Both foils are made with 300-Å polyimide as support structure and then coated with 50 Å of aluminum on both sides to make it conductive. The total foil thickness is equivalent to 7 $\mu m/cm^2$. With both aluminum front and back foils, the expected TOF efficiency of a 170 keV proton beam was expected to be ~10%. However, less than 1% was measured on the engineering unit.

The discrepancy between the measured and expected efficiency is hypothesized to be related to an oxidation layer on the thin aluminum coating on the foil. When aluminum is exposed to air, an oxide layer immediately forms. This oxide layer can be up to tens of Å in thickness, the same order of thickness as the aluminum coating. With such a relative thick oxide layer, the foil becomes an insulator, which affects both i) the secondary electron production at the surface; and ii) the electrostatic optics within PEPSSI. These effects will be more pronounced for backward scattered (those secondary electron that travels opposite from the ion beam) electrons. We believe this is the reason why the measured stop efficiency is much lower than the expected value.

### 4.4.2 PEPSSI Start and EPS Stop Foil

The decision was made to replace the PEPSSI flight unit back foil with a spare MESSENGER EPS back foil, which is constructed with 500 Å of polyamide and 100 Å of palladium, while retaining the aluminum-coated front foil. This combination of foils has been calibrated fully in the flight unit that was then mounted on the New Horizons spacecraft in December 2004. The measured TOF efficiency of this combination for 170 keV protons is ~4%, which is dominated by the lower-than-expected start-foil efficiency since we still retain the original PEPSSI aluminum start foil.

### 4.4.3 Engineering Model Plans

Calibration and characterization of the PEPSSI instrument is an ongoing activity that has two major thrusts: 1) continued characterization and testing of the flight-like engineering model, and 2) extensive in-flight calibrations.

The engineering model is essentially complete with a full complement of SSDs. Preliminary testing has already begun with one SSD pair to verify the carbon foil performance. After the full complement of SSDs are installed on the unit, it ill then be extensively exercised in both the ball jar vacuum chamber with radiation sources and with accelerator beams to help determine the right set of parameters for optimum efficiency, TOF and energy measurement performance (MCP voltage, CFD bias, CFD discriminators, Anode discriminators, etc). In addition, to characterize better the relative efficiencies of different mass species, work will investigate possible cross-talk in the valid-event identification circuitry that may occur when multiple energies and species are measured at the same time.

### 4.5. In-flight Calibration

Energetic particles in interplanetary space will be used to test the mass classification algorithm and verify the content of the various look-up tables. Because the interplanetary environment is variable, it is highly desirable to have the PEPSSI instrument operating substantially longer than that planned for the checkout period. Interplanetary energetic particles tend to vary on the time scale of a solar rotation: ~27 days. It is highly desirable to operate the PEPSSI sensor for at least 27 days during each yearly checkout prior to the Pluto flyby in order to optimize the Pluto encounter measurements. The Jupiter flyby provided a key opportunity for checking out multiple instrument features as noted below.

### 4.5.1 Flight Performance

Following the launch of New Horizons on 19 Jan 2006, PEPSSI commissioning began with initial checkout on 20 Feb, a month later. Activities of spacecraft instruments are keyed to a Science Activity Plan (SAP) number. A total of 23 SAPs were executed in 2006. These included tests, the one-time door opening (3 May 2006), determination of thresholds, detector configuration, efficiency tests, and mapping of the orientation of the sensor head by monitoring when the Sun was in the sensor field of view. These tests are listed in Table XXVIII.

**Table XXVIII. PEPSSI Commissioning Activities**

| SAP No. | Stol / Seq | Date Performed | Title | Description | Test Successful? |
|---|---|---|---|---|---|
| 014 | Stol | 20-Feb-06 | Initial checkout / functional | Run same test as pre launch ground CPT to verify the instrument is fully operational and has survived launch | Yes |
| 004 | Stol | 22-Feb-06 | Table load | Load new science classification tables | Yes |
| 005 | Stol | 22-Feb-06 | Heater test | Test flight cruise heater | Yes |
| 006 | Stol | 1-Mar-06 | SSD checkout | Power on bias voltage and monitor SSDs for 24 hours to characterize background with doors closed | Yes |
| 012 | Stol | 2-Mar-06 | Power off | Standard power off procedure, used for all commissioning activities. | Yes |
| 016 | Stol | 27-Apr-06 | Table load | Load new science classification tables | Yes |
| 001 | Stol | 2-May-06 | Power on for door opening | Power on and establish SSD background counts to observe difference when door is opened. | Yes |
| 007 | Stol | 3-May-06 | Open door | Open door, and verify with SSD counts, S/C primary current circuit (fire twice), S/C inertial data, and SDC sensor | Yes |
| 008 | Stol | 16-May-06 | Initial HV ramp | Ramp up HV to 2000V, and manually verify current and voltage after each step | Test aborted due to Sun in FOV, only ramped to Sun safe levels. This was expected before running the test |
| 008c | Stol | 6-Jun-06 | Load HV macros | New macros step HV in controlled sequence so a S/C reset can not occur between sending a HV command and re-enabling HV safing | Yes |
| 008a | Stol | 6-Jun-06 | Initial HV ramp | Repeat 008 with new macro and S/C pointing so sun is not in FOV | Yes |
| 017 | Stol | 21-Jun-06 | 2000V threshold scan | Determine optimal thresholds for most channels at 2000V, also stayed on for 24 hours | Yes |
| 016 | Stol | 13-Jul-06 | Table load | Load new science classification tables | Yes |
| 018 | Stol | 14-Jul-06 | Sun scan test | Scan when the sun is in FOV by monitoring count rates and current when S/C is in spin mode | Yes |
| 019 | Stol | 14-Jul-06 | MCP plate scrub | Power MCPs to low voltage with sun in FOV for 2 weeks to scrub plates | Yes |
| 018 | Stol | 27-Jul-06 | Sun scan test | Scan when the sun is in FOV by monitoring count rates and current when S/C is in spin mode | Yes |

| SAP No. | Stol / Seq | Date Performed | Title | Description | Test Successful? |
|---|---|---|---|---|---|
| 015 | Stol | 11-Aug-06 | Load macro 11 | Load macro 11 to fully initialize PEPSSI at 2000V, use data from 017 to set all thresholds | Yes |
| 018 | Stol | 11-Aug-06 | Sun scan test | Scan when the sun is in FOV by monitoring count rates and current when S/C is in spin mode | Yes |
| 018 | Seq | 26-Aug-06 | Sun scan test | Scan when the sun is in FOV by monitoring count rates and current when S/C is in spin mode | Yes |
| 020 | Seq | 8-Sep-06 | 2150 V Threshold Scan | Found optimal thresholds for all channels at 2150V | Yes |
| 022 | Seq | 9-Sep-06 | Enhanced Threshold Scan | Additional scans at 2000V to determine thresholds for some detectors that were not determined during 017 | Yes |
| 021 | Stol | 9/25/06 | Load and test macro 12 | Load macro 12, verify it loaded correctly, then run and stay on for 24 hours | Yes |
| 027 | | 11/23/06 | Efficiency test | | Yes |

PEPSSI was powered on again at 00:00 UTC 6 Jan 2007 and run through 19:20 UTC on 8 Jan; the instrument was powered on again at 00:09 UTC 9 Jan and run for seven and a half hours. On Jan 10 at 08:00 UTC PEPSSI was again turned on and run through 14 Jan 16:30 UTC. Each exercise was used to change instrument configuration for checkouts as Jupiter was approached. From 16 Jan 01:00 UTC through 20 Jan 23:00, 22 Jan 07:00 through 24 Jan 03:30, 24 Jan 06:00 though 31 Jan 05:00, and 31 Jan 07:00 though 10 Feb 12:30 further testing was conducted, including the uploading of new table parameters for the Jupiter flyby.

4.5.2 Jupiter Flyby

Coincident with the Jupiter flyby various setting changes were made to check the instrument response. Early on, PEPSSI high voltage was turned down to protect the microchannel plates from direct sunlight. The flyby provided significant science at Jupiter, especially on the Jovian magnetotail, while also providing additional data on instrument performance and sensitivity. The scope of both efforts is shown in Fig. 37 that shows both energy spectra and tests that were conducted between days 40 and 140 (9 Feb

– 20 May 2007). The instrument was finally fully powered down for the first New Horizons hibernation period on 21 Jun (day 172) at 19:40 UTC. Analysis of both science and instrument calibration data from this period is ongoing.

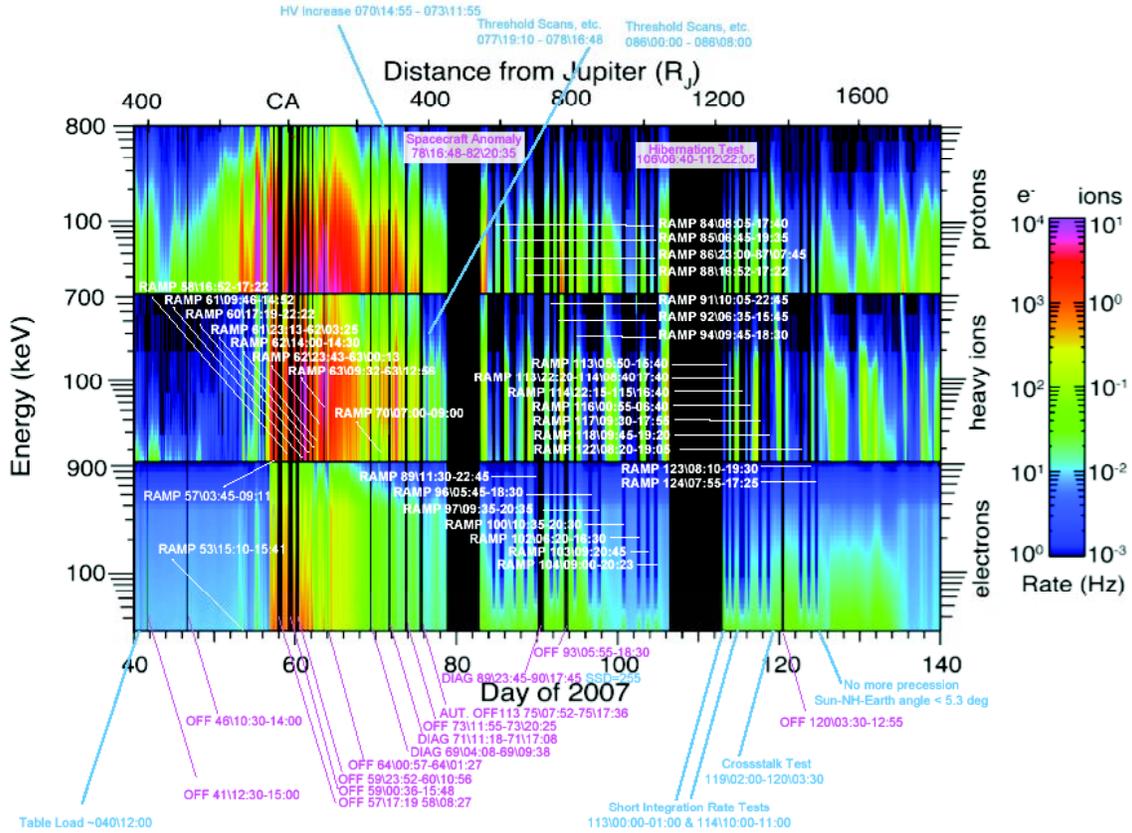

**Fig. 37.** Energy spectrograms from the electron, heavy ion, and proton channels during the Jupiter encounter showing variations from before the last inbound magnetopause crossing until after the first outbound crossing. Labels indicate instrument activity and show the scope of this effort, both for diagnostics and for instrument health (the high voltage on the MCPs was ramped down when sunlight could enter the instrument aperture. This is not an issue at Pluto).

## 5 Operations and Science

### 5.1. Instrument Operations

After power-up and software boot, PEPSSI operates in either one of two operational modes – Standby and Science. Power requirements for each mode differ and depend upon the state of the HVPS as well as the rate at which data is collected. The not to exceed power allocation for PEPSSI is 2.55 watts.

The modes and the current best estimate (CBE) of power for each mode are:

• Boot/Standby: Power on, HVPS disabled (no science data).

    - Housekeeping status and instrument checkout. ----- 1.84 watts

• Science: Power on, HVPS enabled, science data sent to the C&DH.

    - Nominal Event Rate (Pluto flyby)-------------------- 2.18 watts

    - High Event Rate (Jupiter flyby) ---------------------- 2.32 watts

5.1.1    Cover Release In Space

The NH spacecraft was launched into heliocentric orbit with the PEPSSI instrument aperture covers closed. To prevent thermal damage to entry foils in the sensor module, the covers were not to be opened until the distance from the Sun to the NH spacecraft was least 2 AU. With no specific housekeeping flag to indicate the instrument covers are open the bias supply of the HVPS was activated and science data enabled. Open covers are evidenced by detection of background particle energy events. This operation showed that the covers were indeed open, but it also led to the discovery of the misalignment of the instrument on the spacecraft deck.

5.1.2    Idle State (HVPS Disabled)

At instrument power- up, the HVPS is always in the disabled state. The instrument idles indefinitely, waiting for command input from the spacecraft. In this state housekeeping data is output at a preset rate, and output science data packets contain only filler data. Prior to science data collection, the HVPS must be enabled and brought on-line.

5.1.3    HVPS Activation

To activate the HVPS, a preconfigured sequence of commands must be executed to enable HVPS operation, adjust the power supply clock rates, ramp up the output voltages, and set HVPS alarm limits. Once the HVPS is operating and functioning at the appropriate voltage levels, flowed science data is valid.

Following launch into space, prior to first activation of the HVPS, at least 14 days must be allowed to elapse. This is to allow sufficient time for outgassing.

Following HVPS activation, a number of HVPS parameters may be adjusted by command. These include commands to adjust the levels of the SSD bias voltage and the high voltage (HV) outputs. The HV voltages are adjusted during the cruise phase to compensate for known aging of the MCP electron gain characteristics. By adjusting to a slightly higher voltage, constant MCP gain (and hence TOF system characteristics) may be maintained. Similarly, the SSD bias voltage may be periodically adjusted to compensate for SSD aging characteristics.

HV and SSD bias supply clock frequencies may also be adjusted by command. As the HVPS components age, HVPS efficiency will be monitored. Slight adjustments to the HVPS clock frequencies may be made to retain power efficiency.

5.1.4   Science Mode Operation

In Science mode, particle TOF-plus-Energy, electron energy, and TOF-only measurements are made and output on a regular basis. Specific science mode data products are defined above. The rates at which science data products are collected are programmable (by adjusting the data integration time interval) and the range of data rates possible is outlined in prior sections that specify PEPSSI telemetry. Regardless of science data rates, the data collection intervals are synchronized to the MET 1PPS input from the spacecraft.

During PEPSSI science mode operation, the PEPSSI instrument requires few if any commands. However, there are a number of instrument operating adjustments, in addition to the data integration interval, that can be made by command. These adjustments are described below. These adjustments, if required, are typically made during early in-space operations. During cruise mode operations, and prior to any major instrument data collection campaign, instrument operation is monitored to determine if any further adjustments are required.

*5.1.4.1 Energy Leading Edge Discriminator (LED) Threshold*

PEPSSI implements a Leading Edge Discriminator (LED) for each of the 12 SSD energy channels. Only particle energy events that exceed the LED threshold are counted and included in the telemetry science data output. The threshold for each LED is adjustable by instrument command over a range up to 250 keV and to a quantization of 1 keV.

The purpose of the LED is to provide a means to detect particles of minimum energy while minimizing false counts due to system noise. During ground calibration, the LED threshold of each channel is adjusted so that the false count rate is low, much less than one event per second, while maintaining the capability to detect particles at or below the energy thresholds required.

During initial in-space PEPSSI operations, the energy LED thresholds may be tweaked slightly to achieve optimum performance. After the initial tweaking, the threshold of any particular LED will be adjusted only if increased noise is detected in that channel.

*5.1.4.2 Stop Anode Light and Heavy Discriminator Thresholds*

PEPSSI applies two discriminators to the TOF stop-anode signal. The discriminator thresholds are adjustable over a range of signal input levels by instrument command.

For TOF-only measurements, the magnitude of the respective anode signal is a coarse indicator of particle mass. The threshold levels are adjusted so that, for TOF-only measurements, the particles may be differentiated on the basis of light or heavy mass. The threshold level of one discriminator is set to a low value, low enough to detect $H^+$ ions, the other is set to a higher threshold, to detect CNO and higher mass particles. One is referred to as the light discriminator, the other the heavy discriminator.

Due to time constraints, the light and heavy discriminator thresholds were not calibrated and adjusted prior to launch. The thresholds may be adjusted for optimum performance as part of ongoing flight operations. After initial tweaking, it is anticipated light and heavy thresholds will rarely if ever be changed; thresholds will be readjusted only in response to detected changes in TOF system noise or front-end sensitivity.

*5.1.4.3 Constant Fraction Discriminator (CFD) Thresholds*

The composite TOF-start and TOF-stop signals are each passed through a CFD to generate digital level signals. Digital level TOF-start and TOF-stop pulses are required for input to the TOF chip. The CFD thresholds are adjustable over a range of values by instrument command; only TOF anode signals that exceed the CFD thresholds will result in a TOF measurement.

The CFD thresholds determine the lower signal threshold for TOF measurements. It is set during ground calibration to as low a threshold possible consistent with acceptable false TOF detection rates due to TOF system noise. During ground calibration, the threshold of each CFD was adjusted so that the false TOF event rate is low, much less than one event per second, while maintaining the capability to detect particles at or below the energy thresholds previously. During initial in-space operations and checkout of PEPSSI, the CFD thresholds were tweaked to achieve optimum performance. After the initial tweaking, the CFD thresholds will be adjusted only if increased noise is detected in the TOF system.

*5.1.4.4 Start Anode Discriminator Thresholds*

The PEPSSI design includes leading edge discriminators for each of the six start anode signals. The discriminator thresholds are adjustable over a range of values by instrument command. The discriminator outputs are used to determine the direction (i.e., sector) of the respective TOF measurement.

Only TOF anode signals that exceed the discriminator threshold will provide a signal output indicative of particle direction. During ground calibration, the threshold of each discriminator is adjusted to be consistent with CFD thresholds. During in-space operations, adjustments may be required if the CFD thresholds are changed.

*5.1.4.5 Multiple Hit Check Enable/Disable*

As heritage from the MESSENGER energetic particle detector design, PEPSSI includes a multiple-energy-event detect/reject feature. This feature, enabled by command, causes the instrument to reject (not count) multiple energy events that occur closely in time. Multiple events occurring in close proximity may cause degradation in the accuracy of energy measurements, and the intent is to provide a means to prevent these types of events from being counted. If this feature is disabled, and multiple energy events occur closely in time, the multiple events are counted as a single event.

Similar to the PEPSSI multiple energy event detect/reject feature, PEPSSI also includes a capability to reject multiple TOF start anode events that occur closely in time. This capability may be enabled or disabled by instrument command. Start pulses occurring closely in time can degrade TOF measurements; again the intent is to provide a means to prevent 'compromised' measurement events from being counted.

In the Pluto environment, it is expected the particle event rate will be low, on the order of 10 to 100 events per second. The probability of two events occurring closely in time is therefore very low. Given the expected low particle event rate near Pluto, and therefore the increased value of individual particle events, it is currently planned that multiple hit check features will be disabled during Pluto operations.

5.1.5 Power Down Operation

PEPSSI is designed so that power to the instrument may be switched off while the HVPS is operating. No damage to the instrument can occur. However, for normal operations, the instrument HVPS is ramped down in voltage and turned off in a controlled sequence prior to switching off power to the instrument.

**5.2. Data and Data Archiving**

The New Horizons Science Operations Center (SOC) is part of the ground system that processes data returned from the New Horizons planetary spacecraft. Data down-linked from the spacecraft in raw packetized form are retrieved by the SOC from the Mission Operations Center (MOC) along with navigation and related ancillary data. The SOC generates the higher level data products used by the instrument teams and science teams. In addition, the SOC performs archiving of data to the Planetary Data System (PDS). The science data processing component of the SOC is called the SOC pipeline.

The SOC pipeline is divided into three main parts: the Level 1 pipeline segment, the Pre-Level 2 pipeline segment, and the Level 2/3 segment. Pipeline processing is carried out sequentially. Results of the Level 1 pipeline are provided as inputs to the instrument Pre-Level 2 pipeline segment. The Level 2/3 segment combines the pre-Level 2 temporary data files into easily used (Level 2) and science/calibrated (Level 3) formats. The instrument pipeline generates Level 2 and Level 3 results that the SOC forwards to the PDS archiving process. All levels of science data files (Level 1, Pre-Level 2, Level 2, and Level 3) are stored in FITS format.

### 5.2.1 Level 1

The Level 1 data product is a FITS format data file and all data is contained in FITS extension Header Data Units (HDUs). Each HDU contains a PEPSSI science data telemetry block. The non-status HDUs are stored as 1-D FITS images containing the binary data of a decommutated science data telemetry block. Status extensions have been "unrolled" into a normal 2-D FITS binary table with columns for each uncalibrated status quantity and a separate row for each measurement time. Level 1 files are not present in the PDS archive. Data from different telemetry classifications is placed in different files (Table XXIX).

**Table XXIX. Data File Layout**

| File Type | HDU Type |
|---|---|
| N1 | Primary HDU: High Priority Telemetry Block |
|  | Extension 1: PHA Telemetry Block |
|  | Extension 2: Status Telemetry Block |
| N2 | Primary HDU: Medium Priority Telemetry Block |
|  | Extension 1: PHA Telemetry Block |
|  | Extension 2: Status Telemetry Block |
| N3 | Primary HDU: Low Priority PHA Telemetry Block |

### 5.2.2 Level 2

The Level 2/3 pipeline is run on the SOC processing station to transform Level 1 decommutated data into Level 2 and Level 3 calibrated science data. The instrument pipeline creates PDS standard, Level 2 and Level 3 provisional products in FITS format.

The Level 2 files "unroll" the data in the Level 1 files into flat, single datum per cell, tables that are easily human readable using standard FITS tools like fv (http://fv.gsfc.nasa.gov).

In the Level 2 files, all of the data for a single UTC day of observation is present in a single file. Each type of PEPSSI data: High Priority Rate Data (N1), Low Priority Rate Data (N2), Status Data, and the various forms of PHA data, in both normal and diagnostic modes is placed in a different FITS binary table with associated header keywords to create a separate FITS extension or header data unit (HDU).

*5.2.2.1  Summary of the Level 2 Header Data Units (HDUs)*

In PEPSSI Level 2 files, each HDU represents a different type of data, there is only one HDU of each type (EXTNAME).  An HDU will only be present if there is data of that type taken during the time period covered by that file.  Each file contains exactly one UTC observing day worth of data.  The Primary HDU contains no data, only informational header keywords identifying mission info, observational start time, and information about the file creation (date, software version, etc.).

The available HDUs are:

D_N1, D_N1_STATUS, D_N2, D_N2_STATUS, N1, N1_STATUS, N2, N2_STATUS, PHA_DIAG, PHA_ELECTRON, PHA_LOW_ION, PHA_HIGH_ION

The different extensions are described below.

When a charged particle enters the PEPSSI detector, we measure an Energy and/or a Time of Flight (TOF).  Since we don't have enough bandwidth to telemeter all of our events, we use a round robin priority scheme to decide which PHA events to discard and which to telemeter.  All events are counted in various histograms in the Rate data.

*5.2.2.2  PHA HDUs*

The four PHA HDUs: PHA_DIAG, PHA_ELECTRON, PHA_LOW_ION, and PHA_HIGH_ION contain columns for the end time of the accumulating interval, the uncalibrated Energy and Time of Flight values and the detectors involved in measuring a given event (e.g. which solid state energy detector fired and which start anode fired).

Each row represents a separate charged particle event.

*5.2.2.3  Rate HDUs*

The N1 and N2 (and D_N1 and D_N2) extensions contain several types of "Rate" data. The Rate data is accumulated in histograms which are then dumped at set intervals. For N1 data, usually the histograms are accumulated for 600 seconds. For N2 data, the accumulation time is usually 60 seconds except for the first hour of the day when it is 15 seconds. This can be changed, but during normal observing it is almost always true:

B Rates:  The number of high energy ion events in the various "Rate Boxes". For example: a Rate labeled B01S03 represents Protons which deposited between 60 and 94 analog to digital units (ADUs) of energy in the solid state detector.

> C Rates:  The contents of various hardware counters
>
> HK Rates: Various housekeeping quantities such as power levels  and descriminator thresholds
>
> J Rates:  Software counters that represent overall quantities like total number of Electron Events.
>
> L Rates:  The number of low energy (TOF-only) ion events sorted into Boxes.
>
> R Rates:  The number of electron events.

All of the rate column contents are specified in detail in the comment field of their respective TFORM keywords in the FITS Header.

The N1 and D_N1 data are identical in format, the D_N1 data is merely taken when the instrument is in diagnostic mode. The definitions of some of the Rate Boxes are different in diagnostic mode and normal mode. N2 (and D_N2) are identical to their N1 counterparts except that they are typically sampled much more frequently (every 15 or 60 seconds) and only a subset of the L and C rates are present.

*5.2.2.4  Status HDUs*

All the STATUS HDUs contain the same quantities for their respective coverage periods and are the same values as in the Level 1 files calibrated to physically meaningful values.

The Level 2 data files represent all of the scientifically useable data from the PEPSSI instrument and are meant to be the starting point for science analysis of that data.

### 5.2.3 Level 3

The Level 3 data is a calibrated, scientifically useful subset of the Level 2 data. As with the Level 2 data, each Level 3 file covers all of a single day of observation. There are three basic types of data in the L3 files: Quick-Look, flux-calibrated Rate Data, and calibrated PHA data. No Diagnostic mode data is present in the L3 files.

The Level 3 files are meant to be, as much as possible, self-documenting. All calibration constants, calibration formulas, and physical units are present in the FITS header in an easily readable format.

The image in the primary array of the L3 file is a rate-weighted 2-D histogram of the PHA data for that day binned in calibrated deposited energy. The priority scheme distorts ion abundances, so we correct for that by using a "rate-weighted" rather than a single count histogram.

The next five HDUs: SPEC_Protons, SPEC_Helium, SPEC_Heavies, SPEC_Electrons, SPEC_LowIon, contain quick-look spectrograms of their respective species. These spectrograms present counts/second N2 data, averaged over 60 second intervals and summed over all incidence directions.

The FLUX HDU contains calibrated fluxes, uncertainties, and raw counts/sec rates for all of the High Energy Ion and Electron N2 Rate data. There is also an accumulation time column (**DT**) and three timing columns. Separate calibrations are given for different ion species for some of the rate boxes if the composition in that rate box is complicated (e.g. both oxygen and sulfur in a single box).

The three PHA extensions: PHA_ELECTRON, PHA_LOW_ION, and PHA_HIGH_ION contain the PHA event data telemetered in the N2 data. Each row represents a single PHA event. Cross-talk events are excluded. Quantities of limited usefulness (such as Heavy Ion Discriminator triggers) are excluded. Calibrated Deposited Energy and/or Time of Flight values are given. The linear calibration constants and formulas are in the FITS headers. A **Speed** column is calculated from the Time of Flight. The Rate Box classification for each event is given in the **Rate_Box** column.

The PHA_HIGH_ION HDU contains additional columns: **H_Incident_Energy, He_Incident_Energy, O_Incident_Energy,** and **S_Incident_Energy** columns contain the calculated Incident energy assuming that the event is of that (H, He, O, or S) species. The Rate_Normalized_Weight column has removed Priority Group artifacts from the PHA data. This column is usually used in making histograms of the High Energy Ion PHA data.

5.2.4   MIDL

PEPSSI data can be viewed using a software suite known as the Mission Independent Data Layer (MIDL). A full complement of analysis tools is available: line plots of individual particle rate channels, energy spectra and spectrograms, and histograms (one-dimensional and two-dimensional) of the PHA event data. MIDL also incorporates and makes available spacecraft ephemeris and pointing information vital to interpreting the *in situ* PEPSSI data. Each of the plotting tools also serves as a data access mechanism, because the data values used to make any plot can be extracted and saved separately. All discovery, downloading, and local caching of science and ancillary data are handled automatically, providing a seamless view of the entire PEPSSI dataset.

MIDL reads the Level 2 (or Pre-Level 2) data in FITS format as produced by the SOC, and applies the appropriate calibrations, allowing the user to select from a variety of units in which to view the data – everything from raw counts to fully calibrated intensity. Detailed calibration options are available for members of the instrument team, with reasonable defaults provided for non-expert users.

Created prior to PEPSSI, MIDL uses a modular approach in which plotting tools are isolated from dataset-specific read routines. New datasets can take advantage of the existing plotting tools by simply providing the appropriate set of read routines. Most plotting tools used by PEPSSI existed in MIDL prior to their use with PEPSSI. The exception is the PHA histogram tools, which were specifically developed for PEPSSI (but could also be used by other missions with PHA data). The original version of MIDL was funded as a research effort by NASA's Applied Information Science Research (AISR) program but has now been used as a primary analysis suite for three energetic particle instruments (others are Geotail/EPIC and Cassini/MIMI).

Written in Java, MIDL can be launched (via Java Web Start) as a client program on any modern operating system from any web browser.

## 6 Conclusion

The PEPSSI instrument on the New Horizons spacecraft will measure the outermost reach of Pluto's escaping atmosphere at the time of the flyby in 2015. Recent observations (Elliot *et al.* 2007) suggest that the atmospheric escape rate may not be significantly diminished from its value near Pluto's perihelion in 1988. Observations at comets and at Jupiter with PEPSSI indicate that the interaction region and pickup ions from Pluto's atmosphere should be easily discernable and diagnostic of the magnitude and extent of the interaction of Pluto with the solar wind.

The PEPSSI instrument is another step in an evolving suite of combined energy/time-of-flight instruments for probing the properties of multispecies energetic particles across the solar system, and, potentially, beyond. The origins of the these capabilities in a compact, low-mass, low-power design trace back to early planning for the inclusion of particle instruments on a Pluto flyby mission. The development and inclusion of PEPSSI on this mission to Pluto and the Kuiper Belt objects beyond can also be seen as a significant achievement of NASA's PIDDP efforts. With the successful measurements made near Jupiter and down the Jovian magnetotail, the instrument is fully checked out and calibrated for use during the encounter with Pluto in 2015.

With a trajectory near the same ecliptic longitude of Voyager 2, New Horizons has the capability to provide a platform for monitoring the solar wind flow and energetic particle propagation toward the heliopause during the regular annual checkouts and comparing conditions with those at Voyager 2 (Fig. 38). In addition to probing the puzzle of Pluto's escaping atmosphere and its interaction with the solar wind in 2015, as well as potentially looking near a Kuiper Belt Object for cometary activity, PEPSSI will thus continue our research into how the solar system interacts with its environment of the very local interstellar medium.

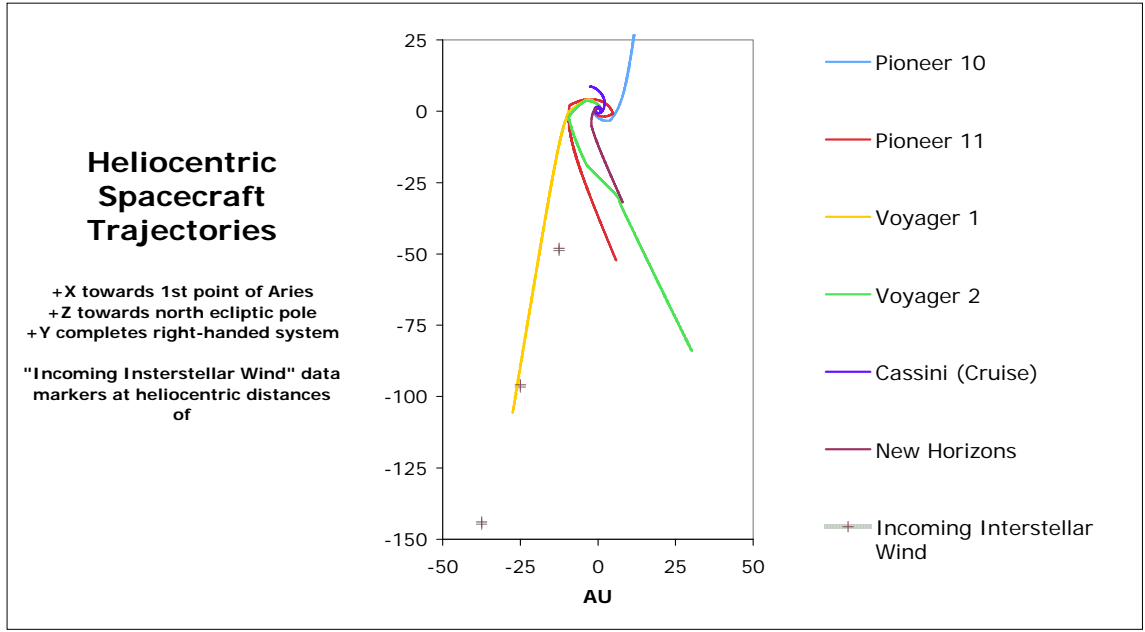

**Fig. 38.** Trajectories of all solar-system escaping spacecraft as projected into the plane of the ecliptic along with that for Cassini during its cruise to Saturn (for scale). Voyager 2 and New Horizons are escaping the solar system at roughly the same longitude and close to the flow direction of the very local interstellar medium.

# 7 Acknowledgements

We would like to thank the support of the engineering support and technical services and support staffs at APL that made the design, fabrications, and the delivery of the PEPSSI instrument to the New Horizons spacecraft possible. This work has been supported as part of the New Horizons effort under NASA Contract NAS5-97271.